\documentclass[letterpaper,12pt]{article}
\usepackage{graphicx} %
\usepackage[left=1in,right=1in,top=1in,bottom=1in]{geometry}
\usepackage[doublespacing]{setspace}
\usepackage[english]{babel}
\usepackage[square,numbers]{natbib}
\usepackage{enumitem}
\usepackage{placeins}
\usepackage{adjustbox}
\usepackage{hyperref}
\usepackage{placeins}
\usepackage{adjustbox}
\usepackage{multirow}
\usepackage{array}
\usepackage{makecell}
\usepackage{longtable}
\usepackage{caption}
\usepackage{booktabs}
\usepackage{xurl}

\newcolumntype{C}[1]{>{\centering\arraybackslash}m{#1}}
\newcolumntype{M}[1]{>{\raggedright\arraybackslash}m{#1}}

\title{\vspace{-.5in}\singlespacing LLMs in social services: How does chatbot accuracy affect human accuracy?}
\author{Jennah Gosciak\thanks{J.G. (Author One) contributed equally to this work
with E.G. (Author Two).} \thanks{Department of Information Science, Cornell Tech} \and Eric Giannella\footnotemark[1] \thanks{Better Government Lab, Georgetown University}  \and Zhaowen Guo\footnotemark[3] \and Michael Chen\thanks{Nava Labs} \and Allison Koenecke\footnotemark[2]}
\date{\today}

\begin{document}

\thispagestyle{empty}
\maketitle
\vspace{-.3in}
\begin{center}
\end{center}

\begin{singlespacing}

\begin{abstract}
Social service programs like the Supplemental Nutrition Assistance Program (SNAP, or food stamps) have eligibility rules that can be challenging to understand. For nonprofit caseworkers who often support clients in navigating a dozen or more complex programs, LLM-based chatbots may offer a means to provide better, faster help to clients whose situations may be less common. In this paper, we measure the potential effects of LLM-based chatbot suggestions on caseworkers' ability to provide accurate guidance.
We first created a 770-question multiple-choice benchmark dataset of difficult, but realistic questions that a caseworker might receive.
Next, using these benchmark questions and corresponding expert-verified answers, we conducted a randomized experiment with caseworkers recruited from nonprofit outreach organizations in Los Angeles.
Caseworkers in the control condition did not see chatbot suggestions and had a mean accuracy of 49\%. Caseworkers in the treatment condition saw chatbot suggestions that we artificially varied to range in aggregate accuracy from low (53\%) to high (100\%).
Caseworker performance significantly improves as chatbot quality improves:
high-quality chatbots (96-100\% accurate) improved caseworker accuracy by 27 percentage points.
At the question-level, incorrect chatbot suggestions substantially reduce caseworker accuracy, with a two-thirds reduction on easy questions where the control group performed best (without chatbot suggestions).
Finally, improvements in caseworker accuracy level off as chatbot accuracy increases, a phenomenon that we call the \emph{``AI underreliance plateau},'' which is a concern for real-world deployment and highlights the importance of evaluating human-in-the-loop tools with their users.
\end{abstract}

\section{Introduction}
Social service delivery in the United States (U.S.) encompasses a range of government programs -- such as the Supplemental Nutrition Assistance Program (SNAP) and Medicaid -- which provide low-income households with access to critical resources (such as food and healthcare).
Incorporating Large Language Models (LLMs) into systems offers a broad array of opportunities for improvements to social service delivery, from automation of certain manual tasks to answering questions on complex government policies.
Still, the trade-offs of many of these applications remain unknown or difficult to quantify.
Governments and nonprofits seeking to adopt LLM tools for public-sector applications need to sort through performance claims, which are often based on automated benchmarking that may bear little resemblance to real-world tasks~\citep{hauser2025evaluating}.

Governments are already adopting LLM-tools for a variety of purposes -- from translation~\citep{machine_translation_nj} and summarization~\citep{poppy_ca} to agency-specific chatbots~\citep{dhs_chatbot, fda_chatbot}. While some government officials advocate for deploying this technology widely in the public-sector~\citep{propublica_article, ca_press_release, federal_ai_action_plan, federal_govt_memo}, there have also been documented negative examples: failed attempts to modernize eligibility processes~\citep{eubanks2018automating}, refusal to use LLMs for translation given concerns related to accuracy and quality~\citep{bay_area_translation}, and several high-profile cases of LLM hallucinations in policy and legal research~\citep{magesh2025hallucination, federal_govt_memo, deloitte_report}. 
In this work, we focus on the use of LLM-based chatbots for answering questions related to SNAP (formerly known as food stamps), a U.S. government program that provides low-income households with financial assistance for food purchases.
Households qualify for SNAP by completing a lengthy application questionnaire and meeting eligibility criteria that they must verify with documentation.
Since government policy rules can be complex and challenging to navigate, potential applicants often seek out support from \emph{caseworkers}, professionals who have broad knowledge of these programs~\cite{nava_casestudy}.

A key uncertainty is whether, if used as intended, LLMs would increase the quality and reliability of services that are delivered by caseworkers to SNAP applicants (who we refer to henceforth as ``clients''). 
This study contributes a principled approach to defining and quantifying these trade-offs.
We examine whether chatbot suggestions of varying accuracy improve caseworker accuracy on a set of realistic benchmark questions.
Our approach can be used to evaluate LLM-based tools in other industry and public sector settings where trained professionals, such as caseworkers, may reasonably turn to LLM tools for support with challenging and unfamiliar tasks.\footnote{Replication materials, including data and code, along with copies of the survey instruments can be found at this anonymous GitHub link \href{https://anonymous.4open.science/r/chatbots-social-services-94EB}{https://anonymous.4open.science/r/chatbots-social-services-94EB}.}

There is a significant body of work studying the performance differences in human-LLM collaborations relative to a human-only or LLM-only baseline, ranging in fields from medicine ~\citep{cabitza2021studying, cabitza2023rams, reverberi2022experimental, senoner2024explainable, liu2025human, bean2025reliability} to software engineering~\citep{vaithilingam2022expectation, campero2022test, mozannar2024reading, peng2023impact, cui2024productivity}. However, little research has focused on human experts in the nonprofit or public sector space.

\begin{figure*}[hbtp!]
    \centering\includegraphics[width=\linewidth]{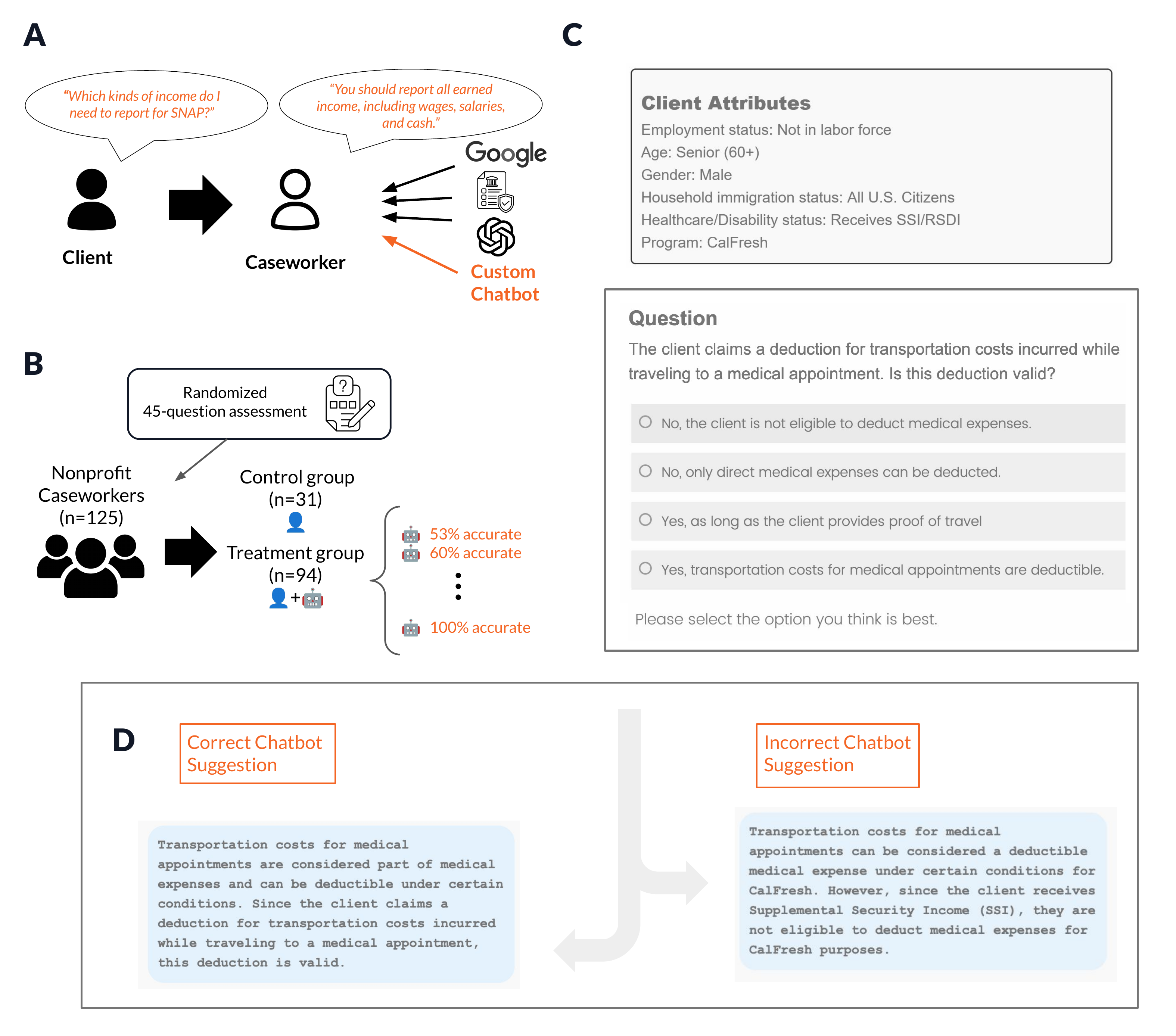}
    \caption{\footnotesize{(A) A typical client-caseworker interaction, in which a caseworker can use several different tools (including Google, policy manuals, ChatGPT, and a custom chatbot) to answer questions related to SNAP eligibility and reporting. We are interested in the effect of custom chatbot suggestions (orange). (B) Experimental design involving a control group which only sees the question text shown in (C) with no chatbot suggestions, and a treatment group assigned to see the questions in (C) with one of the simulated chatbot suggestions shown in (D). Overall chatbot performance levels were artificially varied across treatment arms. There were 10 different levels of aggregate simulated chatbot accuracy ranging from 53\% accurate (21 incorrect suggestions) to 100\% accurate (0 incorrect suggestions). (C) Example of an assessment question (drawn from questions in the benchmark dataset). Questions consist of a client background panel (Top) with five different client attributes that can have up to two distinct values and a multiple-choice question (Bottom) with four or five multiple choice options.
    (D) Examples of correct and incorrect chatbot suggestions. We can toggle whether a suggestion is correct or incorrect for each question, allowing us to vary the overall chatbot accuracy for a given assessment.
    Additionally, on four assessment questions, like the one shown here, the correct answer depends on a unique combination of client attributes.
    While in this example, the correct answer is that the client can deduct transportation costs for medical appointments, if the client attributes were instead ``Age: Adult (22-59)'' and ``Healthcare/Disability status: Not legally disabled,'' the correct answer would be that the client is ineligible to deduct medical expenses.
    Our approach tests the effect of correct and incorrect suggestions for both versions of the question.
    }}
    \label{fig:contributions_framework}
\end{figure*}

Our study is directly inspired by a real example  of an LLM chatbot
designed to support caseworkers in this setting. 
As part of the recruitment for our study, we partnered with Nava, a Public Benefit Corporation, which developed a custom LLM-based chatbot to assist caseworkers in Los Angeles~\citep{nava_pilot}.
Shown in Figure~\ref{fig:contributions_framework}A, caseworkers represent an important human point of contact in the long and complex SNAP application process. They assist clients with assessing whether it makes sense to apply and may then help them to submit an application. Eligibility rules and application requirements involve in-depth knowledge of federal and state policies, and how they interact with client attributes (such as age or disability status).

While caseworkers may be able to reasonably answer most questions, chatbots can fill an important gap: they can provide caseworkers with support on highly specialized and challenging questions. Unlike eligibility workers in government agencies, who decide whether to deny or accept applications for SNAP, these caseworkers must cover a wide range of social services programs, thereby making it extremely difficult to become an expert in each.  
We focus on caseworkers as our users of interest, and not their clients, for two reasons. First, caseworkers are trained professionals who can prevent inaccurate or harmful answers from reaching downstream clients.
Second, they represent an important part of what is otherwise a fundamentally \emph{human} process; many individuals benefit from human outreach and encouragement to seek available programs and a small share of individuals may even require human assistance to apply at all.

Off-the-shelf LLMs can answer many of the questions clients may have accurately. One example of a question caseworkers might encounter is the following: -- \emph{``Your client mentions they received a cash gift from a family member. How should this be classified in terms of income reporting?''} GPT-based chatbot responses, even with slight variations in prompt and temperature, will generally select the correct response (e.g., \emph{``It should be counted as income''}). However, they could still yield incorrect responses that indicate cash gifts do not need to be reported.

The conventional method for evaluating chatbot performance on SNAP-related questions would be to report chatbot performance on a benchmark dataset, which is often generated for a specific application.
In contrast, we aim to study the efficacy of such chatbots if used \emph{in practice} by real-world caseworkers---which better reflects the human-computer interactions that are socially consequential for these applications. 

To perform our \emph{human-in-the-loop evaluation} of chatbot responses, we first develop a comprehensive and expert-vetted 770-question benchmark dataset for evaluating accuracy with respect to applying SNAP rules. Question topics are based on analysis of SNAP Quality Control (QC) Data,\footnote{\href{https://snapqcdata.net/}{https://snapqcdata.net/}} which is a dataset from the U.S. Department of Agriculture Food and Nutrition Service on SNAP errors. It contains detailed data about cases reviewed from state-level SNAP agencies, and it reflects common errors made in the SNAP benefits process, such as unreported sources of income or misapplied expense deductions.
We generate representative questions by prompting GPT-4 with scenarios, based in part on the SNAP QC data and with input from SNAP QC experts, who were employed as quality control auditors for SNAP in two states (more information on this process is included in the SI Appendix along with the distribution of error categories across questions in our assessment in SI Appendix Table S3).
We then worked with the SNAP QC experts to independently verify answers to questions (each question is confirmed by two different experts).
All questions are focused on California's SNAP program, CalFresh, and are formatted as multiple choice questions with 4-5 answer options (only one of which is correct).
Upon publication, we will publicly release our benchmark dataset.
Further details on this dataset, as well as robustness checks on using multiple choice versus open-ended answers, are in the SI.

We additionally allow questions to vary by ``client attributes'' -- which are listed in the upper card in Figure~\ref{fig:contributions_framework}C. 
Each of the $770$ questions can result in several potential variants by changing hypothetical client attributes such as age or gender.
For $25$ of the $770$ questions, the correct answer option depends on specific combinations of client attributes.
These questions require caseworkers to consider how client attributes interact with policy rules.
For example, consider the question posed in Figure~\ref{fig:contributions_framework}C for a senior client receiving either Social Security Income (SSI) or Retirement, Survivors, and Disability Insurance (RSDI): \emph{``The client claims a deduction for transportation costs incurred while traveling to a medical appointment. Is this deduction valid?''}
The correct answer is listed in Figure~\ref{fig:contributions_framework}D, which is that transportation costs are deductible. If the client were instead a woman in the labor force, the correct answer would remain the same. However, if the client were a non-senior and not legally disabled, then the correct answer would be different: medical expenses would not be deductible.

We use our benchmark dataset to evaluate how well caseworkers---with or without chatbot suggestions of varying accuracy---answer these SNAP questions. Between May - July 2025, we recruited $125$ caseworkers to take a 45-question assessment on Qualtrics (with questions randomly drawn from a subset of our 770-question benchmark dataset, using a process described in the Materials and Methods section and the SI Appendix). Caseworkers had an average of 4.06 years of experience, and were requested to take the assessment without using external tools for assistance such as chatbots or Google; details on recruitment and caseworker demographics can be found in the Materials and Methods section and SI Appendix Tables S6-S8. 

Our experimental design is depicted in Figure~\ref{fig:contributions_framework}B. Participants in the control group saw a client attribute card and a SNAP-related question with no chatbot suggestion (e.g., only seeing  Figure~\ref{fig:contributions_framework}C). Participants in the treatment group additionally received one chatbot suggestion alongside each assessment question (e.g., seeing both Figure~\ref{fig:contributions_framework}C and one of either the corresponding correct or incorrect suggestion, as shown in Figure~\ref{fig:contributions_framework}D). 

Notably, our experiment relied on artificially varying the accuracy of the chatbot in aggregate (over the 45 questions each caseworker sees).
Rather than seeing ``live'' chatbot suggestions, we instead hard-coded correct or incorrect suggestions for each question (as depicted in Figure~\ref{fig:contributions_framework}D), which were verified by QC experts (as discussed in the the Materials and Methods section).
We tested 10 different levels of question accuracy across the 94 participants in the treatment group. For example, in our study, 7 participants received assistance from a mediocre chatbot that answered 33/45 questions correctly (73\% accuracy), whereas 19 participants saw a perfect chatbot that answered 45/45 questions correctly (100\% accuracy).
Artificially varying the accuracy of chatbot suggestions ensures our study's findings are relevant even as the technical performance of LLMs improves over time. In addition, it allows us to assess shifts in caseworker behavior, such as identifying changepoints at which users may stop trusting the LLM-based system~\citep{gaur2016effects}.
This experimental design can inform a broader practical evaluation framework for LLM tools that researchers and government agencies can leverage for decision-making around LLM adoption.

Questions in the assessment also vary by difficulty. We provisionally assigned difficulty labels (easy, medium, or hard) to questions when developing the benchmark dataset based on SNAP QC expert opinion. We used these to ensure that caseworkers saw a balanced set of questions in terms of difficulty on each assessment. After data collection finished, we then re-classified questions as easy, medium, or hard based on the percentile ranking of question-level accuracy from caseworkers in the control group (who did not see any chatbot suggestions); the difficulty-related results that we present are based on this second classification of difficulty.\footnote{Changes in difficulty labels primarily affected medium difficulty questions. 58\% of medium questions shifted to easy or hard questions, while 61\% of easy questions remained easy and 52\% of hard questions remained hard. Due to small sample size concerns, we only changed the difficulty labels for questions where more than 7 caseworkers in the control group answered the question.} A comparison of these changes appears in SI Figure S3.

Our primary research question is: (1) does chatbot assistance improve caseworker performance on a benchmark task, relative to a control group of caseworkers?
We further answer more granular research questions:
(2) Does higher chatbot accuracy increase human caseworker accuracy? 
(3) To what extent do correct vs. incorrect suggestions affect caseworker accuracy at the question-level?
(4) Do these effects differ based on question type (e.g., across difficulty, or client attributes)?
(5) Does chatbot assistance reduce administrative burdens, as measured by (a) time to answer questions and (b) self-reported effort and difficulty?
(6) Are caseworkers able to discern chatbot quality?

\section*{Results}

Answering each of the research questions posed above, we now present the results from our randomized experiment for all participants (n=125).\footnote{We included an attention check in our study (further details in the Materials and Methods and shown in SI Appendix Figure S4), which 63\% of participants passed. We present results for all participants in the main text, and results for the subset of participants who passed the attention check (n=79) in the SI Appendix. This choice is due to our belief that including participants who failed the attention check better reflects the competing demands placed on caseworkers' attention and is perhaps more reflective of their actual working conditions. Furthermore, results for both samples (all participants, and the subset who passed the attention check) are directionally similar; while the set of participants who failed the attention check did have lower accuracy on average, their accuracy levels were not statistically significantly different from those who passed the attention check (see SI Appendix Table S12).\label{attentioncheck_footnote}}

\subsection*{Chatbot suggestions improve caseworker accuracy on benchmark questions}
Caseworkers shown \textit{any} accuracy of chatbot suggestions performed better compared to those answering the assessment questions without chatbot suggestions, on average. As shown in Figure~\ref{fig:accuracy_panel}A, chatbot suggestions led to a mean 21 percentage point increase in caseworker participant-level accuracy compared to the control group, where experienced caseworkers scored 49\% on average.
It is worth noting that the assessment questions were meant to be difficult, as our benchmark dataset is based on SNAP Quality Control data.
For comparison, random guessing would result in only 20-25\% accuracy on average.
The question topics reflect errors made by eligibility workers in government agencies, who have more training on SNAP and who tend to focus on a much more limited set of programs.
While most caseworkers ($>$ 75\%) reported that they were at least slightly familiar with the question material, caseworkers also indicated on average that only around 50\% of questions reflected day-to-day situations with clients.

\subsection*{The effect of chatbot suggestions depends on chatbot quality} The largest increases in accuracy  occur with high-accuracy chatbot suggestions. For example, as shown in Figure~\ref{fig:accuracy_panel}A, chatbots with $96\%$-$100\%$ chatbot accuracy (only 0-2 incorrect suggestions across the entire assessment) are associated with a 27 percentage point increase in caseworker participant-level accuracy on average relative to the control group performance of 49\% ($p<0.001$). In contrast, a low-accuracy chatbot with 53-73\% accuracy (21-12 incorrect suggestions overall) is associated with a much smaller improvement in caseworker participant-level accuracy of 8 percentage points ($p<0.01$).

\subsection*{Caseworker accuracy improves until reaching the \emph{``AI underreliance plateau''}} 
\begin{figure*}[htbp!]
    \centering
    \includegraphics[width=\linewidth]{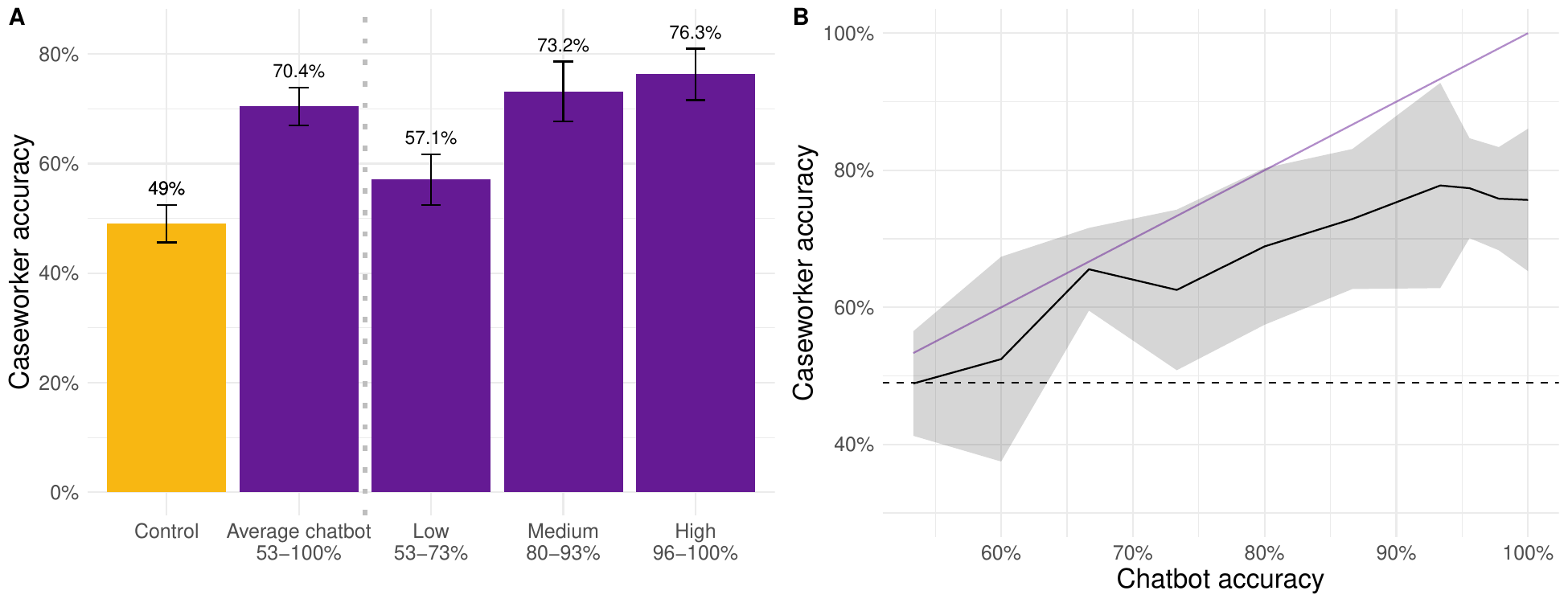}
    \caption{(A) Caseworker participant-level accuracy for any chatbot on average (``Average chatbot'') and by chatbot quality: low (53-73\% accurate), medium (80-93\% accurate), and high quality (96-100\%). We assign these qualitative labels of chatbot quality to specific accuracy ranges to ensure sample sizes are reasonably balanced across groups: 26 caseworkers saw low-quality chatbots, 17 saw medium quality chatbots, and 51 caseworkers saw high quality chatbots.
    Overall, caseworker accuracy improves with chatbot suggestions by 21 percentage points, regardless of chatbot quality. However, high quality chatbots lead to the largest improvements in accuracy.
    (B) Human accuracy increases alongside chatbot accuracy up to a point, resulting in an AI underreliance plateau around $80\%$ accuracy. The dashed horizontal line shows accuracy from participants in the control group, which is roughly comparable to the worst-performing chatbot. The black line illustrates human accuracy if respondents followed $100\%$ of the chatbot suggestions. We include the 95\% confidence interval for each estimate in light gray. Due to small sample sizes for some chatbot accuracy conditions, we estimate 95\% confidence intervals using the student t-distribution.}
    \label{fig:accuracy_panel}
\end{figure*}

Caseworker accuracy improvements plateau, particularly at the highest levels of chatbot accuracy (greater than 90\%).
Figure~\ref{fig:accuracy_panel}B presents the average caseworker participant-level accuracy for all levels of chatbot accuracy, ranging from 53\% to 100\%.
The dotted horizontal line illustrates the baseline performance of caseworkers who do not see any chatbot suggestions. The purple line is the accuracy caseworkers could have achieved if they followed all of the chatbot suggestions exactly.
While caseworker accuracy increases linearly alongside chatbot accuracy, it starts to lag at around 80\% chatbot accuracy and there is a pronounced gap between the two for chatbot accuracies greater than 90\%.

We refer to this phenomenon as the \emph{AI underreliance plateau}, i.e. when human performance consistently falls below the quality of readily available AI help.
Caseworkers do not always follow the chatbot suggestions even when correct, perhaps due to issues of trust and an inability to discern quality. 
Furthermore, it indicates the potential diminishing returns to improvements in chatbot accuracy, particularly on questions where caseworkers perform poorly on their own.
As we show with LOESS plots in SI Appendix Figure S5, the AI underreliance plateau is most pronounced on medium and hard questions. For example, average caseworker accuracy on difficult questions is 65\% when viewing chatbot suggestions that are 100\% correct -- an improvement over the control-group accuracy of 22\%, but still a 35 percentage point gap from the accuracy that could be achieved by following all of the chatbot suggestions.  We discuss potential drivers, and implications, of the AI underreliance plateau in the Discussion section.
 
The onset of the AI underreliance plateau roughly aligns with the accuracy levels of real-world chatbots.  To contextualize our findings, we evaluated the accuracy of Nava's chatbot on all multiple-choice questions used in our study. Overall accuracy is 85\%. If we estimate Nava's chatbot accuracy based on the 45 questions that appear in each assessment set (i.e., simulating actual chatbot answers to the same assessments that the caseworkers took), then we find that the chatbot accuracy can range from 76\% to 82\%.

\subsection*{The negative impact of an incorrect suggestion can be substantial}
In addition to the participant-level results on caseworker accuracy, presented above, we examine the effect of different types of chatbot suggestions (i.e., incorrect, correct, or no chatbot suggestions) on caseworker accuracy at the question level.
To ensure responses are comparable, we restrict our analysis to questions for which LLMs generated incorrect suggestions and that appeared in our study with an incorrect suggestion for at least some participants.
This subset of questions (n=35) skews more difficult, containing four easy questions, 11 medium questions, and 20 hard questions.
As a result, the accuracy of caseworkers in the control group is lower here than the control group accuracy of 49\%, shown in Figure~\ref{fig:accuracy_panel}. 
We explain our methodology and estimation approach further in the Materials and Methods section, and we present an expanded set of regression results in SI Appendix Tables 17-22.

As shown in Figure~\ref{fig:question_level_effect}, an incorrect chatbot suggestion leads to a mean 18 percentage point decrease in caseworker accuracy relative to the control group (where caseworker accuracy was on average 38\% without chatbot suggestions). One reason for this trend may be that caseworkers are unable to assess chatbot quality themselves; indeed, the Pearson correlation coefficient between caseworkers' perception of chatbot accuracy and actual chatbot accuracy is 0.04. When shown an incorrect chatbot suggestion, caseworkers may not be confident enough in their own knowledge to override the chatbot, even though they might have answered correctly on their own.

\subsection*{Incorrect suggestions are most harmful on easy questions; correct suggestions have benefits on hard questions}
\begin{figure*}[htbp]
    \centering
    \includegraphics[width=\linewidth]{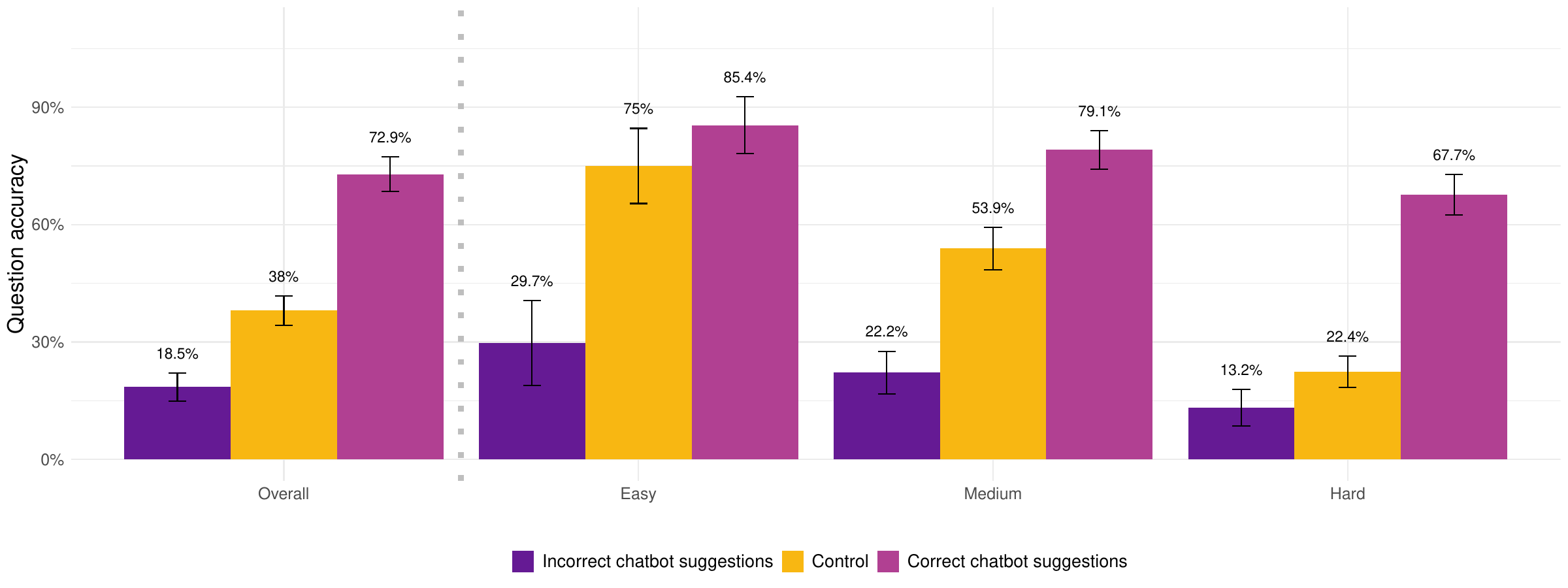}
    \caption{Incorrect chatbot suggestions lead to the largest reductions in caseworker accuracy on easy and medium questions relative to accuracy in the control group, wherein no chatbot suggestions are seen. At the same time, correct chatbot suggestions can be beneficial, particularly on hard questions where they increase caseworker accuracy by 45 percentage points on average. Question difficulty appears on the x-axis and caseworker question-level accuracy on the y-axis. 
    We denote 95\% confidence intervals with error bars using cluster robust standard errors. We define question-level accuracy as the percentage of correct caseworker responses by chatbot suggestion type: incorrect, correct, or no suggestions (purple, violet, and  yellow in the figure).
    To improve the comparison of our estimates, we only include questions that ever appear in an assessment with an incorrect chatbot suggestion (35 questions in total). These questions appear to be harder on average for caseworkers in the control group, hence the control group accuracy is lower compared to the participant-level estimates in Figure~\ref{fig:accuracy_panel}.}
\label{fig:question_level_effect}
\end{figure*}

Figure~\ref{fig:question_level_effect} additionally presents the question-level accuracy by question difficulty. Question-level accuracy appears on the y-axis, question difficulty appears on the x-axis, and the type of chatbot suggestion (correct, incorrect, or a question with no suggestions) is indicated by bar color. 
Regardless of question difficulty, incorrect chatbot suggestions reduce caseworker accuracy. However, they are especially harmful on easy questions where most participants in the control group would have answered the question correctly on their own (accuracy $\geq 75\%$).
 When shown incorrect chatbot suggestions on easy questions, caseworker accuracy falls by approximately two-thirds relative to the control group.

Effects from correct chatbot suggestions were approximately mirrored -- they helped the most with hard questions where accuracy in the control group was lowest. For example, seeing a correct chatbot suggestion increases caseworker question-level accuracy from 38\% in the control group to 73\% overall (an increase of around $35$ percentage points). On hard questions, a correct chatbot suggestion increases accuracy from 22\% to 68\% (an increase of 45 percentage points).

\subsection*{Varying client attributes does not meaningfully affect accuracy}

Client-level attributes appear in the background panel for each of the assessment questions (see Figure~\ref{fig:contributions_framework}C). 
Changing client attributes leads to two possible subcases.
For 38 out of the 45 questions in each assessment, questions could vary by either age or gender. For these questions, changing the specific client attribute would not affect the correct answer.
However, for four out of the 45 questions, we considered that changing some client attributes could affect the correct answer. For example, as in Figure~\ref{fig:contributions_framework}D, changing a client's age to ``Adult'' and Healthcare/Disability Status to ``Not Legally Disabled'' would mean that the client is no longer eligible for medical expense deductions.

Restricting to questions where the correct answer does not depend on client attributes, we examine whether caseworker accuracy changes depending on the specific values of a hypothetical client's age or gender. For example, are caseworkers more accurate for questions concerning ``Male'' applicants compared to ``Female'' applicants? Or are caseworkers more accurate for questions concerning ``Adult'' clients compared to ``Senior'' clients?
Varying the gender or age of the hypothetical client has no significant effect on caseworker accuracy, regardless of whether caseworkers saw chatbot suggestions or not, as shown in SI Appendix Figure S6.

 However, questions where the correct answer depends on client attributes tend to be more difficult overall.
 For these questions, caseworker accuracy was on average lower, compared to questions where the correct answer did not depend on client attributes. These questions appear to be harder generally; accuracy was lower for both caseworkers who saw chatbot suggestions and those who did not (see SI Appendix Figure S7).
This is an example of the difficulty caseworkers face in navigating the complex interactions between SNAP policy rules and client demographics. 

\subsection*{Chatbot suggestions do not appear to reduce, or exacerbate, administrative burdens}

We assess the relationship between chatbot suggestions and administrative burdens based on differences in (1) caseworker question-level duration and (2) a post-assessment survey question on self-reported mental effort. We find that chatbot suggestions do not appear to increase or decrease significantly the median time to answer questions on average, and there are small (not significant) decreases in mental effort for participants who see chatbot suggestions relative to participants who do not.

\subsubsection*{Chatbot suggestions are not associated with large differences in the time to answer questions} 

\begin{figure}[ht!]
    \centering
    \includegraphics[width=0.75\linewidth]{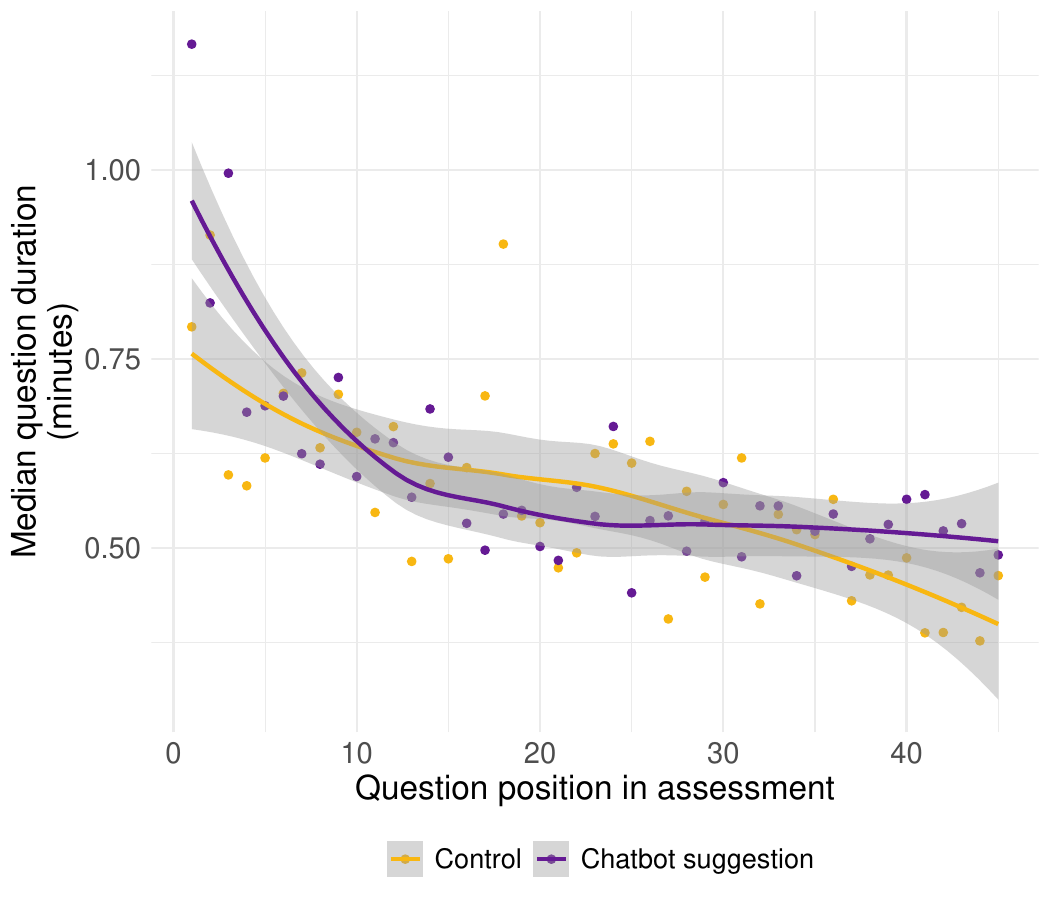}
    \caption{Median caseworker time to answer each question is similar with and without chatbot suggestions, and decreases over the course of the assessment. The position in the survey (from 1 to 45) is on the x-axis. The median question duration is on the y-axis. We use the median question duration, as large outliers may skew results. The line color indicates whether participants saw LLM suggestions (treatment group) or no LLM suggestions (control group). Additionally, we include a trend line with 95\% confidence intervals.
    Trends are similar for caseworkers who saw chatbot suggestions and those who did not, as demonstrated by the overlapping lines.
}
    \label{fig:lineplot_duration}
\end{figure}

Figure~\ref{fig:lineplot_duration} depicts the median question-level duration depending on whether participants saw either a correct or incorrect chatbot suggestion, or no suggestion at all. The median question duration is shown on the y-axis, and the question position in the survey (ranging from 1 to 45) appears on the x-axis.
Though chatbot suggestions can improve accuracy, the time taken to read chatbot suggestions appears comparable to the time otherwise spent answering the question without chatbot suggestions.
We do not observe strong evidence of significant differences in the median question-level duration, as demonstrated by the overlapping lines in the figure for most question positions. 
Overall, the average median duration was 37 seconds for caseworkers who saw chatbot suggestions compared to 38 seconds for those who did not.

There is, however, a decreasing trend in median question-level duration over the course of the entire assessment, consistent with prior literature on survey takers~\citep{galesic2009effects}.
Caseworkers may experience survey fatigue, particularly for questions that appear toward the end of the assessment.
These decreases are present both for caseworkers who saw chatbot suggestions and those who did not, though there are noticeable differences between the two groups in the beginning and end of the assessment. In particular, caseworkers with chatbot suggestions appear to have a steeper learning curve up until question 10 and they also experience less fatigue in later questions (after question 35).
Differences in question duration align with slight variations in caseworker accuracy over the course of the assessment. As shown in SI Appendix Figure S12, though caseworker accuracy is comparable at first, increases in question position  lead to small increases in accuracy on average for caseworkers who saw chatbot suggestions, and small decreases for those who did not.

\subsubsection*{Chatbot suggestions do not significantly affect self-reported mental effort}
At the end of the assessment, we asked participants to report the mental effort they invested. Participants who saw chatbot suggestions consistently reported lower levels of mental effort as shown in SI Appendix Figure S8, though differences are not statistically significant based on a chi-squared test for independence.
For example, 56\% of participants who saw chatbot suggestions reported investing high or very high mental effort compared to 61\% of participants in the control group. Additionally, none of the particpants in the control group reported investing low or very low mental effort compared to 11\% of participants who saw chatbot suggestions.
Though the findings are not conclusive, they suggest that chatbot suggestions may potentially lead to slight decreases in mental effort, thus reducing administrative burdens.

\subsection*{Caseworkers struggle to accurately discern chatbot quality}
We now present additional findings that illustrate how caseworkers may struggle to evaluate chatbot quality. These findings underscore the importance of behavioral dynamics such as caseworker perceptions of chatbot accuracy and the tendency to agree with the chatbot suggestions.

\subsubsection*{Caseworker characteristics do not meaningfully affect performance on the assessment}

Caseworker characteristics such as years of experience, the level of self-reported familiarity with the question material, or frequency of self-reported chatbot usage are all only weakly associated with improvements in caseworker accuracy. Instead, the characteristic most predictive of assessment performance is a participant's self-reported perception of chatbot accuracy ($\beta = 0.269, p < 0.001$). This positive relationship holds, even when controlling for other respondent-level characteristics such as years of experience or self-reported chatbot usage at the 5\% level of significance (see SI Appendix Table S14). Self-reported familiarity with question material is also associated with an average increase in caseworker accuracy, though the effects are not statistically significant when restricting to caseworkers who pass the attention check.

\subsubsection*{Caseworkers' perception of chatbot accuracy is uncorrelated with true chatbot accuracy}
As shown in SI Appendix Figure S13, the Pearson correlation coefficient between accuracy perception and chatbot accuracy is 0.04.
In other words, caseworkers' perception of accuracy largely diverges from the actual chatbot accuracy they saw.
This pattern suggests that caseworkers may struggle to evaluate the quality of the chatbot suggestions, either based on prior experience or domain knowledge.
It also explains why caseworkers'  perception of chatbot accuracy is positively and significantly associated with increases in caseworker participant-level accuracy.
Since caseworker accuracy in the control group is comparable to the lowest performing chatbot accuracy (53\%),
following the chatbot suggestions more often, regardless of quality, tends to improve performance.

\subsubsection*{Level of agreement with chatbot suggestions is stable and does not adjust to chatbot quality}
We also assess the likelihood that participants follow the chatbot suggestions based on both chatbot quality and prior agreement rates as shown in SI Appendix Figures S9-S11. In aggregate, caseworkers self-reported that they followed the chatbot suggestions roughly 70\% of the time. 
We do not observe any significant differences in agreement based on chatbot quality or question position in the survey.
However, we do find that greater past agreement with the chatbot suggestions on average  increases the likelihood of future agreement ($\beta = 0.007, \ p< 0.001$; see Table S16 in the SI Appendix). 
These findings suggest that decisions about whether to agree with the chatbot are based primarily on participant behavior -- their overall propensity to agree with the chatbot -- rather than the ability to distinguish between correct or incorrect chatbot suggestions.

\section*{Discussion}

In this section, we first discuss our findings in connection to related literature. We then provide design recommendations for chatbot developers and other decision-makers who may be interested in using chatbots for similar human-in-the-loop processes.

Prior work has found that access to AI tools can increase human productivity~\citep{brynjolfsson2025generative, noy2023experimental}. This is consistent with our finding that chatbot suggestions do help caseworkers, significantly improving their accuracy compared to caseworkers who do not see any chatbot suggestions.
Our findings hold even for medium-quality chatbots in the range of the Nava chatbot and off the shelf chatbots; as shown in Figure ~\ref{fig:accuracy_panel}, an 80-93\% accurate chatbot improves caseworker accuracy by 24 percentage points on average.

However, we also observe an AI underreliance plateau. When paired with high quality chatbot suggestions, caseworkers are not as accurate as they could be if they followed the chatbot suggestions more often.
We believe this gap in accuracy stems from a lack of trust. To address this, chatbot developers may want to provide participants with an estimate of the overall chatbot accuracy or use retrieval augmented generation (RAG) models that include verbatim policy citations. Motivated caseworkers can then verify that the information is true.
Prior work has shown that including citations can increase participants' self-reported trust~\cite{ding2025citations}. 
In follow-up work, we are exploring whether citations can also improve caseworker accuracy.

Issues related to trust are well-documented in the literature. 
Prior work has demonstrated that individuals may be less likely to trust statements they suspect were generated by AI, and that trust in AI-generated statements can vary depending on slight differences in the context or information provided~\citep{buchanan2024people}. Prior work on the use of chatbots among SNAP applicants (as opposed to caseworkers) has highlighted the importance of trust in mediating interactions with chatbots for SNAP. 
While LLMs present several opportunities for reducing administrative burdens in the SNAP application process, issues of trust may arise related to concerns of AI competence, integrity, and benevolence~\citep{jo2025ai}. 

Our work similarly evaluates user trust in the context of AI competence, as measured by chatbot accuracy.
We observe that caseworkers may not trust the chatbot, even when accurate, as they are not able to judge the quality of the chatbot suggestions.
Caseworkers are frequently unable to discern when the chatbot suggestions are incorrect, even on relatively easy questions. Conversely, when presented with correct chatbot suggestions, caseworkers do not always follow the suggestions -- leading to the AI underreliance plateau that we describe above.

Prior work has shown that human-AI collaborations tend to perform worse than either the best performing humans or AI alone~\cite{vaccaro2024combinations}. One reason is the inability of humans to correctly determine when to trust the AI tool. For example, Cabrera et al. \cite{cabrera2023improving} similarly observe a pattern where accuracy increases with the use of an AI tool, but only when humans have an accurate ``mental model'' of the AI's performance. 
We observe this phenomenon in our study. Participants perform worse than the chatbot because they are unable to tell when its suggestions are correct or not.

To address these challenges, we recommend further exploration of design fixes like including warnings about the chatbot's limitations (similar to prior work on misinformation~\citep{greene2021quantifying, williams2024misinformation, van2023can} and algorithmic risk assessment~\citep{engel2021machine} or providing well-calibrated confidence scores. However, in the case of warnings, prior work has found limited evidence of their effectiveness and developers should be wary of spillover effects where warnings might decrease trust in accurate chatbot suggestions, further exacerbating the problem~\citep{engel2021machine, williams2024misinformation, van2023can}.
Similarly, while interventions such as confidence scores may improve trust, they can be ineffective in improving accuracy when both humans and LLMs err in similar ways~\citep{zhang2020effect}.
Verbatim citations (e.g., from policy documents) may be a helpful intervention, as they provide a pathway for participants to find and verify the correct answer themselves. However, the success of this intervention also relies on caseworker resources, time, and motivation, which may be limited in real-world settings. We find evidence of substantial disengagement in our study, as almost 37\% of recruited participants failed a difficult instruction manipulation check requiring high attention to detail.\textsuperscript{\ddag}

Apart from these interventions, chatbot developers may want to consider the broader social system in which a chatbot is deployed. In the case of public benefits, some questions may be more consequential for individual client outcomes (e.g., questions related to eligibility).
For these questions, perhaps caseworkers should not use a chatbot at all, and properly cite or vet the information they provide. In other settings, with less consequential questions, it may be acceptable for caseworkers to use the chatbot suggestions in their work. A careful chatbot roll-out should identify the contextual factors that make chatbot use permissible and provide formal guidance to caseworkers.

Including chatbot suggestions did not affect the time involved in taking the assessment. Though our study reflects an ideal setting in which caseworkers do not have to iteratively prompt the chatbot themselves, we do realistically mimic the typing effect of a chatbot that might slow them down if reading carefully. The characters in the chatbot suggestions appeared gradually when each question loaded, as if a chatbot was producing the response in real-time.  Caseworkers had the option to click ``Fast forward,'' which would skip the typing effect. However, few caseworkers skipped through the typing effect and we do not observe any significant increases in question-level duration when including chatbot suggestions.\footnote{Caseworkers had the option to fast forward through the typing effect, which could speed up the time to read the question, though few made use of this. Caseworkers clicked fast forward in only 13\% of questions.} This finding illustrates that interventions to reduce the interactional complexity of chatbots (e.g., developing collaborative databases of effective prompts for a range of questions) is a promising direction for minimizing impacts on caseworker time and resources. 

Lastly, a direction for future work is to consider the role of benchmark datasets, like the SNAP-related benchmark data we develop, in training and auditing chatbots.

\subsection*{Limitations}
First, an important consideration is that our study reflects human caseworker accuracy under almost ideal  conditions of access to chatbot suggestions. For example, we do not account for the time required to prompt a chatbot to return a helpful response and we selected the highest quality chatbot suggestions through an iterative, human review process with SNAP experts. As a result, our findings speak to the \emph{potential} for a range of chatbot products to improve social service delivery, rather than being an evaluation of present-day chatbot capabilities.

Next, our study measures human accuracy as it is affected by chatbot accuracy, as opposed to broader societal outcomes. For example, instead of measuring caseworker accuracy, we could measure client-level outcomes such as SNAP take-up. Those outcomes are very far downstream of these interactions and there are numerous intervening factors. For example, caseworkers provide information to clients, which might involve telling some clients that they don't qualify for SNAP or working with clients who are already receiving benefits.

Another concern is whether the questions in our study are truly representative of real-world caseworker interactions. We note that we focus on SNAP quality control errors in the benchmark dataset, which we did to create assessment questions that would be realistic, but difficult enough to allow for variation (as evidenced by the 49\% accuracy of the control group). On average, caseworkers reported that about half the assessment questions reflected day-to-day situations with clients, which aligns with control-group performance. While 75\% indicated that they were at least slightly familiar with the question content, only 4\% indicated they were very familiar (SI Appendix Tables S7-S8). LLM-based tools like chatbots are often used to extend the capabilities of professionals to help them better handle topics that may be less familiar - a software developer unfamiliar with a particular programming language may rely on a coding agent for support, for example. We believe these types of use cases will also be important in social services work such as benefits outreach, where staff are already spread thin and burnout is high. 

We chose multiple-choice questions as opposed to open-ended questions with free-text responses, which might approximate real-world interactions more closely. Multiple choice questions simplified the human expert review process and easily standardized responses from caseworkers.
However, to assess the strenth of this comparison, we evaluated the accuracy of Nava's chatbot on a subset (n=89) of both the multiple choice assessment questions and corresponding open-ended versions of the same questions.
Nava's chatbot achieved 85\% accuracy on the multiple-choice questions, and performed worse on the free-text response questions with 76\% accuracy. In part, the drop in accuracy may stem from the ambiguous and open-ended nature of the questions, which can lead to answers with irrelevant details. These differences underscore how our findings capture the effect of chatbot suggestions under ideal conditions, which may inflate the potential benefits over what would arise in real-world use. 

Finally, our work is limited in scope. We focus on caseworkers in California and our questions are specific to the California implementation of SNAP. SNAP rules, in turn, can change over time. Our benchmark dataset may require future updates to account for changes in federal policy. For example, recent legislation in the U.S. like the 2025-2026 congressional bill H.R. 1~\citep{hr1} removes SNAP eligibility for many immigrant groups, such as asylees and refugees. Similarly, our findings may not be directly generalizable to other contexts, such as settings where caseworkers assist with different social service programs (e.g., Medicaid).

\subsection*{Conclusion}
Our work points to the importance of improving practical methods for conducting AI evaluations, particularly in the \emph{context of use}. Many approaches to AI evaluation focus on issues of standardization and validity~\citep{roose_nyt, wallach2025position, dow2024dimensions}. Accounting for system context and holistically evaluating tools on realistic decision tasks are both areas of critical importance~\citep{hutchinson2022evaluation, lai2023towards}, particularly for improving human-AI evaluations. 
Our approach addresses many of these limitations, with a replicable, stakeholder-driven evaluation framework. We ground our evaluation in a real-world application, develop high-quality benchmark questions through an iterative expert review process, and measure impacts on the targeted end-users of the tool -- human caseworkers.

We also expand on the limited evaluation tools available to governments for assessing the use of AI tools used in the public sector.
At present, there are few benchmark datasets and tools oriented toward the use of LLMs in the public sector~\citep{liu2025evaluation}, procurement systems in governments are ill-equipped to deal with emerging AI uses~\citep{johnson2025legacy}, and greater empirical work is needed to measure the impact of specific AI applications~\citep{zuiderwijk2021implications}. 

Overall, our results suggest that incorporating LLMs into outreach for social safety net programs could potentially have benefits if rolled out carefully in a vetted human-in-the-loop system. A key warning from this study was that individuals will follow incorrect suggestions on questions they otherwise would have answered correctly. 
Yet, on balance, if we consider potential advancements in real-world chatbot quality, the much larger problem is that correct chatbot suggestions went unused, which we call the AI underreliance plateau. 
Understanding the reasons for this plateau and interventions that might reduce it is an important direction for future research.
To underscore a key insight from our study, conducting evaluations with human users is critical. Improvements in chatbot accuracy alone, as simulated in our online assessment, do not directly translate to improvements in human accuracy.

\section*{Materials and Methods}
\subsection*{Benchmark dataset with human-expert review}
\label{sec:benchmark_dataset}
A critical component of our work is the creation of a comprehensive and expert-vetted 770-question benchmarking dataset, based on representative data on SNAP quality control errors.
As shown in Figure~\ref{fig:contributions_framework}C, 
each question in the benchmark dataset has the following features: a client background panel with $5$ attributes, a specific question posed by this hypothetical client about CalFresh (the name for SNAP in California), 4-5 multiple choice options, and one correct multiple-choice answer.
This setup allows us to vary client attributes (leading to many possible question permutations) and test whether respondents or LLMs disproportionately answer questions incorrectly for certain demographics of potential SNAP clients.
We also ensure that, for a subset of questions in the benchmark data, the correct answer can change depending on a unique combination of attributes (e.g., self-employed vs. employed by others), allowing researchers to measure differences in accuracy when client attributes are consequential. Answering these questions requires knowledge about how client attributes interact with policy rules.
Questions for the benchmark dataset were generated via GPT-4 prompting with a structured prompt (shown in the SI Appendix) that specified the LLM’s role, task, and instructions, and provided several multiple-choice question examples. SNAP QC experts then reviewed questions for relevance and clarity, made revisions, and identified correct answer options, which were verified by at least two SNAP QC experts. More details on this process appear in the SI Appendix.

\subsection*{Assessment Design}
\label{sec:assessment_design}
We generate four unique multiple-choice assessments (which we will refer to as ``assessment sets''), each containing $45$ questions from the original benchmark dataset. We randomly selected questions ensuring an approximate balance of topics (see SI Appendix Table S3). For each question, we varied at least one client attribute in the client background panel.\footnote{We did this for all but $3$ questions in the survey, which are fixed with a consistent client background panel across all assessment versions.} Refer to the SI Appendix for more details; the Qualtrics logic and questions are available in an anonymous GitHub repository.\footnote{\href{https://anonymous.4open.science/r/chatbots-social-services-94EB}{https://anonymous.4open.science/r/chatbots-social-services-94EB}}

For each of the four assessment sets, we produced two versions by varying the attributes in the client background panel. Each version contains the same set of questions. For $38$ out of $45$ questions, we varied either age or gender -- switching the default attribute value to an alternative. For simplicity, each attribute only has two possible values: age is either ``Adult (22-59)'' or ``Senior (60+)''; gender is either ``Male'' or ``Female.'' We ensure that changing age or gender does not invalidate the question or change the correct answer.\footnote{In general, all questions use gender-neutral language and pronouns, except for two questions, which we identified after the completion of the experiment as containing she/her pronouns in the question and answer options (even if the question referred to a hypothetical male client). We exclude these two questions from analyses that focus on gender.}

An additional four questions in the assessment came from the subset of benchmark questions where the correct answer depends on a unique combination of client attributes. For these questions, we produced two versions, varying the attributes that meaningfully affect the answer. For example, the answer to an employment-related question may change if the client is self-employed or employed by others. Lastly, we selected three questions to appear with a fixed client background panel across all assessment sets and versions. We present caseworker question-level accuracy for these three questions, overall and for each question individually, in SI Appendix Figures S14 and S15.

Due to stratification by difficulty and limitations in the number of questions with incorrect chatbot suggestions for both question versions with original and varied client attributes, we were not able to use all questions in the benchmark dataset. Additionally, there is a high amount of question overlap across assessment sets (see SI Appendix Table S2). There are $89$ unique questions cross all four sets. Aside from questions that we intentionally included in all four sets (three questions with fixed client attributes; four questions where the correct answer depends on client attributes), $41$ $(50\%)$ of the questions appear in more than one question set and $9$ $(11\%)$ appear in all four.

\subsection*{Validation of chatbot suggestions}
\label{sec:llm_validation}
We generated chatbot suggestions for each unique question and client background panel by prompting different LLM models (GPT-4o and GPT-4o-mini with the system prompt from Nava's chatbot) at a range of temperature settings (0, 0.5, 1, 1.5).

We submitted questions via batch processing and automated scoring by requesting the LLM provide a multiple-choice letter option along with its response.
The prompt text included the following: ``Select one correct answer. Please provide your rationale without any reference to the multiple choice letters, followed by \#\#\#, and then provide only the single letter choice that you believe is correct.'' We then extracted the multiple-choice letter option from the LLM's response using regular expressions and graded the response by comparing to the correct answer option.

For many questions, this process resulted in numerous chatbot suggestions of varying quality. Some responses were short and repeated the question and answer text verbatim. Other responses introduced new information. To select the final set of chatbot suggestions that we show to participants in the study, each question underwent an iterative, manual review process from SNAP experts and members of the research team. In cases where there were four distinct chatbot suggestions or less, SNAP experts reviewed all of the chatbot suggestions. In other cases with more than four distinct chatbot suggestions, we selected four representative suggestions using the following approach. First, we used the OpenAI ``text-embedding-3-small'' model to produce embeddings for all of the chatbot suggestions. Second, we computed cosine distances between the embeddings. Third, we clustered the chatbot suggestions into four distinct groups (separately for correct and incorrect suggestions, and for each question and combination of client attributes). We selected the cluster medoids to serve as the four representative chatbot suggestions. Two SNAP experts reviewed and rated the four chatbot suggestions, along with members of the research team. In some cases, feedback from the SNAP experts resulted in small modifications to the question and required re-prompting the LLM to obtain more accurate or realistic responses.
Specifically, in several cases, we removed references to the multiple choice letter that appeared in the question used to prompt the LLM, as participants in our study did not see any multiple choice letters.
In the first week of data collection, we noticed that one chatbot suggestion contained a stale reference to a multiple choice letter (i.e., the text ``The correct answer is B''). We subsequently removed this text from the chatbot suggestion. It only affected four participants in one question.

This iterative review process led to a final selection of the most helpful \emph{correct} chatbot suggestion and, when available, the most plausible \emph{incorrect} suggestion for each question and unique set of client attributes. 
Ultimately, we obtained realistic correct chatbot suggestions for all 89 questions (both for the original question and at least one version with a varied client attribute) that appear in the assessment and incorrect chatbot suggestions for 35.\footnote{Since we did not directly ask the LLM to provide us with incorrect suggestions, there were some questions where the LLM only produced correct suggestions. At most, $21$ out of $45$ questions in each assessment appeared with an incorrect suggestion. Additionally, for some questions, we may have obtained incorrect chatbot suggestions initially but opted not to use them in the assessment: either to ensure  balance in question difficulty and topic, or because we did not obtain incorrect suggestions for multiple versions of the question where age or gender were varied.}

\subsection*{Assessment Distribution}
We conducted our assessments using Qualtrics.
For each question, we stored all question-related information such as the question text, client attributes, the correct LLM suggestions, and the incorrect LLM suggestions (when available) as embedded data variables. We distributed the assessment via email after interested participants filled out a short intake survey. This ensured that we could send each participant a specific assessment that was either (1) a control assessment with no chatbot conditions, or (2) a treatment assessment with a randomly assigned number of incorrect LLM suggestions.

To simulate the effect of a chatbot, we added JavaScript and HTML to each assessment question so that the question-text realistically appeared with a typing effect when a new page loaded.\footnote{Some JavaScript code for Qualtrics was generated using Claude Sonnet 4.5, and was both reviewed and tested before use in the survey.} Participants had the option to fast forward through the typing, so it did not necessarily impact the time to answer the question, though few participants actually made use of the fast forward button ($<$ 13\% of questions).
The typing effect omitted line breaks, which affected between 2-6\% of questions (n = 1-3) on each assessment. Furthermore, we tested whether there was any association between line breaks and a correct answer, and found no statistically significant relationship.

We randomized the order of the answer options for all questions and participants only saw the corresponding text associated with each answer option (i.e., there were no multiple choice letters).

\subsection*{Recruitment}

We recruited 125 caseworkers familiar with CalFresh from the partner nonprofits who work with Amplifi, a nonprofit that manages technological tools to support caseworkers. We coordinated with Amplifi to send out recruitment emails and excluded participants who had no self-reported knowledge of CalFresh. We randomly assigned 31 caseworkers to the control condition in which they saw no chatbot suggestions and 94 to the treatment condition in which they saw chatbot suggestions of varying accuracy.

To test participant engagement, we included an instruction manipulation check in the middle of the survey~\citep{oppenheimer2009instructional}. The attention check question asks participants to select an incorrect answer option that participants would likely not select otherwise.
The difference in accuracy between participants who passed the attention check and those who failed is not statistically significant. 
As a result, we present results for all participants in our main analysis. We include an additional set of results in the SI Appendix that only uses responses from participants who passed the attention check (approximately $63$\% of participants). The results from this analysis are directionally similar to our main results.

Of the 125 participants, 41 ($33$\%) participated in Nava's chatbot pilot.
The majority of participants ($79$\%) knew CalFresh at least moderately well, and on average participants had four years of experience. For those assigned to treatment conditions, participants reported that on average they used chatbot suggestions $70$\% of the time. Most participants had either positive or neutral views about chatbots.  Only a small percentage ($10$\%) reported negative or very negative views about chatbots. The majority of participants ($73$\%) reported that they were at least somewhat familiar with the material in the assessment. Refer to Tables S6-S8 in the SI Appendix for complete descriptive information on participant-level characteristics.

\subsection*{Experimental Design and  Random Assignment}
We evaluate caseworker accuracy on a $45$-question multiple-choice assessment (described above) in a randomized experiment.
The design of the experiment is as follows. We randomly assigned caseworkers to different experiment arms.
In the control arm of the experiment ($25$\% of the sample size), participants completed the $45$-question assessment without any chatbot suggestions. In the treatment arm of the experiment, participants were provided with realistic chatbot suggestions that could be correct or incorrect.
We toggled the accuracy of the chatbot suggestions for each question so that participants saw a range ($53\%$-$100\%$) of overall chatbot suggestion accuracies on the whole assessment. The number of incorrect suggestions was stratified by difficulty (using provisional human expert-labeled difficulty values).
Aside from high-accuracy chatbot conditions (0, 1, or 2 incorrect chatbot suggestions), we otherwise incremented the number of incorrect suggestions in batches of 3, adding an incorrect suggestion for an easy, medium, and difficult question each time. This process resulted in $10$ possible treatment arms: 53\% \ (21 incorrect), 60\% \ (18 incorrect), \ldots, 93\% \ (3 incorrect), 96\% \ (2 incorrect), 98\% \ (1 incorrect), 100\% \ (0 incorrect). We oversampled on high-accuracy treatment arms, assigning participants to the $96$\%-$100$\% accuracy arms with higher probability. The number of caseworkers assigned to each treatment arm is shown in SI Appendix Table S5.
Additional details on the power analysis used to determine the sample size and a more in-depth explanation of the randomization process can be found in the SI Appendix. 

\subsection*{Analysis}
We conduct our analysis at two levels: the participant level, where the unit of analysis is a single caseworker, and the question level, where the unit of analysis is a caseworker's answer to a single question. For the participant level results, we estimate the overall effect of chatbot suggestions on participant accuracy (percent of correct answers out of 45) using a linear model with robust standard errors. Results for the two main results appear in Table~\ref{tab:effects}; the same results are presented in Figure~\ref{fig:accuracy_panel}. We also compute simple mean estimates of caseworker accuracy for all levels of chatbot accuracy,  as shown in Figure~\ref{fig:accuracy_panel}B. We use the student t-distribution to estimate 95\% confidence intervals.

We measure caseworker accuracy at the question level by calculating the percent of questions that caseworkers answered correctly (shown in Figure~\ref{fig:question_level_effect}). We compute 95\% confidence intervals for these percentages using estimates from linear models (shown in SI Appendix Table 17). We demonstrate the robustness of our results to different model specifications in SI Appendix Tables 17-22. For the question-level analyses by chatbot suggestion type (shown in Figure~\ref{fig:question_level_effect}), we only use questions that ever appear in the assessment with an incorrect suggestion (35 questions in total). These questions are on average harder for caseworkers, as demonstrated by the lower control group accuracy (38\%) relative to the participant-level control group accuracy (49\%). We present regression results for all questions and only those that appear with an incorrect suggestion in SI Appendix Tables 17-22.

\begin{table}[!htbp] \centering 
\adjustbox{max width=\textwidth}{
\begin{tabular}{@{\extracolsep{5pt}}lcc} 
\\[-1.8ex]\hline 
\hline \\[-1.8ex] 
 & \multicolumn{2}{c}{\textit{Dependent variable:}} \\ 
\cline{2-3} 
\\[-1.8ex] & \multicolumn{2}{c}{Caseworker participant-level accuracy} \\ 
\\[-1.8ex] & (1) & (2)\\ 
\hline \\[-1.8ex] 
Chatbot suggestions & 21.393$^{***}$ &  \\ 
  & (2.443) &  \\ 
  & & \\ 
 Low quality (53-73\% accurate) &  & 8.062$^{**}$ \\ 
  &  & (2.906) \\ 
  & & \\ 
Medium quality (80-93\% accurate) &  & 24.170$^{***}$ \\ 
  &  & (3.249) \\ 
  & & \\ 
High quality (96-100\% accurate) &  & 27.264$^{***}$ \\ 
  &  & (2.936) \\ 
  & & \\ 
 Constant & 49.032$^{***}$ & 49.032$^{***}$ \\ 
  & (1.713) & (1.727) \\ 
  & & \\ 
\hline \\[-1.8ex] 
Observations & 125 & 125 \\ 
R$^{2}$ & 0.268 & 0.431 \\ 
Adjusted R$^{2}$ & 0.262 & 0.417 \\ 
Residual Std. Error & 15.401 (df = 123) & 13.686 (df = 121) \\ 
F Statistic & 44.981$^{***}$ (df = 1; 123) & 30.571$^{***}$ (df = 3; 121) \\ 
\hline 
\hline \\[-1.8ex] 
\textit{Note:}  & \multicolumn{2}{r}{$^{*}$p$<$0.05; $^{**}$p$<$0.01; $^{***}$p$<$0.001} \\ 
\end{tabular} 
}
  \caption{Results for effects of chatbot suggestions at the participant-level. We estimate all effects with robust standard errors. Column (1) is the overall treatment effect with a single covariate indicating whether caseworkers saw chatbot suggestions or not (``Treatment'').
  The coefficient on chatbot suggestions indicates that chatbot suggestions led to a 21.4 percentage point increase in accuracy on average.
  Column (2) is the effect of chatbot suggestions by  low, medium, and high quality chatbots. For example, a ``High quality chatbot'' is associated with a 27 percentage point increase in caseworker accuracy on average, relative to the accuracy of caseworkers in the control group.
  } 
  \label{tab:effects} 
\end{table}

\subsection*{Funding and Acknowledgements}
This material is based upon work supported by the National Science Foundation under Grant Number 2427748. Any opinions, findings, and conclusions or recommendations expressed in this material are those of the authors and do not necessarily reflect the views of the National Science Foundation.

\bibliography{references.bib}

\FloatBarrier
\newpage 
\renewcommand{\thefigure}{S\arabic{figure}}
\setcounter{figure}{0}

\renewcommand{\thetable}{S\arabic{table}}
\setcounter{table}{0}

\appendix
\section{Question Generation Based on SNAP Quality Control Errors}
\label{sec:question_generation}
In this section, we provide more information on the process for generating questions for the benchmark dataset. First, we developed the SNAP Quality Control (QC) Error Viewer\footnote{\href{https://zwguo.shinyapps.io/snap-qc/}{https://zwguo.shinyapps.io/snap-qc/}} to identify commonly occurring errors and demographic groups that are disproportionately affected if there are any. 
This tool includes three key dimensions of errors: Error Elements (which part of the case was wrong, e.g. shelter deductions); Error Nature (how the error occurred, e.g. incorrect inclusion/exclusion of deductions); Error Type (why the error happened, categorized as client, technical, or agency errors).

For our analysis, we focused solely on agency errors, which are directly attributable to caseworker actions or procedural missteps. To ensure the results were representative of long-term trends, we examined data from 2017 to 2022.

Using the QC Error Viewer, we filtered errors based on a frequency threshold (e.g., occurring in more than 200 instances) and focused on agency errors, including ``reported information disregarded or not applied”, ``policy incorrectly applied'', ``agency failed to verify required information'', ``agency failed to follow up on inconsistent or incomplete information'', and ``agency failed to follow up on impending changes''. 

\begin{figure}[htbp!]
    \centering
    \includegraphics[width=\linewidth]{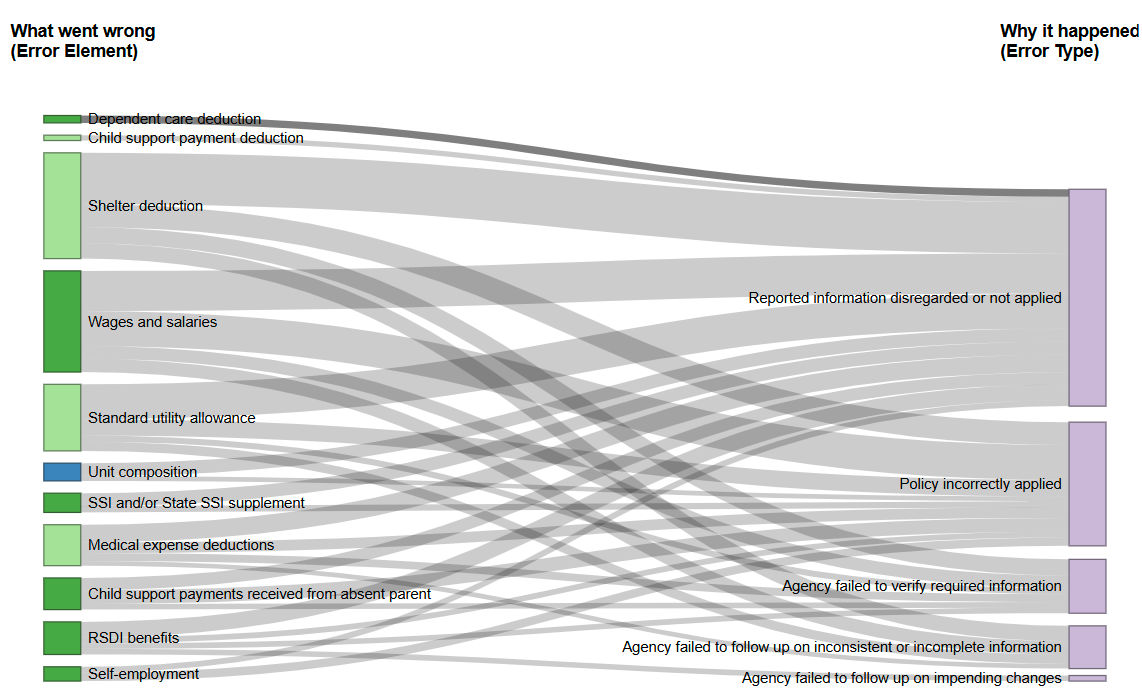}
    \caption{An example from the SNAP QC Error Viewer, an application that visualizes quality control errors in SNAP cases based on nationally representative data. This view allow us to examine the kinds of errors that frequently occur and potential reasons for the error (such as ``Agency failed to verify required information'').}
    \label{fig:snap_qc_errors}
\end{figure}

As shown in Figure~\ref{fig:snap_qc_errors}, this approach allowed us to identify the most significant error elements and error natures that warrant inclusion in our test questions.
After identifying these errors, we analyzed whether specific demographic groups were disproportionately affected. Our focus was on employment status, age, disability status, and homelessness status, as these factors are particularly relevant to SNAP applications.

\begin{figure*}[t!]
        \centering
        \includegraphics[width=\linewidth]{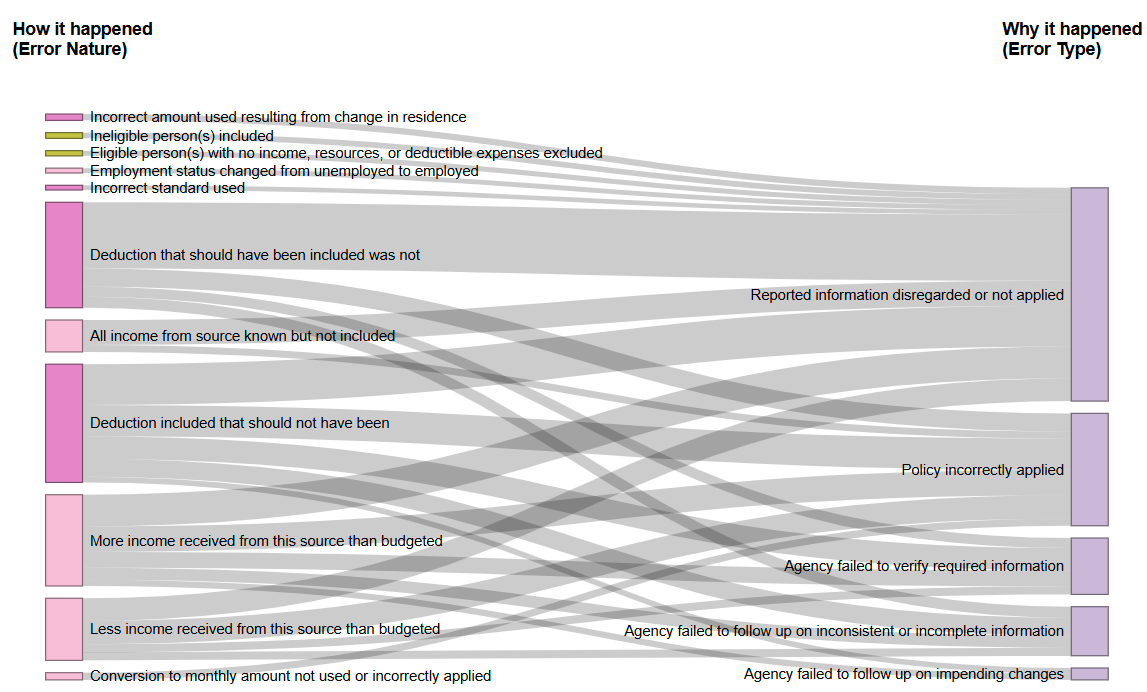}
        \caption{A second example from the SNAP QC Error Viewer shows the relationship between how an error happened and why it happened. For example, we can see that discrepancies in income reporting may arise when ``Reported information [is] disregarded or not applied.''}
    \label{fig:snap_qc_errors2}
\end{figure*}

We  then incorporated these findings into a structured prompt (shown below) to generate test questions using GPT-4. This prompt was designed to incorporate key findings and ensure the questions reflected real-world scenarios encountered by caseworkers.
We generated test questions using a structured prompt that instructed the LLM to act as a test designer creating training questions for SNAP caseworkers processing SNAP applications. For each SNAP QC error element (the policy component where the error occurred) and error nature (a description of how the error occurred), the model was prompted to generate five single-choice scenario-based questions.

The prompt included task instructions and a set of few-shot examples that defined the expected structure, tone, and difficulty of the questions. Each example illustrated a realistic client scenario followed by multiple answer options with only one correct response. These examples were written and refined using SNAP QC policy knowledge to ensure policy alignment and realistic caseworker decision contexts. Generation was performed using a deterministic setting (temperature = 0) to maintain consistent formatting and reduce variation across batches.

SNAP QC experts reviewed the generated questions to assess their difficulty, revise the wording of both questions and answer options, and determine their suitability for use. For error combinations that produced low-quality questions, QC experts wrote additional example scenarios that were incorporated into the prompt as new few-shot examples. This iterative process refined the prompts and improved the quality and consistency of the generated questions.

\subsection{Structured prompt for generating test questions}
\label{sec:structured_prompt}

\begin{verbatim}
You are a test designer tasked with creating five single-choice questions 
to train caseworkers who process applications.

Follow these guidelines:
- Frame each question as a real-world scenario where a client seeks assistance.
- Use neutral language, excluding any demographic or program-specific terms.
- Ensure each question has only one correct answer.
- End each question with "[Select one correct answer.]"

Here are example questions to guide the style and structure:

Example 1:
Error Element: Student status
Error Nature: Ineligible persons included

Question 1: A client is a full-time college student and cares for a 
3-year-old child while attending school. Is the client eligible to be 
included in the household? [Select one correct answer.]

A. No – full-time students are automatically ineligible
B. Yes – she is exempt from the student rule because she cares for a child under 6
C. No – she is eligible only if she is employed
D. Yes – all college students qualify if they have dependents

Question 2: A client’s 19-year-old dependent lives at home, is unemployed, 
and attends online classes full-time at a university. Is the dependent 
an eligible household member? [Select one correct answer.]

A. Yes – they are unemployed and live at home
B. Yes – they are taking classes online and not attending in person
C. No – they are considered an ineligible student
D. Yes – dependents under 22 are always eligible

Question 3: A client is a full-time college student and works 15 hours 
per week at a part-time job. Does this client qualify for benefits? 
[Select one correct answer.]

A. Yes – they meet the work requirement for students
B. No – full-time students are always ineligible
C. Yes – but only if their income is below the threshold
D. No – students must work at least 20 hours per week to qualify

Question 4: A client indicates they are enrolled in a part-time program 
but also works full-time. How does this affect their eligibility as a 
student? [Select one correct answer.]

A. They are eligible as long as they take at least one class
B. They are not considered a full-time student and may not qualify
for certain benefits
C. Their work status does not affect their student eligibility
D. They must provide proof of their work hours to qualify
E. They can only be considered a student if they have a scholarship

Question 5: A client is a full-time graduate student living off-campus 
and financially independent. They do not work and are not disabled. 
Do they qualify as an eligible household member? [Select one correct answer.]

A. Yes - all graduate students qualify regardless of their employment status
B. No - full-time students without an exemption do not qualify
C. Yes - as long as they live off-campus and are financially independent
D. No - only undergraduate students have eligibility restrictions

Now, generate five single-choice questions based on:

Error Element: [ERROR ELEMENT]
Error Nature: [ERROR NATURE]

\end{verbatim}

\begin{longtable}[c]{|M{2in}|M{4in}|}
\hline
\textbf{Error element or nature}                                              & \textbf{Demographics} \\
\hline 
Shelter deduction                                                             & \begin{tabular}[c]{@{}l@{}}Seniors (above 60)\\ Not in labor force and not looking for work\\ Disability\end{tabular}                        \\
\hline 
Wages and salaries  
& \begin{tabular}[c]{@{}l@{}}Early midlife (25-39)\\ Employed by other\end{tabular}                                                            \\ \hline
Standard utility allowance                                                    & \begin{tabular}[c]{@{}l@{}}Not in labor force and not looking for work\\ Disability\end{tabular}  \\ \hline 
Medical expense deductions                                                    & \begin{tabular}[c]{@{}l@{}}Seniors (above 60)\\ Disability\end{tabular}                                                                      \\ \hline
RSDI benefits                                                                 & \begin{tabular}[c]{@{}l@{}}Not in labor force and not looking for work\\ Seniors (above 60)\\ Disability\end{tabular}                        \\ \hline
Child support payments received from absent parent                            & \begin{tabular}[c]{@{}l@{}}Employed by other\\ Early midlife (25-39)\end{tabular}                                                            \\ \hline
SSI and/or State SSI supplement                                               & \begin{tabular}[c]{@{}l@{}}Not in labor force and not looking for work\\ Late midlife (40-59)\\ Seniors (above 60)\\ Disability\end{tabular} \\ \hline
Self-employment                                                               & \begin{tabular}[c]{@{}l@{}}Early midlife (25-39)\\ Late midlife (40-59)\\ Self-employed, nonfarming\end{tabular}                             \\ \hline
Unit composition                                                              & \begin{tabular}[c]{@{}l@{}}Early midlife (25-39)\\ Unemployed and looking for work\end{tabular}                                              \\ \hline
Dependent care deduction                                                      & \begin{tabular}[c]{@{}l@{}}Early midlife (25-39)\\ Employed by other\end{tabular}                                                            \\ \hline
Child support payment deduction                                               & \begin{tabular}[c]{@{}l@{}}Employed by other\\ Early midlife (25-39)\\ Late midlife (40-59)\\ Disability\end{tabular}                        \\ \hline
Deduction that should have been included was not                              & \begin{tabular}[c]{@{}l@{}}Not in labor force and not looking for work\\ Seniors (above 60)\\ Disability\end{tabular}                        \\ \hline
Deduction included that should not have been                                  & \begin{tabular}[c]{@{}l@{}}Not in labor force and not looking for work\\ Seniors (above 60)\\ Disability\end{tabular}                        \\ \hline
More income received from this source than budgeted                           & \begin{tabular}[c]{@{}l@{}}Early midlife (25-39)\\ Employed by other\end{tabular}                                                            \\ \hline
Less income received from this source than budgeted                           & \begin{tabular}[c]{@{}l@{}}Employed by other\\ Early midlife (25-39)\end{tabular}                                                            \\ \hline
All income from source known but not included                                 & \begin{tabular}[c]{@{}l@{}}Not in labor force and not looking for work\\ Disability\end{tabular}                                             \\ \hline
Conversion to monthly amount not used or incorrectly applied                  & \begin{tabular}[c]{@{}l@{}}Early midlife\\ Employed by other\end{tabular}                                                                    \\ \hline
Incorrect amount used resulting from change in residence                      & \begin{tabular}[c]{@{}l@{}}Not in labor force and not looking for work\\ Late midlife (40-59)\\ Seniors (above 60)\\ Disability\end{tabular} \\ \hline
Ineligible person(s) included                                                 & \begin{tabular}[c]{@{}l@{}}Unemployed and looking for work\\ Late midlife (40-59)\end{tabular}                                               \\ \hline
Eligible person(s) with no income, resources, or deductible expenses excluded & \begin{tabular}[c]{@{}l@{}}Unemployed and looking for work\\ Young adults (below 24)\\ Early midlife (25-39)\end{tabular}                    \\ \hline
Employment status changed from unemployed to employed                         & \begin{tabular}[c]{@{}l@{}}Young adults (below 24)Early midlife (25-39)\\ Employed by other\end{tabular}                                     \\ \hline
Incorrect standard used                                                       & \begin{tabular}[c]{@{}l@{}}Disability\\ Seniors (above 60)\\ Not in labor force and not looking for work\end{tabular}  \\ \hline             
\end{longtable}

\rmfamily

\FloatBarrier
\subsection{Comparison between multiple-choice and free-response questions}
As discussed in the main text, multiple-choice questions may not fully approximate the kinds of interactions that occur in real-world settings. Often, caseworkers may ask a chatbot open-ended, free-response questions. We chose to use multiple-choice questions as they produce standardized responses that are easier to grade. However, to evaluate how multiple-choice questions may compare to realistic open-ended questions, we compared the chatbot accuracy on both setes of questions using Nava's chatbot.

We selected at least one version of each unique question that appears in our study (n=89). In other words, we did not allow questions to vary by client background attributes. We then converted each multiple-choice question to a comparable free-response question. In some cases, this process involved revising the question text substantively. However, in general, we simply removed the multiple choice answer options.

Testing Nava's chatbot on both the multiple-choice and free-response questions, we find that Nava's chatbot obtains 85\% accuracy on the multiple-choice questions and 76\% accuracy on the free-response questions. To grade the chatbot answers to the multiple-choice questions, we use the standard approach detailed in the Materials and Methods. To grade the chatbot answers to the free-response questions, we prompted GPT-4o with the following prompt ``You are helping classify free-text responses by mapping them to a single multiple choice option. Given a multiple choice question and a free-text response, please select a single multiple choice letter that is most similar to the free text response. Return none if there are no similar options.'' A member of the research team then manually reviewed all of the multiple choice answer options provided, leading to minor corrections in some cases.

The difference in chatbot accuracy between the multiple-choice and free-response questions reflects the challenges associated with prompting chatbots to obtain helpful or relevant information. In many cases, we graded a free-response question as incorrect because the chatbot focused on minor, irrelevant information that didn't answer the question. While there are differences in chatbot accuracy between the two types of questions, we believe this difference underscores the fact that our study examines human behavior under ideal chatbot conditions. Considering the practical challenges associated with obtaining useful chatbot responses is not within our scope.

\FloatBarrier
\section{Background on Nava's Chatbot Design}
We use Nava's custom chatbot as a motivating example for our study.
Nava engaged in an iterative design process including over 75 research sessions with caseworkers, subject matter experts, and families.
The chatbot uses different LLM base models (such as GPT-4o, GPT-4o-mini, and Google Gemini) along with retrieval augmented generation (RAG) to produce responses that contain: (1) a summarized LLM response and (2) direct policy citations from SNAP policy documents.
To efficiently retrieve relevant information from SNAP policy documents, the chatbot developers use an embedding model to compare caseworker's questions to different subsets of lengthy policy documents (often involving thousands of pages). To test the chatbot's effectiveness before expanding to more users, Nava partnered with the non-profit Amplifi, which provides software tools to assist caseworkers. 
Our simulated chatbot suggestions differ from the design of Nava's chatbot. Importantly, our suggestions do not include verbatim policy citations. Our study reflects the potential impacts of chatbot suggestions by themselves on caseworker accuracy.

\FloatBarrier
\section{Assessment Design and Question Characteristics}
\label{sec:si_assessment_design}

In this section, we provide additional details about the process for creating each of the four multiple-choice assessments that participants saw during the experiment.

Table~\ref{tab:question-overlap} summarizes the amount of question overlap across each of the four assessments.
We selected a subset of questions ($n=89$) from the full benchmark dataset to use in the Qualtrics study (the process is described in the section on Assessment Design in the Materials and Methods).
For a question to be eligible for inclusion in our study, we must have been able to obtain correct responses via LLM prompting. We also needed valid incorrect LLM responses for a subset of questions, as caseworkers could see up to 21 possible incorrect suggestions on a given assessment.
Subject to these constraints, we also ensured that questions on each assessment were evenly split by difficulty levels and sought to include a range of error categories within each difficulty level (as shown in Table~\ref{table:categories-across-sets}).

This process led us to select four sets of 45 questions, comprising 89 unique questions overall.
Each of the 89 questions appeared in around 2.02 assessments on average (out of the four total assessments). Some questions appeared in all four assessments.
For seven questions, this was intentional. We selected the following seven questions to appear in all four assessments: three questions with the same set of fixed client attributes and four questions where the correct answer depends on a unique combination of client attributes.

\begin{table}[htbp!]
    \centering
    \begin{tabular}{p{4in}C{1.5in}}
      &  Percent of questions (n=89) \\
      \toprule
     Percent of questions appearing in $> 1$ assessment & 54\%  \\
     Percent of questions appearing in all assessments & 18\% \\
     \midrule
     Average overlap in assessments & 2.02 assessments \\
     \bottomrule
    \end{tabular}
    \caption{Percent of questions that appear in multiple assessments. Since there were only 89 in the entire study, some questions appear in multiple assessments. 54\% of questions appear in more than one assessment; 18\% of questions appear in all four assessments. Some of these questions were chosen deliberately. For example, we chose 3 questions to appear in all four assessments with the same client attributes. The four questions where the correct answer depends on client attributes also appear in all four assessments. The average overlap in assessments (i.e., the average number of assessments in which each question appeared) was 2.02 assessments.}
    \label{tab:question-overlap}
\end{table}

Table~\ref{table:categories-across-sets} shows the distribution of error categories across the four different assessments we generated. While ensuring that assessments were balanced in terms of difficulty level, we sought to approximately balance questions by category. Some categories appear less frequently because there were comparatively fewer question for this category at a range of difficulty levels.

\begin{figure}[htbp!]
    \centering
    \includegraphics[width=0.7\linewidth]{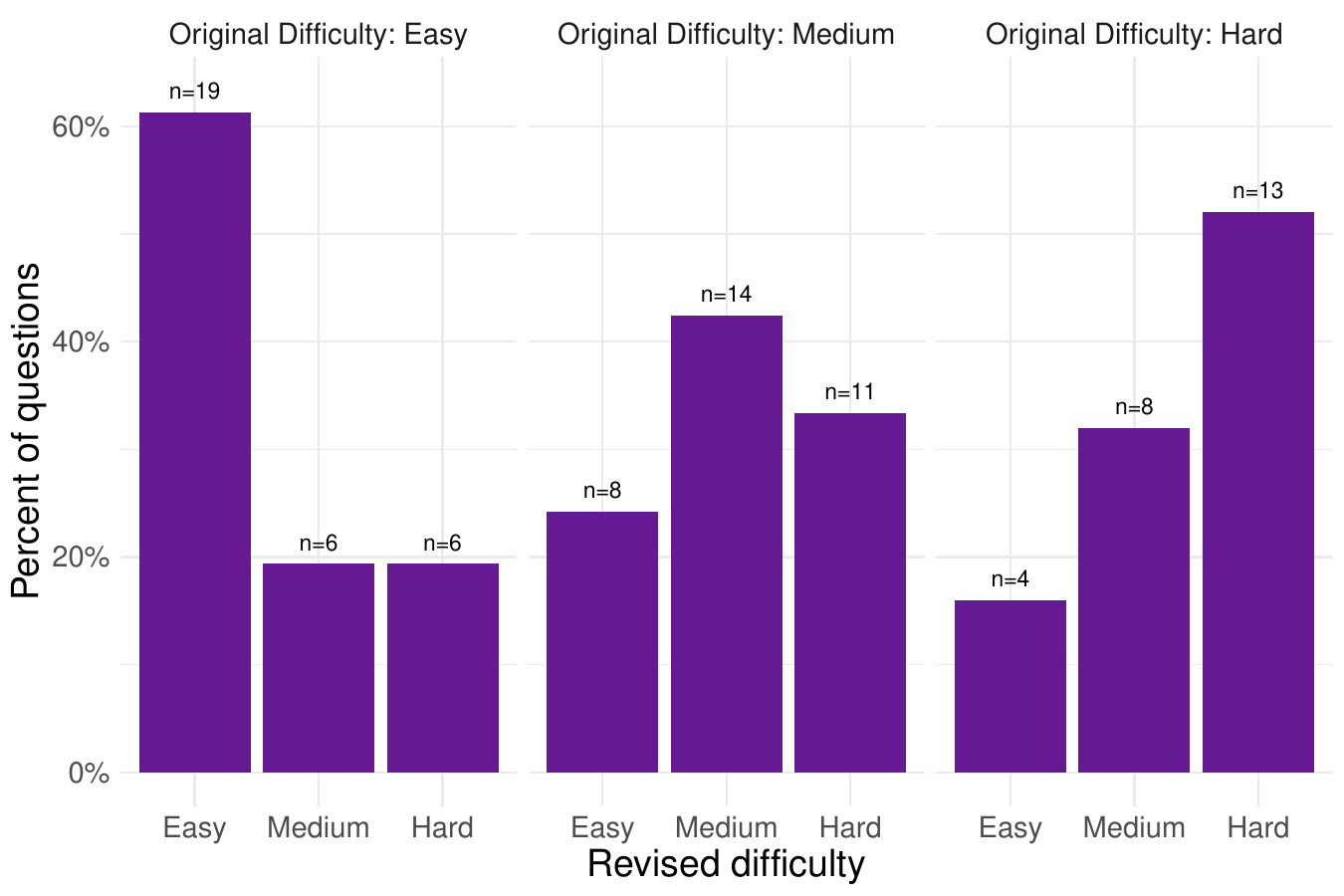}
    \caption{Comparison of original difficulty labels to revised difficulty labels, based on question-level control group performance. To select questions for each assessment, we stratified by difficulty. We used provisional difficulty labels from SNAP QC experts, determined in the the review process of the benchmark dataset questions. However, at the conclusion of the experiment, we assigned new difficulty labels based on the performance of caseworkers in the control group, who did not see any chatbot suggestions. Difficulty values changed in about half of the questions. On the x-axis, we show the revised difficulty labels. In each of the three facets, we show the original difficulty labels (assigned by SNAP QC experts). If none of the difficulty labels changed, we would expect to see bars at 100\% in the Easy, Medium, and Hard categories for each of their respective facets.}
    \label{fig:difficulty_comparison}
\end{figure}

Figure~\ref{fig:difficulty_comparison} compares changes in the difficulty labels assigned to each question. We used provisional difficulty labels based on review from SNAP QC experts. After collecting responses, we developed new difficulty labels based on the performance of caseworkers in the control group, who did not see any chatbot suggestions. We tested two methods for this approach: (1) using a k-means algorithm to cluster questions into three groups based on the average caseworker question-level accuracy and (2) grouping questions into three groups based on the tertiles of the average caseworker question-level accuracy. There was little difference in the two approaches; we chose the second approach.

\newpage 
\begin{table}[htbp!]
\centering 
\begin{tabular}{lC{0.3in}C{0.3in}C{0.3in}C{0.3in}|C{0.3in}}
Category & \multicolumn{4}{c}{Number of questions} \\
& \multicolumn{4}{c}{Assessment number:} \\
 & 1 & 2 & 3 & 4 & Total\\
\hline
Child support payment deduction & 1 & 0 & 0 & 1 & 1\\
Child support payments received from absent parent & 3 & 1 & 2 & 1 & 5\\
Citizenship and noncitizenship status & 1 & 1 & 1 & 2 & 4\\
Combined gross income & 1 & 0 & 1 & 0 & 1\\
Combined net income & 1 & 0 & 1 & 0 & 2\\
Contributions & 4 & 5 & 4 & 4 & 7\\
Dependent care deduction & 1 & 1 & 1 & 1 & 1\\
Medical expense deductions & 5 & 5 & 6 & 5 & 11\\
Other earned income & 1 & 1 & 0 & 1 & 1\\
Other unearned income & 1 & 1 & 2 & 2 & 3\\
RSDI benefits & 2 & 2 & 1 & 1 & 2\\
Recipient disqualification & 2 & 1 & 1 & 2 & 4\\
Reporting systems & 2 & 2 & 2 & 1 & 3\\
SSI and/or State SSI supplement & 4 & 2 & 3 & 3 & 5\\
Self-employment & 4 & 4 & 4 & 3 & 8\\
Shelter deduction & 2 & 3 & 3 & 2 & 5\\
Social Security Number & 1 & 1 & 1 & 1 & 1\\
Standard utility allowance & 2 & 5 & 3 & 4 & 7\\
Student status & 1 & 0 & 1 & 1 & 1\\
Unit composition & 1 & 2 & 2 & 2 & 5\\
Wages and salaries & 5 & 6 & 4 & 5 & 8\\
Other government benefits & 0 & 1 & 1 & 2 & 2\\
Unemployment compensation & 0 & 1 & 1 & 1 & 2\\
\midrule
Total & 45 & 45  & 45 & 45 & 89 \\
\bottomrule
\end{tabular}
\caption{Distribution of categories across the four different multiple-choice assessments in our study. Some topics, such as ``Medical expense deductions'' or ``Wages and salaries'' appear more frequently because there are comparatively more unique questions on these topics. Each of the four assessments has 45 questions in total, and there were 89 questions in the entire study.}
\label{table:categories-across-sets}
\end{table}

\begin{table}[htbp!]
    \centering
    \begin{tabular}{lp{2.2in}C{0.3in}C{0.3in}C{0.3in}C{0.3in}|C{0.3in}}
& & \multicolumn{4}{c}{Number of questions} \\
& & \multicolumn{4}{c}{Assessment number:} \\
Client attribute & Value & 1 & 2 & 3 & 4 & Total\\
\hline
 & Adult (22-59) & 34 & 34 & 35 & 37 & 71\\
\multirow{-2}{*}{\raggedright\arraybackslash Age} & Senior (60+) & 27 & 29 & 31 & 29 & 63\\
 & All U.S. Citizens & 44 & 44 & 44 & 44 & 86\\
\multirow{-2}{*}{\raggedright\arraybackslash Citizenship} & Mixed household (eligible citizens and ineligible non-citizens) & 1 & 1 & 1 & 1 & 3\\
 & Not legally disabled & 27 & 26 & 28 & 28 & 54\\
\multirow{-2}{*}{\raggedright\arraybackslash Disability} & Receives SSI/RSDI & 20 & 21 & 19 & 19 & 37\\
 & Employed by others & 18 & 12 & 12 & 14 & 26\\
 & Not in labor force & 23 & 29 & 29 & 28 & 55\\
\multirow{-3}{*}{\raggedright\arraybackslash Employment} & Self-employed & 6 & 6 & 6 & 5 & 10\\
 & Female & 36 & 34 & 34 & 32 & 73\\
\multirow{-2}{*}{\raggedright\arraybackslash Gender} & Male & 33 & 33 & 30 & 32 & 66\\
\midrule
& Client attribute varies by age or gender; answer is fixed & 38 & 38 & 38 & 38 & 82\\
& Client attribute varies; correct answer changes & 4 & 4 & 4 & 4 & 4\\
\multirow{-4}{*}{\raggedright\arraybackslash Question variations} & Client attributes do not change & 3 & 3 & 3 & 3 & 3\\
\bottomrule
\end{tabular}
    \caption{Each of the four assessments is comprised of 45 multiple-choice questions. We produce two versions of each assessment where we vary at least one hypothetical client attribute for a subset of questions. In 38 out of the 45 questions, we vary either age or gender, which has no effect on the correct answer. We do not vary questions where age or gender would meaningfully change the content of the question (e.g., a medical deduction question pertaining to a client who is over the age of 60). For an additional four questions, the correct answer depends on specific client attributes. We vary the client attributes that would meaningfully affect the correct answer in this cases (e.g., age or employment status). Lastly, there are three questions that appear in all four versions of the assessment with the same client attributes.}
    \label{tab:placeholder}
\end{table}

\FloatBarrier

\begin{table}[htbp]
    \centering
    \begin{tabular}{l|c}
    Chatbot accuracy (number of incorrect suggestions) & n (\%) \\
    \hline
    53\% (21) & 8 (6)\\
    60\% (18) & 5 (4)\\
    67\% (15) & 6 (5)\\
    73\% (12) & 7 (6)\\
    80\% (9) & 6 (5)\\
    87\% (6) & 5 (4)\\
    93\% (3) & 6 (5)\\
    96\% (2) & 17 (14)\\
    98\% (1) & 15 (12)\\
    100\% (0) & 19 (15)\\
    \hline
       Control & 31 (25)\\
    \end{tabular}
    \caption{Number of participants in each experiment condition. Between 12-15\% of participants were assigned to high-accuracy conditions (96-100\% accurate chatbots). Between 5-8\% of participants were assigned to low and medium accuracy conditions (53-93\% accurate chatbots). 25\% of participants were in the control condition.}
    \label{tab:random_assignment}
\end{table}

\section{Random Assignment}
In this section, we discuss the random assignment process. We randomly assigned participants to (1) a control condition, in which they did not see any chatbot suggestions, or (2) a treatment condition, in which they saw chatbot suggestions of varying overall accuracy. Additionally, we randomly assigned participants to each of the four unique assessment types: $\approx30$ participants to each.

The actual number of participants in the study assigned to each experiment condition is shown in Table~\ref{tab:random_assignment}. We assigned 31 participants (25\%) to the control condition. For participants assigned to see chatbot suggestions, we randomly varied the overall accuracy of the chatbot based on the following number of incorrect suggestions: $0, 1, 2, 3, 6, 9, 12, 15, 18, 21$. We oversampled on assignments to high-accuracy chatbots ( 0, 1, or 2 incorrect suggestions), assigning between 12-15\% of participants to each ($41\%$ in total). We assigned around one third of participants to conditions with $3, \ldots, 21$ incorrect chatbot suggestions. 

\section{Selecting Incorrect Chatbot Suggestions}
\begin{figure}[t!]
    \centering
    \begin{minipage}{\textwidth}
    \centering 
    \includegraphics[width=0.6\linewidth]{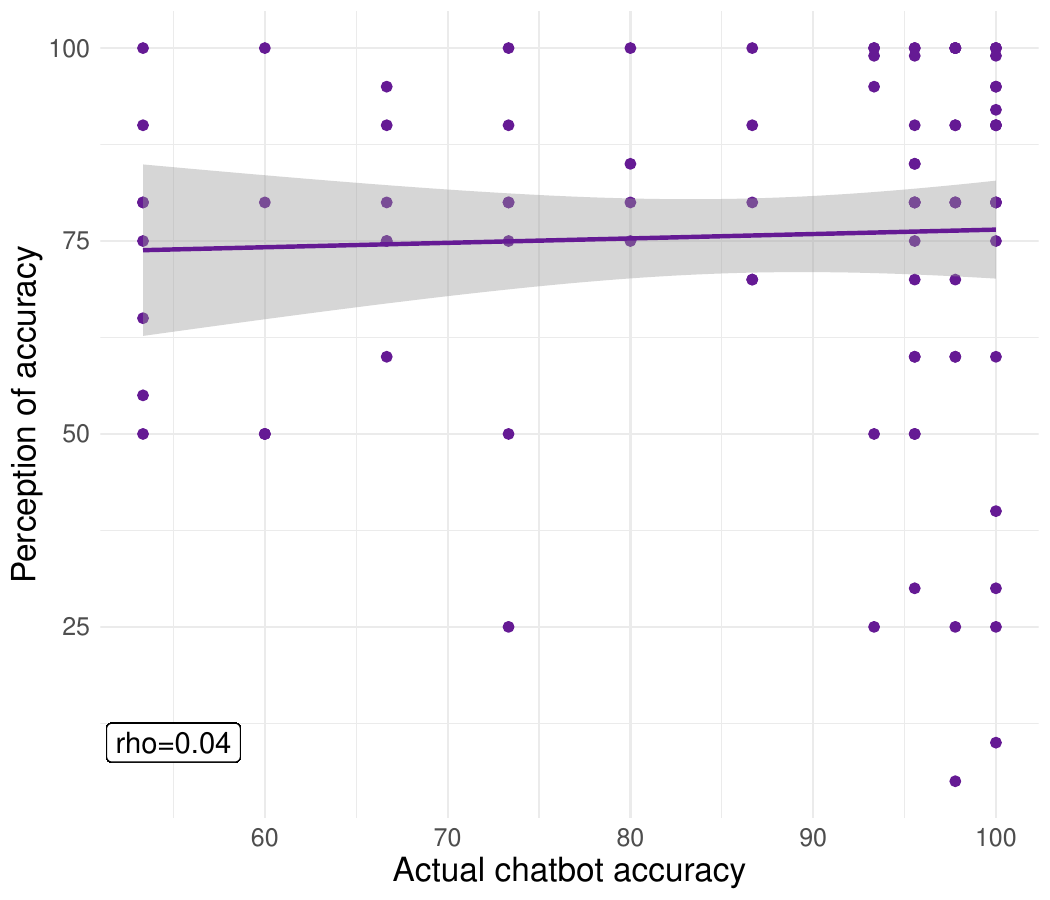}
    \caption{Relationship between caseworker perception of accuracy (y-axis) and actual chatbot accuracy (x-axis). While the two variables should be highly correlated, the Pearson correlation coefficient is 0.04 (annotated in the lower lefthand corner of the figure). Caseworkers appear to struggle in evaluating the accuracy of the chatbot.}
    \label{fig:accuracy_perception_corr_all}
    \end{minipage}
    \begin{minipage}{\textwidth}
    \centering
    \includegraphics[width=0.65\linewidth]{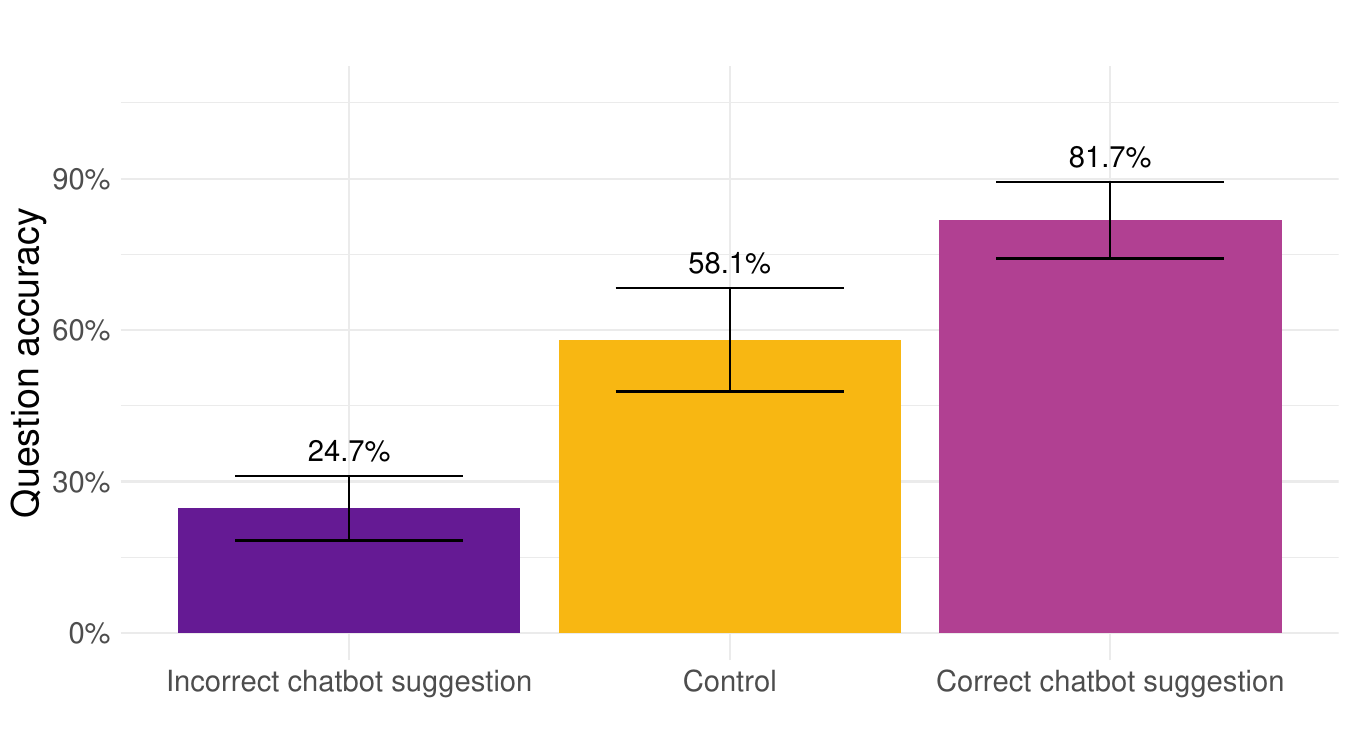}
    \caption{Question-level accuracy is shown for the three questions that appear in all assessments with the same set of client attributes. Similar to Figure~3 in the main text, incorrect chatbot suggestions lead to large reductions in caseworker accuracy relative to accuracy in the control group (which is higher, on average, than the control-group accuracy shown for all questions in Figure~3). Correct suggestions also lead to improvements in caseworker accuracy relative to the control group. Different colors (purple, yellow, and light purple) indicate the type of chatbot suggestion. Caseworker question-level accuracy appears on the y-axis. We include error bars with the 95\% confidence intervals for all estimates.}
    \label{fig:special_qs}
    \end{minipage}
\end{figure}

\begin{figure}[t!]
    \centering
    \includegraphics[width=0.8\linewidth]{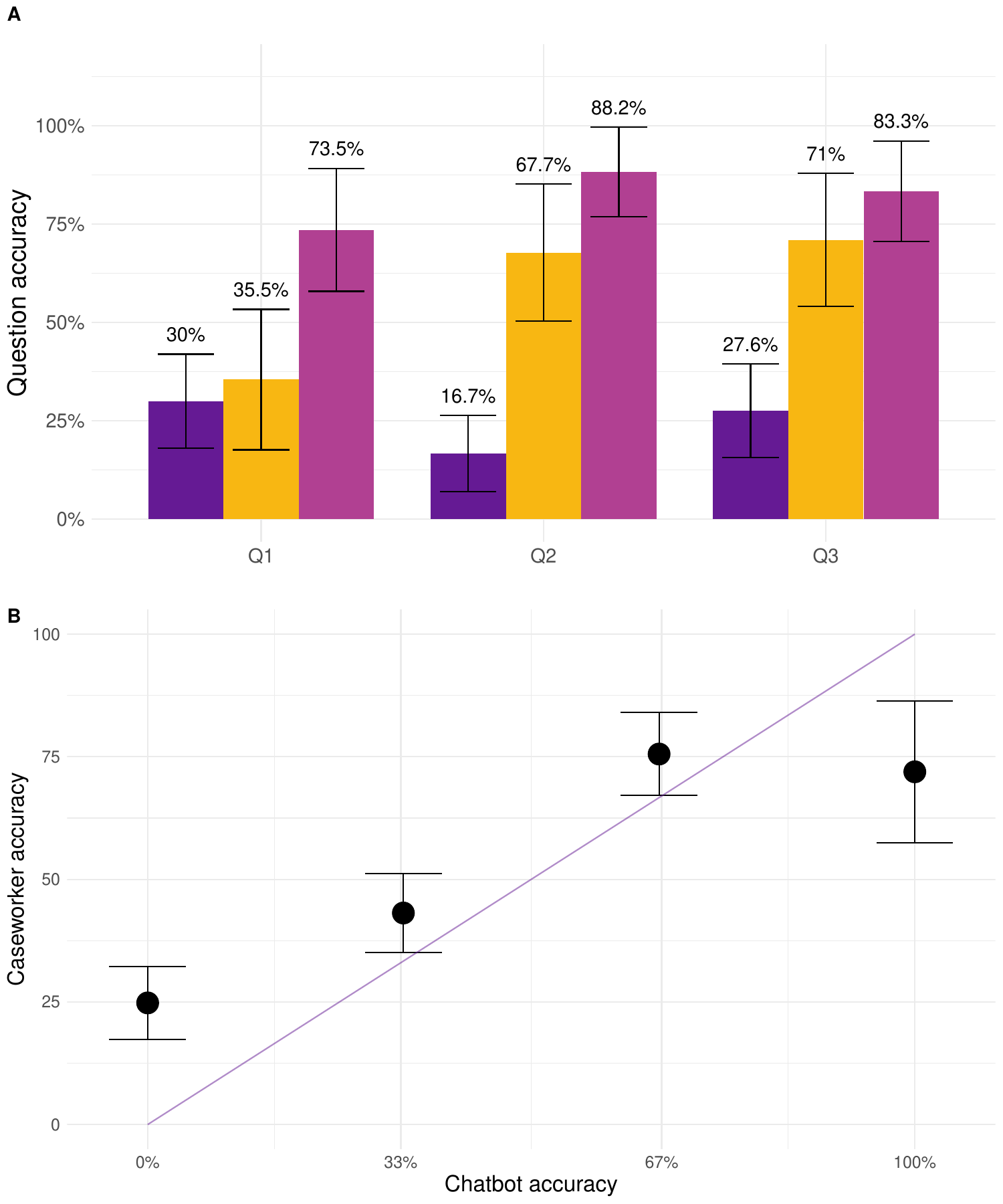}
    \caption{\small{(A) Question-level accuracy is shown for the three questions that appear in all assessments with the same set of client attributes. In contrast to Figure~\ref{fig:special_qs}, accuracy is shown for each specific question (denoted ``Q1,'' ``Q2,'' and ``Q3''). Incorrect chatbot suggestions lead to large reductions in caseworker accuracy relative to accuracy in the control group  for Q2 and Q3, where accuracy in the control group is high. Correct suggestions lead to improvements in caseworker accuracy relative to the control group in Q1, where accuracy in the control group is low. Different colors (purple, yellow, and light purple) indicate the type of chatbot suggestion. Caseworker question-level accuracy appears on the y-axis. We include error bars with the 95\% confidence intervals for all estimates. (B) Caseworker participant-level accuracy is shown only for the three questions that appear in all assessments with the same set of client attributes. The simulated chatbot accuracy for the three questions is shown on the x-axis. Simulated chatbot accuracy ranges from 0\% (all three questions have incorrect chatbot suggestions) to 100\% (all three questions have correct chatbot suggestions). Similar to Figure 2B in the main text, there is slight evidence for a plateau effect between the medium (66.67\%) and high (100\%) accuracy conditions. Error bars denote the 95\% confidence intervals for all estimates.}}
    \label{fig:special_qs_X_question}
\end{figure}

In all four of the assessments, we selected three questions to appear consistently with the same client attributes. Each of these questions (which we will call ``Q1,'' ``Q2,'' and ``Q3'') had both a correct and incorrect chatbot suggestion. Each one had a different difficulty label, based on the provisional difficulty labels from SNAP QC experts (e.g., Q1 was easy, Q3 was medium, and Q3 was hard). We present question-level accuracy results for these three questions (overall and for each individual question) in Figures~\ref{fig:special_qs} and \ref{fig:special_qs_X_question}.

In conditions with only $1$ or $2$ incorrect chatbot suggestions, we randomly assigned participants to see incorrect chatbot suggestions for a subset of Q1-Q3 (e.g., Q1 and Q3 or Q1 and Q2).
For conditions with $3$ incorrect chatbot suggestions, participants saw incorrect chatbot suggestions for all 3 (i.e., Q1, Q2, and Q3).

For conditions where the number of incorrect suggestions was more than 3, we cumulatively added incorrect suggestions, drawing from incorrect suggestions for other questions. We added incorrect suggestions in increments of 3 to ensure that we had an incorrect suggestion for questions from each difficulty level (based on the provisional difficulty labels from SNAP QC experts).

For example, an assessment with 6 incorrect suggestions included the three incorrect suggestions for Q1-Q3 along with 3 new incorrect suggestions for an easy, medium, and hard question.
In other words, questions with an incorrect chatbot suggestion Q1-Q3 would always appear in the $6,\ldots,21$ conditions.

There were 4 questions in our study where the correct answer depends on client attributes. In each of the four assessments, we randomly selected two of these four questions to appear with incorrect suggestions. We included these incorrect suggestions in either the $6$ or $9$ incorrect chatbot suggestion condition, depending on the provisional difficulty labels of the questions.\footnote{In cases where two of the questions had the same difficulty level, we needed to split the questions across the $6$ and $9$ incorrect conditions to preserve stratification by difficulty.}

\section{Incorrect Chatbot Suggestion Position}
Each question in our study could appear in positions 1 through 45.
We placed incorrect suggestions deliberately throughout the assessment with some degree of randomness, to prevent a predictable pattern that participants could anticipate.

We first assigned questions with incorrect suggestions to one of $22$ windows, spacing the incorrect suggestions as evenly as possible across all $22$ windows.
Each window aligned with two positions in the actual assessment (e.g., the first two questions, the third and fourth questions, etc.). Once assigned to a window, each question was randomly assigned to either the first or second position in the window. We used these assignments to determine the final position assignments question of questions with incorrect chatbot suggestions.

For example, if an incorrect chatbot suggestion was assigned to the 22nd window and the 1st position, the question would appear in the 43rd position in the assessment. In conditions with only $1$ incorrect chatbot suggestion, the incorrect suggestion appears in either position 43 or 44. In conditions with $2$ incorrect chatbot suggestions, the incorrect chatbot suggestions appear in either the 21st or 22nd position and in either the 43rd or 44th position. 

For questions that did not have incorrect chatbot suggestions, we randomly assigned them to question positions throughout the survey. As a result, this means that participants who took the same survey -- but with a different number of incorrect chatbot suggestions or in the control group would -- would have seen the quesitons in a different order.

\section{Power Analysis}
\label{sec:power_analysis}

We conducted simulation-based power analysis to determine appropriate sample size and participant allocation strategy. Each scenario included 40 participants per simulation run and was replicated 500 times. In each run, 25\% of participants were assigned to the Human-only control group and the remaining 75\% to the Human+LLM treatment group. All participants answered a total of 45 questions, evenly distributed across difficulty levels (15 easy, 15 medium, and 15 hard). Within this set, 3 questions, one at each difficulty level, were preserved across participants to allow controlled placement of LLM-generated errors.

Participant-level covariates included experience and LLM familiarity. Experience was simulated from a gamma distribution and truncated at a maximum of 5, while LLM familiarity was modeled as a binary variable drawn from a Bernoulli distribution with a 70\% probability of familiarity. Each participant’s baseline accuracy was determined by the difficulty of the question: 0.70 for easy, 0.60 for medium, and 0.50 for hard questions. To reflect realistic variation in performance, we added random noise at both the person level and question level.

Participants in the Human+LLM group were randomly assigned a target number of incorrect chatbot suggestions. 15\% of this group received one error, another 15\% received two errors, and the remaining participants were evenly distributed across conditions with 0, 3, 6, 9, 12, 15, 18, or 21 errors. Incorrect suggestions were distributed evenly across difficulty levels to avoid confounding effects. We intentionally oversampled participants exposed to high-quality chatbot suggestions to ensure that a sufficient number of participants could experience the potential benefits of LLM assistance. At the same time, we introduced variation in LLM correctness across participants to enable identification of behavioral patterns such as over-reliance on inaccurate suggestions and disengagement following repeated errors.

Trust in chatbot suggestions was modeled dynamically based on a combination of participant characteristics and task experience. Each participant was assigned a baseline trust score using a logistic function of LLM familiarity and experience. As participants progressed through the task, a running average of previous correct chatbot suggestions, referred to as trust memory, was updated cumulatively to reflect perceived reliability. The final probability of trusting the LLM on each question was computed as the product of the baseline trust score and trust memory.
If a participant had encountered a number of LLM errors that exceeded a predefined threshold, they were classified as disengaged and automatically rejected all further suggestions. Otherwise, trust decisions were drawn from a Bernoulli distribution using the computed trust probability. This trust modeling approach captures key behavioral dynamics, including adaptive learning from the LLM performance, declining trust following repeated errors, and heterogeneous trust tendencies driven by individual characteristics.

The accuracy of each response was determined by the baseline accuracy associated with question difficulty, adjusted by participant-level and question-level random noise, and whether the participant chose to trust the chatbot suggestion. If the chatbot suggestion was trusted and correct, the response was coded as correct; if the chatbot suggestion was trusted but incorrect, accuracy is penalized, modeled as a reduction of 0.1 from their independently predicted accuracy.
If the suggestion was not trusted, accuracy reflects the participant’s independent baseline, which incorporates the difficulty-specific base rate, adjusted for their experience and random noise at both person and question level.

We assumed an average treatment effect of 0.1 for LLM assistance on accuracy and a threshold effect that emerges when participants are exposed to 18 or more incorrect chatbot suggestions. Subgroup effects were modeled by assigning small to moderate differential impacts (ranging from ±0.05 to ±0.10) on accuracy depending on participant or question attributes. Our power analysis indicates that at least three batches of participants—totaling 120 participants—is required to achieve at least 80\% power across the planned analyses. This includes 30 participants in the Human-only control group (25\%) and at least 90 participants in the Human+LLM treatment group (75\%).

\FloatBarrier
\newpage
\section{Attention check}
\label{sec:attention check}

\begin{figure}[htbp]
    \centering
    \includegraphics[width=0.5\linewidth]{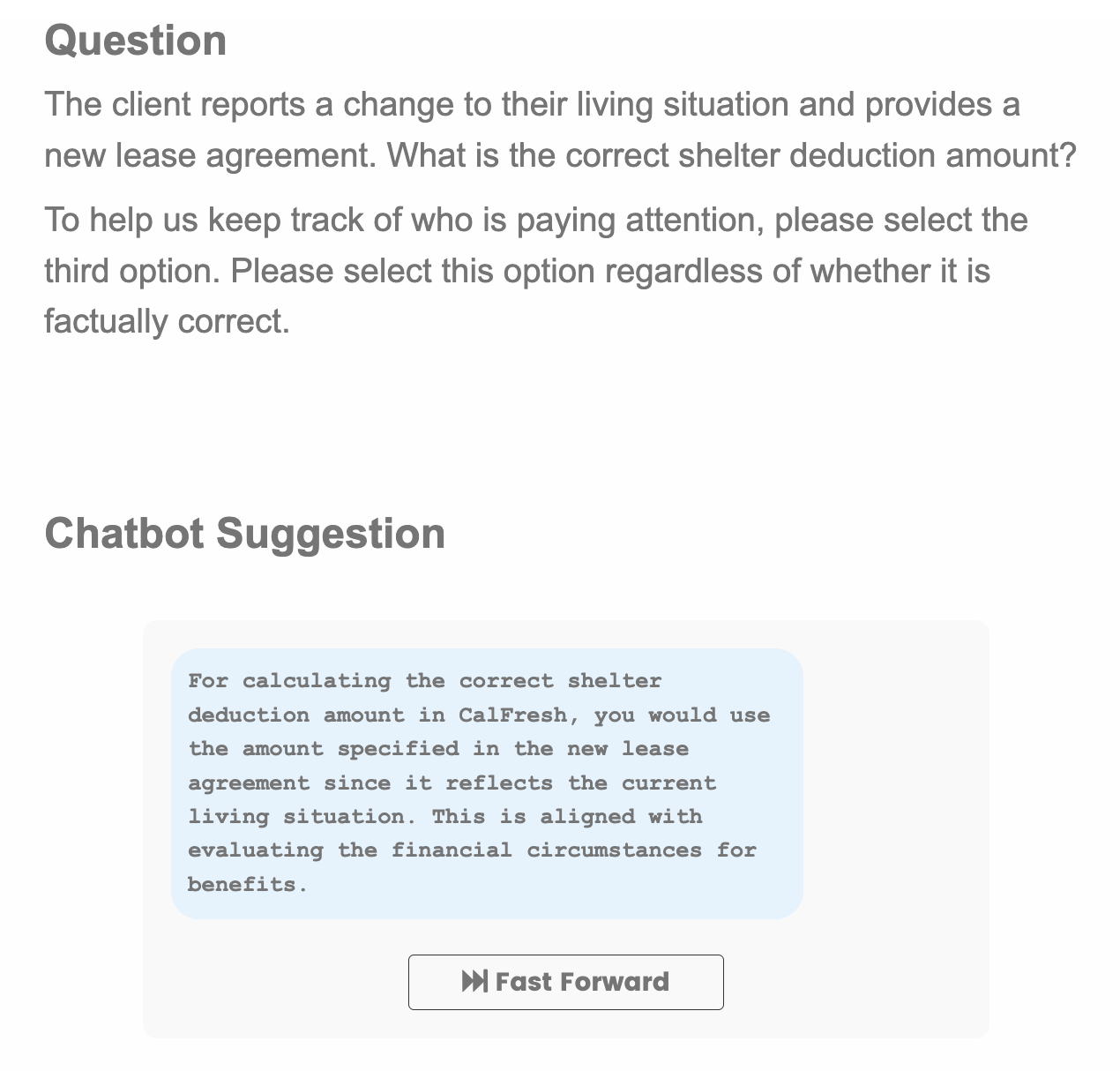}
    \caption{Example of the attention check that we included in the middle of the survey (after question 23). 37\% of participants in our assessment failed the attention check, meaning they did not select the third answer option. We present results for all participants in the main text, and additional results only for those who pass the attention check in the SI Appendix.}
    \label{fig:attention}
\end{figure}

\FloatBarrier

\section{Participant Characteristics}

In this section, we discuss the characteristics of participants in our study. We recruited caseworkers from over 30 nonprofit organizations in California that partner with the non-profit Amplifi. Caseworkers support clients in different languages, primarily English and Spanish (as shown in Table~\ref{tab:languages}). On average, caseworkers had about four years of experience (Table~\ref{tab:continuous_attributes}) and all caseworkers knew CalFresh at least slightly well (Table~\ref{tab:categorical_scale_questions}).

\begin{table}[htbp]
    \centering
    \begin{tabular}{lcc}
    & Passed attention check (n=79) & All participants (n=125) \\
    Languages other than English & n (\%) & n (\%) \\
    \toprule 
    Spanish & 56 (47.86) & 92 (78.63)\\
    Mandarin & 2 (1.71) & 3 (2.56)\\
    Korean & 2 (1.71) & 2 (1.71)\\
    Cantonese & 1 (0.85) & 2 (1.71)\\
    Other & 2 (1.71) & 4 (3.42)\\
    \midrule
    English only & 18 (15.38) & 24 (20.51)\\
    \bottomrule
    \end{tabular}
    \caption{The number and percentage of participants who support clients in languages other than English. Many participants support clients in Spanish. A smaller percentage (<3\%) support clients in Mandarin, Cantonese, or Korean.}
    \label{tab:languages}
\end{table}

\begin{table}[htbp!]
\centering
\adjustbox{max width=\textwidth}{
\begin{tabular}[t]{lcc|cc}
& \multicolumn{2}{c}{Passed attention check (n=79)} & \multicolumn{2}{c}{All Participants (n=125)} \\
  & Median & Mean & Median & Mean \\
\toprule
  Caseworker years of experience & 3
(1-5) & 4.06
(3.06-5.06) & 3
(1-5) & 3.94
(3.21-4.67)\\
  Questions similar to day-to-day situations (\%) & 50
(21.25-70) & 47.77
(41.38-54.16) & 50
(30-80) & 51.38
(46.16-56.6)\\
 Chatbot suggestions perceived as correct (\%) & 80
(70-95) & 77.25
(71.64-82.87) & 80
(60-95) & 75.7
(70.98-80.42)\\
Chatbot suggestions used (\%)  & 80
(52.5-95) & 73.04
(66.73-79.34) & 80
(50-90) & 70.04
(64.64-75.45)\\
\hline
\end{tabular}}
\caption{Distribution of participant characteristics and post-assessment question responses. We present the median and mean values for each of these questions (1) across all participants (n=125), and (2) only participants who passed the attention check (n=79). We include the inter-quartile range for the median values  and the 95\% confidence intervals for the mean values in parentheses.}
\label{tab:continuous_attributes}
\end{table}

\begin{longtable}[c]{llC{1in}|C{1in}}
& & Passed Attention Check (n=79) & All Participants (n=125) \\
\endhead
Question & Answers & n (\%) & n (\%) \\
\hline
\multirow{5}{1.3in}{\raggedright\arraybackslash Program Knowledge of CalFresh}  & Not well at all & 0 (0\%) & 0 (0\%)\\

 & Slightly well & 18 (23.38\%) & 23 (18.85\%)\\

 & Moderately well & 23 (29.87\%) & 43 (35.25\%)\\

 & Very well & 27 (35.06\%) & 39 (31.97\%)\\
& Extremely well & 9 (11.69\%) & 17 (13.93\%)\\
\hline
 & Never & 19 (24.36\%) & 33 (26.61\%)\\

 & Once a month or less & 17 (21.79\%) & 25 (20.16\%)\\

 & A couple times a month & 15 (19.23\%) & 24 (19.35\%)\\

 & At least once a week & 13 (16.67\%) & 22 (17.74\%)\\

 & About once a day & 8 (10.26\%) & 12 (9.68\%)\\

\multirow{-6}{1.3in}{\raggedright\arraybackslash Existing Chatbot Use} & More than once a day & 6 (7.69\%) & 8 (6.45\%)\\
\hline 
& More harm than benefit & 4 (5.13\%) & 7 (5.65\%)\\
 & About an equal mix of benefit and harm & 34 (43.59\%) & 61 (49.19\%)\\
\multirow{-3}{1.3in}{\raggedright\arraybackslash Whether chatbots are harmful or beneficial} & More benefit than harm & 40 (51.28\%) & 56 (45.16\%)\\
\hline
& Very negative & 3 (3.85\%) & 4 (3.23\%)\\

 & Somewhat negative & 7 (8.97\%) & 9 (7.26\%)\\

 & Neither positive nor negative & 27 (34.62\%) & 45 (36.29\%)\\

 & Somewhat positive & 27 (34.62\%) & 45 (36.29\%)\\

\multirow{-5}{*}{\raggedright\arraybackslash Views on chatbots} & Very positive & 14 (17.95\%) & 21 (16.94\%)\\
\hline
&  Somewhat difficult & 22 (28.21\%) & 24 (19.51\%)\\

 & Neither easy nor difficult & 20 (25.64\%) & 37 (30.08\%)\\

 & Somewhat easy & 22 (28.21\%) & 41 (33.33\%)\\

\multirow{-4}{1.3in}{\raggedright\arraybackslash How easy or difficult to find answers to questions} & Very easy & 14 (17.95\%) & 21 (17.07\%)\\
\hline
& Once a month or less & 18 (23.08\%) & 28 (22.76\%)\\

 & A couple times a month & 26 (33.33\%) & 36 (29.27\%)\\

 & At least once a week & 22 (28.21\%) & 37 (30.08\%)\\

 & About once a day & 7 (8.97\%) & 11 (8.94\%)\\

\multirow{-5}{1.3in}{\raggedright\arraybackslash How often you run into questions that you do not know the answer to} & More than once a day & 5 (6.41\%) & 11 (8.94\%)\\
\hline
 & Never & 1 (1.28\%) & 2 (1.63\%)\\

 & Once a month or less & 4 (5.13\%) & 9 (7.32\%)\\

 & A couple times a month & 11 (14.1\%) & 18 (14.63\%)\\

 & At least once a week & 9 (11.54\%) & 17 (13.82\%)\\

 & About once a day & 13 (16.67\%) & 22 (17.89\%)\\

\multirow{-6}{1.3in}{\raggedright\arraybackslash How often do you use: Google} & More than once a day & 40 (51.28\%) & 55 (44.72\%)\\
\hline
 & Never & 6 (7.69\%) & 8 (6.5\%)\\

 & Once a month or less & 5 (6.41\%) & 9 (7.32\%)\\

 & A couple times a month & 14 (17.95\%) & 23 (18.7\%)\\

 & At least once a week & 17 (21.79\%) & 27 (21.95\%)\\

 & About once a day & 18 (23.08\%) & 25 (20.33\%)\\

\multirow{-6}{1.3in}{\raggedright\arraybackslash How often do you use: Slack/Email} & More than once a day & 18 (23.08\%) & 31 (25.2\%)\\
\hline
& Never & 15 (19.23\%) & 19 (15.45\%)\\

 & Once a month or less & 22 (28.21\%) & 34 (27.64\%)\\

 & A couple times a month & 16 (20.51\%) & 25 (20.33\%)\\

 & At least once a week & 16 (20.51\%) & 24 (19.51\%)\\

 & About once a day & 5 (6.41\%) & 9 (7.32\%)\\

\multirow{-6}{1.3in}{\raggedright\arraybackslash How often do you use: Policy Manuals} & More than once a day & 4 (5.13\%) & 12 (9.76\%)\\
\hline
& Never & 26 (33.33\%) & 40 (32.52\%)\\

 & Once a month or less & 14 (17.95\%) & 25 (20.33\%)\\

 & A couple times a month & 13 (16.67\%) & 20 (16.26\%)\\

 & At least once a week & 7 (8.97\%) & 14 (11.38\%)\\

 & About once a day & 9 (11.54\%) & 14 (11.38\%)\\

\multirow{-6}{1.3in}{\raggedright\arraybackslash How often do you use: Chatbots} & More than once a day & 9 (11.54\%) & 10 (8.13\%)\\
\hline
 & Never & 3 (3.85\%) & 3 (2.44\%)\\

 & Once a month or less & 8 (10.26\%) & 13 (10.57\%)\\

 & A couple times a month & 14 (17.95\%) & 20 (16.26\%)\\

 & At least once a week & 16 (20.51\%) & 30 (24.39\%)\\

 & About once a day & 19 (24.36\%) & 25 (20.33\%)\\

\multirow{-6}{1.3in}{\raggedright\arraybackslash How often do you use: In Person} & More than once a day & 18 (23.08\%) & 32 (26.02\%)\\
\hline
& Never & 5 (6.41\%) & 9 (7.32\%)\\

 & Once a month or less & 14 (17.95\%) & 20 (16.26\%)\\

 & A couple times a month & 13 (16.67\%) & 18 (14.63\%)\\

 & At least once a week & 19 (24.36\%) & 32 (26.02\%)\\

 & About once a day & 13 (16.67\%) & 21 (17.07\%)\\

\multirow{-6}{1.3in}{\raggedright\arraybackslash How often do you use: Phone} & More than once a day & 14 (17.95\%) & 23 (18.7\%)\\
\hline
& Very low mental effort & 1 (1.27\%) & 2 (1.6\%)\\

 & Low mental effort & 7 (8.86\%) & 8 (6.4\%)\\

 & Neither low nor high mental effort & 25 (31.65\%) & 45 (36\%)\\

 & High mental effort & 40 (50.63\%) & 59 (47.2\%)\\
\multirow{-5}{1.3in}{\raggedright\arraybackslash Mental Effort} & Very high mental effort & 6 (7.59\%) & 11 (8.8\%)\\
\hline
& Not at all familiar & 2 (2.53\%) & 4 (3.2\%)\\

 & Slightly familiar & 21 (26.58\%) & 29 (23.2\%)\\

 & Somewhat familiar & 33 (41.77\%) & 53 (42.4\%)\\

 & Familiar & 20 (25.32\%) & 32 (25.6\%)\\

\multirow{-5}{1.3in}{\raggedright\arraybackslash Familiarity with questions} & Very familiar & 3 (3.8\%) & 7 (5.6\%)\\
\midrule
\multicolumn{4}{c}{Questions only asked of participants who saw chatbot suggestions (n=94):}\\
\midrule
& Not at all helpful & 0 (0\%) & 0 (0\%)\\
& Slightly helpful & 4 (7.14\%) & 7 (7.45\%)\\

 & Moderately helpful & 11 (19.64\%) & 16 (17.02\%)\\

 & Very helpful & 25 (44.64\%) & 39 (41.49\%)\\

\multirow{-4}{1.3in}{\raggedright\arraybackslash Perception of chatbot helpfulness} & Extremely helpful & 16 (28.57\%) & 32 (34.04\%)\\
\hline
& Not at all interested & 0 (0\%) & 2 (2.13\%)\\

 & Slightly interested & 8 (14.29\%) & 10 (10.64\%)\\

 & Moderately interested & 15 (26.79\%) & 20 (21.28\%)\\

 & Very interested & 22 (39.29\%) & 36 (38.3\%)\\

\multirow{-5}{1.3in}{\raggedright\arraybackslash Interested in using chatbot} & Extremely interested & 11 (19.64\%) & 26 (27.66\%)\\
\bottomrule 
\caption{Participant responses to Likert-scale questions in the assessment on topics related to prior knowledge of CalFresh, attitudes toward chatbots and existing chatbot usage, and methods for answering client questions. There were two questions only asked of participants who saw the chatbot suggestions in the assessment: whether they thought the chatbot suggestions were helpful and if they were interested in using the chatbot in the future. We present the frequency and percentage of responses for each answer option across \emph{all participants (n=125)} and \emph{only participants who passed the attention check (n=79)}.
We briefly discuss trends for participants who passed the attention check (n=79); trends among all participants are comparable.
All participants passing the attention check have at least some knowledge of CalFresh. Most participants reported at least some prior chatbot use (76\%) and have either neutral or positive perceptions of chatbots (87\%). Participants use a variety of resources to answer client questions: Google, Slack/email, phone, and in-person resources are the most frequently used. For example, more than half of participants (68\%) use Google at least once a day. Slightly more than half of participants (58\%) reported that the assessment questions involved high mental effort. Almost all participants (97\%) were at least slightly familiar with the question material. Among questions only asked of those who saw chatbot suggestions: most participants thought the suggestions were at least moderately helpful and were at least moderately interested in using the chatbot in the future.}
\label{tab:categorical_scale_questions}
\end{longtable}

\section{Baseline Equivalence Table}
In this section, we check for any differences in pre-treatment characteristics between caseworkers in the treatment and control groups. We do not observe any statistically significant differences between the two groups for significance thresholds of 0.05 or less.

\begin{table}[htbp!]
    \centering
    \begin{tabular}{lcc}
        & Treatment group & Control group \\
        & (n=94) & (n=31) \\
        \toprule 
         Years of experience & 3.94 & 3.95\\
Supports clients in a language other than English & 78.41\% & 82.76\% \\
\bottomrule 
    \end{tabular}
    \caption{Differences between participants in the treatment and control groups for (1) average baseline years of experience and (2) the percent of caseworkers who support clients in a language other than English. We test for statistical significance using t-tests. The differences between the two groups are not statistically significant at the 0.05 or less.}
    \label{tab:bequiv_continuous}
\end{table}

\begin{longtable}[c]{llC{1in}|C{1in}}
\sffamily 
& & Treatment group (n=94) & Control group (n=31) \\
\endhead
Question & Answers & n (\%) & n (\%) \\
\hline
 & Slightly well & 5 (16.13\%) & 18 (19.78\%)\\

 & Moderately well & 9 (29.03\%) & 34 (37.36\%)\\

 & Very well & 12 (38.71\%) & 27 (29.67\%)\\

\multirow{-4}{1.3in}{\raggedright\arraybackslash Program Knowledge of CalFresh} & Extremely well & 5 (16.13\%) & 12 (13.19\%)\\
\hline
 & Never & 5 (16.13\%) & 28 (30.11\%)\\

 & Once a month or less & 6 (19.35\%) & 19 (20.43\%)\\

 & A couple times a month & 5 (16.13\%) & 19 (20.43\%)\\

 & At least once a week & 6 (19.35\%) & 16 (17.2\%)\\

 & About once a day & 7 (22.58\%) & 5 (5.38\%)\\

\multirow{-6}{1.3in}{\raggedright\arraybackslash Existing Chatbot Use} & More than once a day & 2 (6.45\%) & 6 (6.45\%)\\
\hline
 & More harm than benefit & 2 (6.45\%) & 5 (5.38\%)\\

 & About an equal mix of benefit and harm & 15 (48.39\%) & 46 (49.46\%)\\

\multirow{-3}{1.3in}{\raggedright\arraybackslash Whether chatbots are beneficial or harmful} & More benefit than harm & 14 (45.16\%) & 42 (45.16\%)\\
\hline
 & Very negative & 0 (0\%) & 4 (4.3\%)\\

 & Somewhat negative & 2 (6.45\%) & 7 (7.53\%)\\

 & Neither positive nor negative & 10 (32.26\%) & 35 (37.63\%)\\

 & Somewhat positive & 12 (38.71\%) & 33 (35.48\%)\\

\multirow{-5}{1.3in}{\raggedright\arraybackslash Views on chatbots} & Very positive & 7 (22.58\%) & 14 (15.05\%)\\
\hline
 & Somewhat difficult & 9 (29.03\%) & 15 (16.3\%)\\

 & Neither easy nor difficult & 8 (25.81\%) & 29 (31.52\%)\\

 & Somewhat easy & 10 (32.26\%) & 31 (33.7\%)\\

\multirow{-4}{1.3in}{\raggedright\arraybackslash How easy or difficult to find answers to questions} & Very easy & 4 (12.9\%) & 17 (18.48\%)\\
\hline
 & Once a month or less & 4 (12.9\%) & 24 (26.09\%)\\

 & A couple times a month & 13 (41.94\%) & 23 (25\%)\\

 & At least once a week & 9 (29.03\%) & 28 (30.43\%)\\

 & About once a day & 3 (9.68\%) & 8 (8.7\%)\\

\multirow{-5}{1.3in}{\raggedright\arraybackslash How often you run into questions that you do not know the answer to} & More than once a day & 2 (6.45\%) & 9 (9.78\%)\\
\hline
 & Never & 1 (3.23\%) & 1 (1.09\%)\\

 & Once a month or less & 2 (6.45\%) & 7 (7.61\%)\\

 & A couple times a month & 1 (3.23\%) & 17 (18.48\%)\\

 & At least once a week & 4 (12.9\%) & 13 (14.13\%)\\

 & About once a day & 5 (16.13\%) & 17 (18.48\%)\\

\multirow{-6}{1.3in}{\raggedright\arraybackslash How often do you use: Google} & More than once a day & 18 (58.06\%) & 37 (40.22\%)\\
\hline
 & Never & 0 (0\%) & 8 (8.7\%)\\

 & Once a month or less & 3 (9.68\%) & 6 (6.52\%)\\

 & A couple times a month & 2 (6.45\%) & 21 (22.83\%)\\

 & At least once a week & 11 (35.48\%) & 16 (17.39\%)\\

 & About once a day & 7 (22.58\%) & 18 (19.57\%)\\

\multirow{-6}{1.3in}{\raggedright\arraybackslash How often do you use: Slack/Email} & More than once a day & 8 (25.81\%) & 23 (25\%)\\
\hline 
 & Never & 2 (6.45\%) & 17 (18.48\%)\\

 & Once a month or less & 9 (29.03\%) & 25 (27.17\%)\\

 & A couple times a month & 6 (19.35\%) & 19 (20.65\%)\\

 & At least once a week & 7 (22.58\%) & 17 (18.48\%)\\

 & About once a day & 3 (9.68\%) & 6 (6.52\%)\\

\multirow{-6}{1.3in}{\raggedright\arraybackslash How often do you use: Policy Manuals} & More than once a day & 4 (12.9\%) & 8 (8.7\%)\\
\hline
 & Never & 6 (19.35\%) & 34 (36.96\%)\\

 & Once a month or less & 6 (19.35\%) & 19 (20.65\%)\\

 & A couple times a month & 5 (16.13\%) & 15 (16.3\%)\\

 & At least once a week & 4 (12.9\%) & 10 (10.87\%)\\

 & About once a day & 8 (25.81\%) & 6 (6.52\%)\\

\multirow{-6}{1.3in}{\raggedright\arraybackslash How often do you use: Chatbots} & More than once a day & 2 (6.45\%) & 8 (8.7\%)\\
\hline
 & Never & 0 (0\%) & 3 (3.26\%)\\

 & Once a month or less & 1 (3.23\%) & 12 (13.04\%)\\

 & A couple times a month & 6 (19.35\%) & 14 (15.22\%)\\

 & At least once a week & 5 (16.13\%) & 25 (27.17\%)\\

 & About once a day & 11 (35.48\%) & 14 (15.22\%)\\

\multirow{-6}{1.3in}{\raggedright\arraybackslash How often do you use: In Person} & More than once a day & 8 (25.81\%) & 24 (26.09\%)\\
\hline
 & Never & 2 (6.45\%) & 7 (7.61\%)\\

 & Once a month or less & 4 (12.9\%) & 16 (17.39\%)\\

 & A couple times a month & 2 (6.45\%) & 16 (17.39\%)\\

 & At least once a week & 10 (32.26\%) & 22 (23.91\%)\\

 & About once a day & 9 (29.03\%) & 12 (13.04\%)\\

\multirow{-6}{1.3in}{\raggedright\arraybackslash How often do you use: Phone} & More than once a day & 4 (12.9\%) & 19 (20.65\%)\\
\bottomrule
\caption{Differences between participants in the treatment and control groups for baseline survey questions related to existing program knowledge of CalFresh, perceptions of chatbots, and methods for answering questions related to CalFresh.
We test for statistical significance using Chi-Squared tests of independence. We do not observe any statistically significant relationships across participants in the treatment and control groups for significance thresholds of 0.05 or less.}
\label{tab:bequiv_categorical}
\end{longtable}

\newpage 
\section{Additional Figures and Results from Main Analysis}

\begin{figure*}[htbp!]
    \centering
    \includegraphics[width=\linewidth]{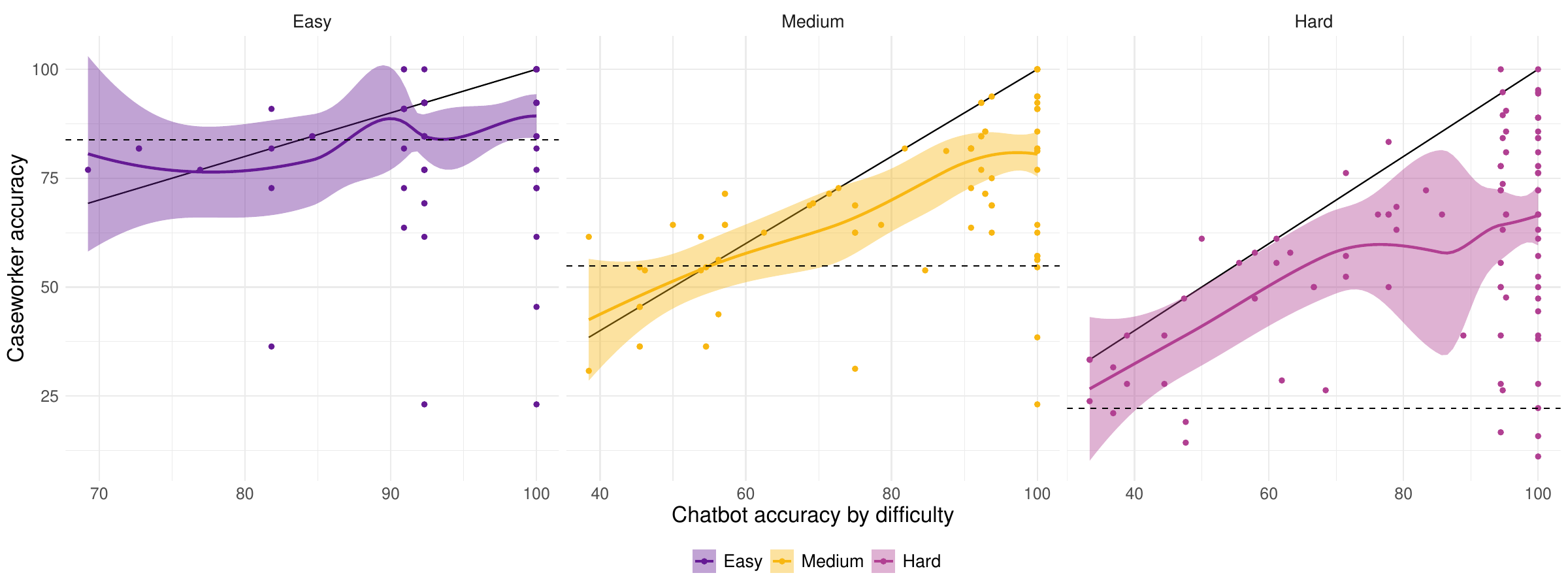}
    \caption{The gap between human and chatbot accuracy is larger on hard and medium difficulty questions. The horizontal dashed lines show accuracy from participants in the control group for questions from each difficulty level. The black lines illustrate human accuracy if respondents followed $100\%$ of the chatbot suggestions. We present both respondent accuracy and linear trend lines with $95$\% confidence intervals. Note: since we labeled question difficulty based on control-group performance, chatbot accuracies are not evenly distributed across difficulty levels. For example, there are fewer incorrect chatbot suggestions for questions labeled as easy, so chatbot accuracy does not fall below 70\%.}
    \label{fig:lineplot_accuracy_x_difficulty}
\end{figure*}

In this section, we include several figures relevant to the discussion of our results in the main text. Figure~\ref{fig:lineplot_accuracy_x_difficulty} presents trends in caseworker participant-level accuracy based on chatbot accuracy for easy, medium, and hard questions separately. The AI underreliance plateau effect that we discuss in the main text persists regardless of difficulty level. However it is most pronounced on hard questions where the gap between the chatbot accuracy and caseworker accuracy widens substantially at higher levels of chatbot accuracy. 

\begin{figure}[htbp!]
\centering
\begin{minipage}{0.75\textwidth}
     \centering
        \includegraphics[width=\linewidth]{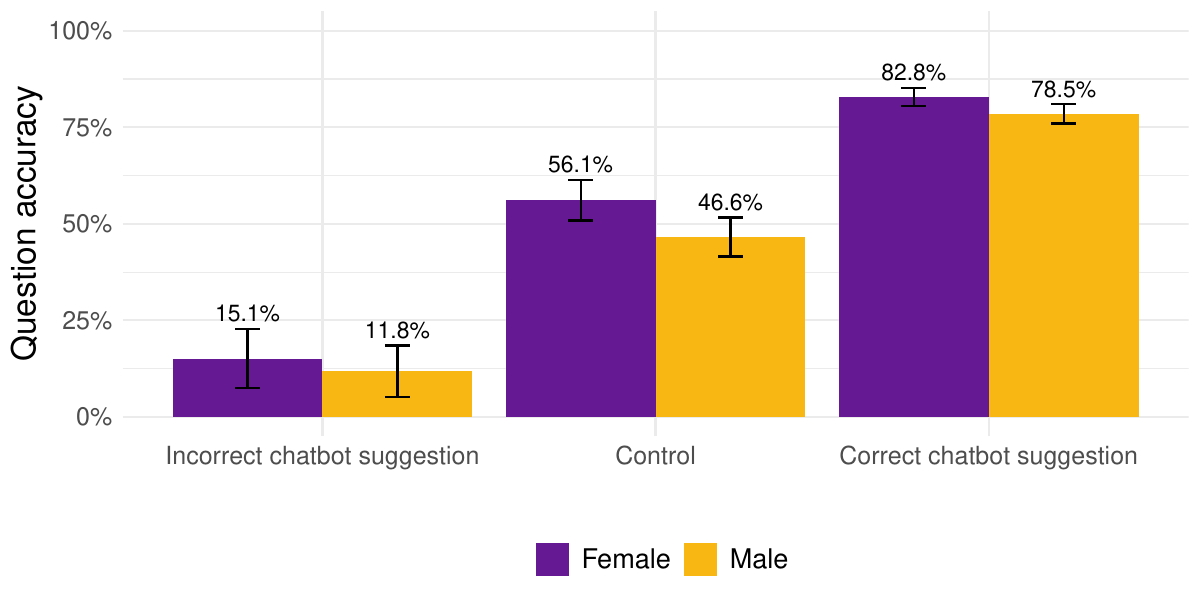}
    \end{minipage}
    \hfill
    \begin{minipage}{0.75\textwidth}
        \centering
        \includegraphics[width=\linewidth]{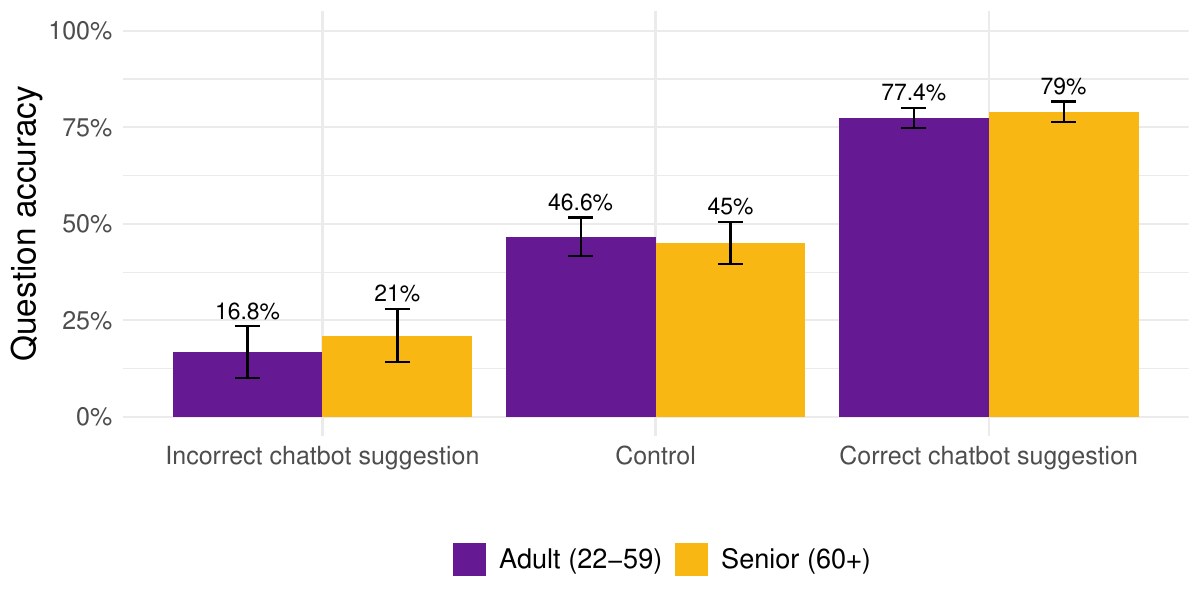}
\end{minipage}
\caption{Varying either age or gender largely has null effects for caseworker question-level accuracy regardless of the treatment condition or suggestion type. In this figure, the type of suggestion appears on the x-axis (correct, incorrect, or a question for the control group). The colors of the bars indicate the varied attribute value (Female versus Male, Adult versus Senior). Caseworker question-level accuracy appears on the y-axis. We provide error bars with 95\% confidence intervals.}
    \label{fig:varying_attributes}
\end{figure}

\begin{figure}[htbp!]
\centering
\begin{minipage}{0.75\textwidth}
        \includegraphics[width=\linewidth]{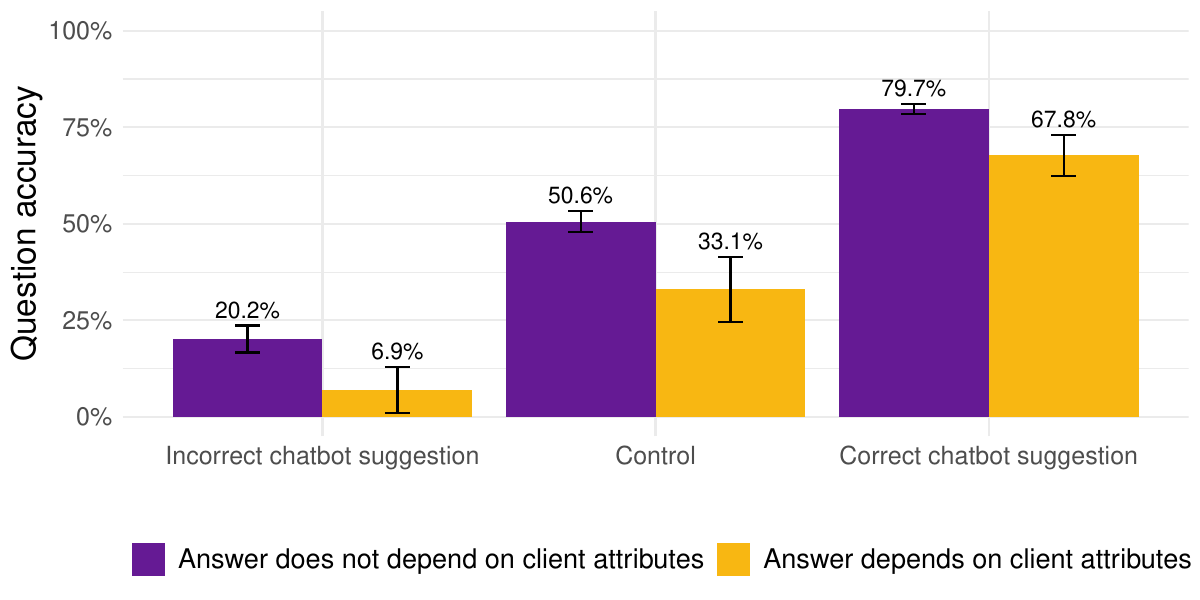}
\end{minipage}
\caption{Questions where the correct answer depends on client attributes were more difficult for caseworkers compared to questions whether the correct answer is independent of client attributes. These differences persist regardless of whether caseworkers saw chatbot suggestions or not, but the difference is largest for questions with an incorrect chatbot suggestion. Chatbot suggestion types appear on the x-axis (correct, incorrect, or a question for the control group). Caseworker question-level accuracy is on the y-axis. We provide error bars with 95\% confidence intervals.}
    \label{fig:preamble_changes}
\begin{minipage}{0.75\textwidth}
    \centering 
    \includegraphics[width=\linewidth]{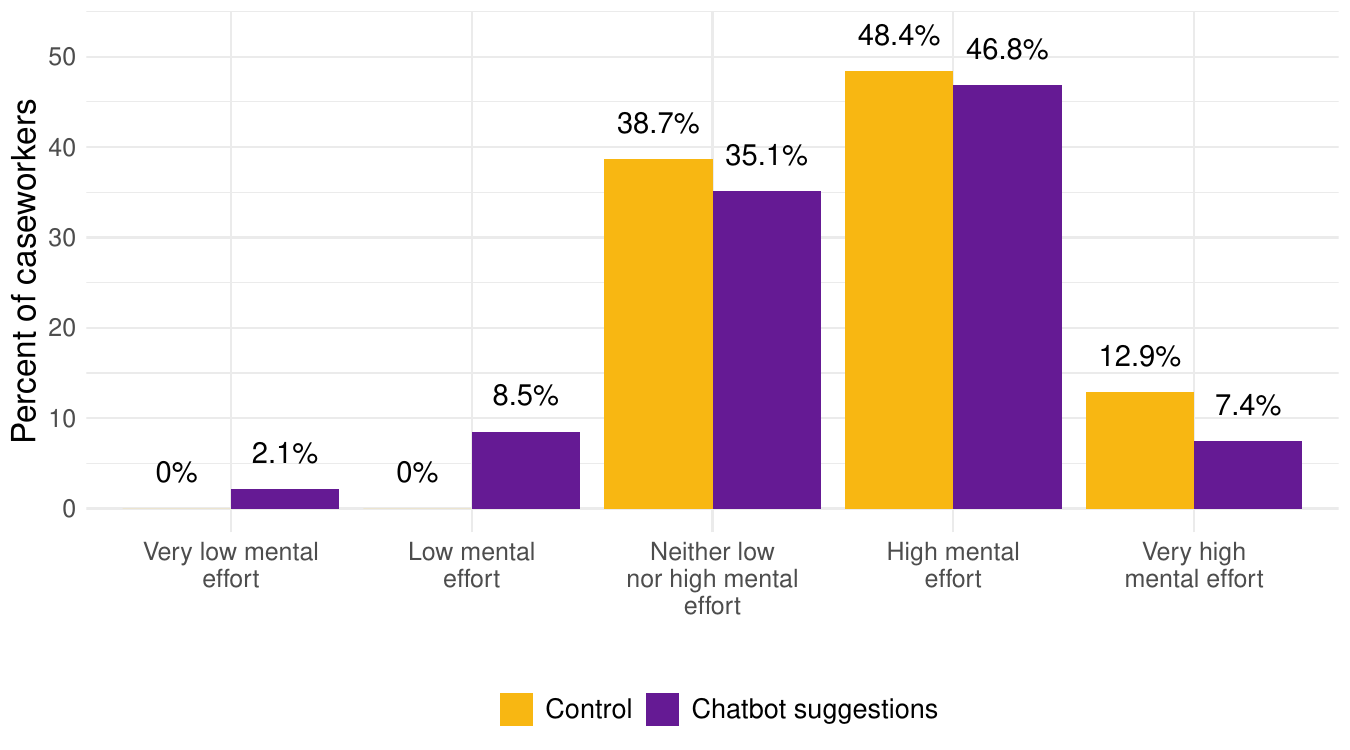}
\end{minipage}
\caption{Differences in the self-reported mental effort by the percentage of caseworker responses to a post-assesment question. After finishing the assessment, caseworkers were asked to select the overall level of mental effort invested in the assessment. Caseworkers who saw chatbot suggestions reported investing high and very high mental effort at slightly higher rates. Some also reported lower levels of mental effort whereas no caseworkers in the control group reported very low and low mental effort. However,  using a chi-squared test for independence, we do not observe any statistically significant differences in the question responses between caseworkers who saw chatbot suggestions and those who did not.}
    \label{fig:compliance_effort_all}
\end{figure}

\begin{figure*}[htbp!]
    \centering
    \begin{minipage}{\textwidth}
    \includegraphics[width=\linewidth]{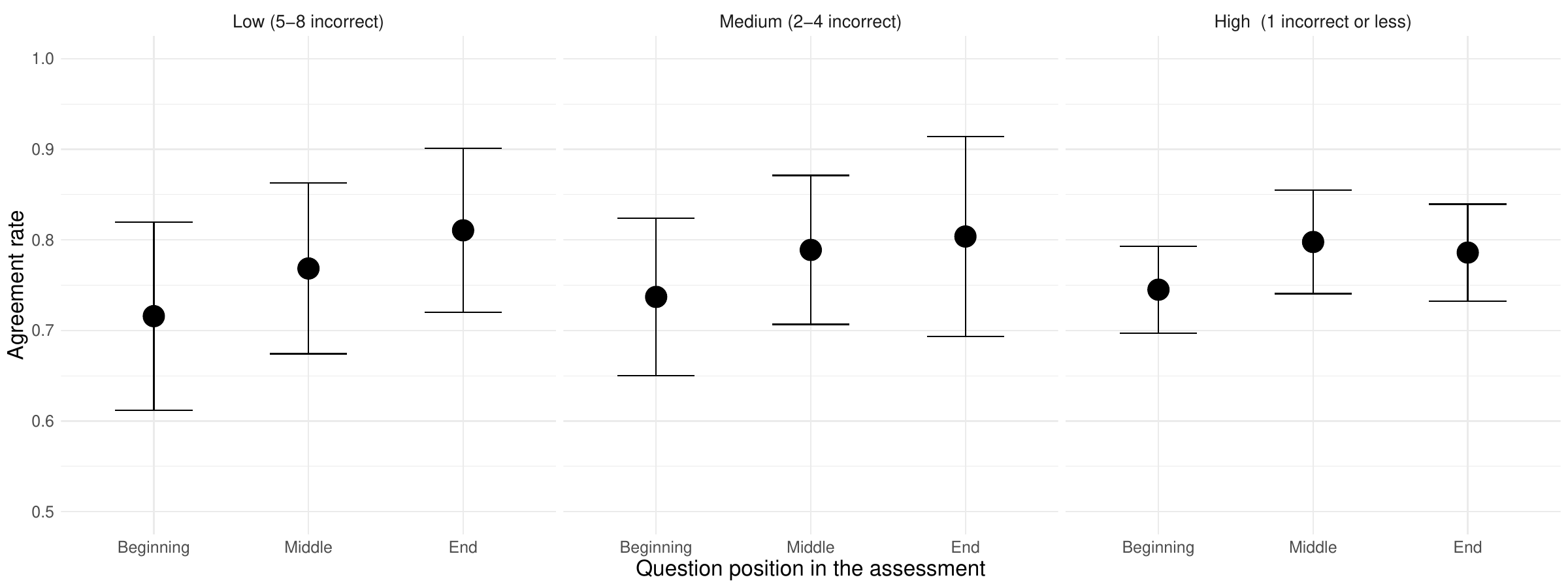}
    \caption{\footnotesize{Caseworker agreement based on the survey section and the number of incorrect suggestions within each section of the survey. A caseworker agrees with the chatbot if they select the answer option corresponding to the chatbot suggestion, regardless of whether the chatbot suggestion is correct or incorrect. The x-axis distinguishes between the beginning (questions 1-15), middle (questions 16-30), and end (questions 31-45) of the survey. On the y-axis, we show the percent of questions where caseworkers \emph{agreed} with the chatbot suggestions. We provide three distinct subplots based on whether a given section was low quality (5-8 incorrect chatbot suggestions), medium quality (2-4 incorrect suggestions), and high quality (less than 1 incorrect suggestion). Notably, we do not observe significant differences in agreement rates.}}
    \label{fig:agreements}
    \end{minipage}
    \vfill 
    \begin{minipage}{\textwidth}
    \centering
    \includegraphics[width=\linewidth]{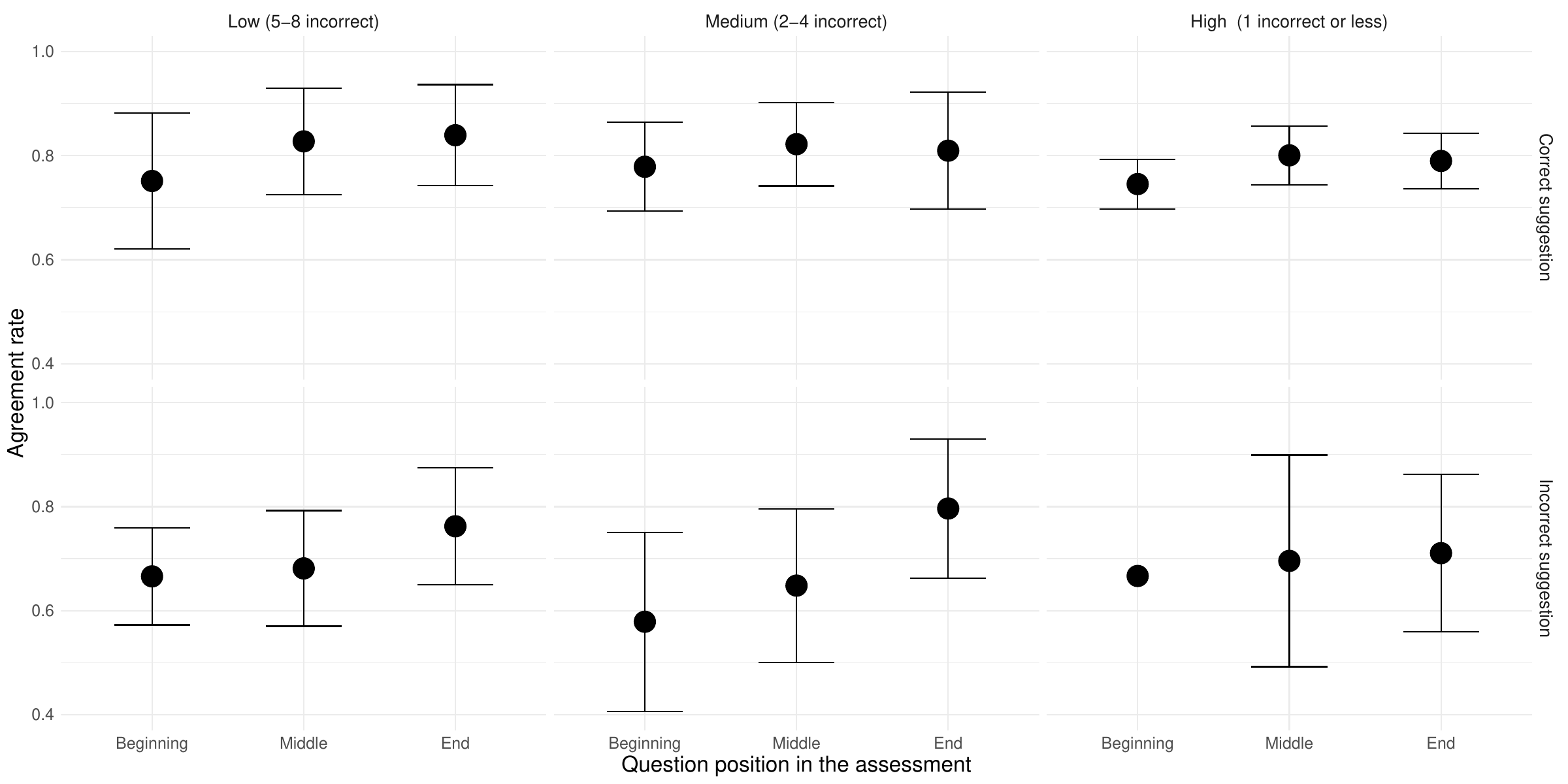}
    \caption{Caseworker agreement based on survey section, suggestion type, and the number of incorrect suggestions shown. The x-axis distinguishes between the beginning (questions 1-15), middle (questions 16-30), and end (questions 31-45) of the survey. On the y-axis, we show the percent of questions where caseworkers \emph{agreed} with the chatbot suggestions. We provide three distinct subplots based on whether a given section was low quality (5-8 incorrect chatbot suggestions), medium quality (2-4 incorrect suggestions), and high quality (less than 1 incorrect suggestion). Notably, we do not observe significant differences in agreement rates. Error bars indicate the 95\% confidence intervals for our estimates; we exclude the error bars in one case where the full length would distort the y-axis.}
    \label{fig:agreementsXtype}
    \end{minipage}
\end{figure*}

\begin{figure}[htbp!]
    \centering
    \begin{minipage}{0.6\textwidth}
        \includegraphics[width=\linewidth]{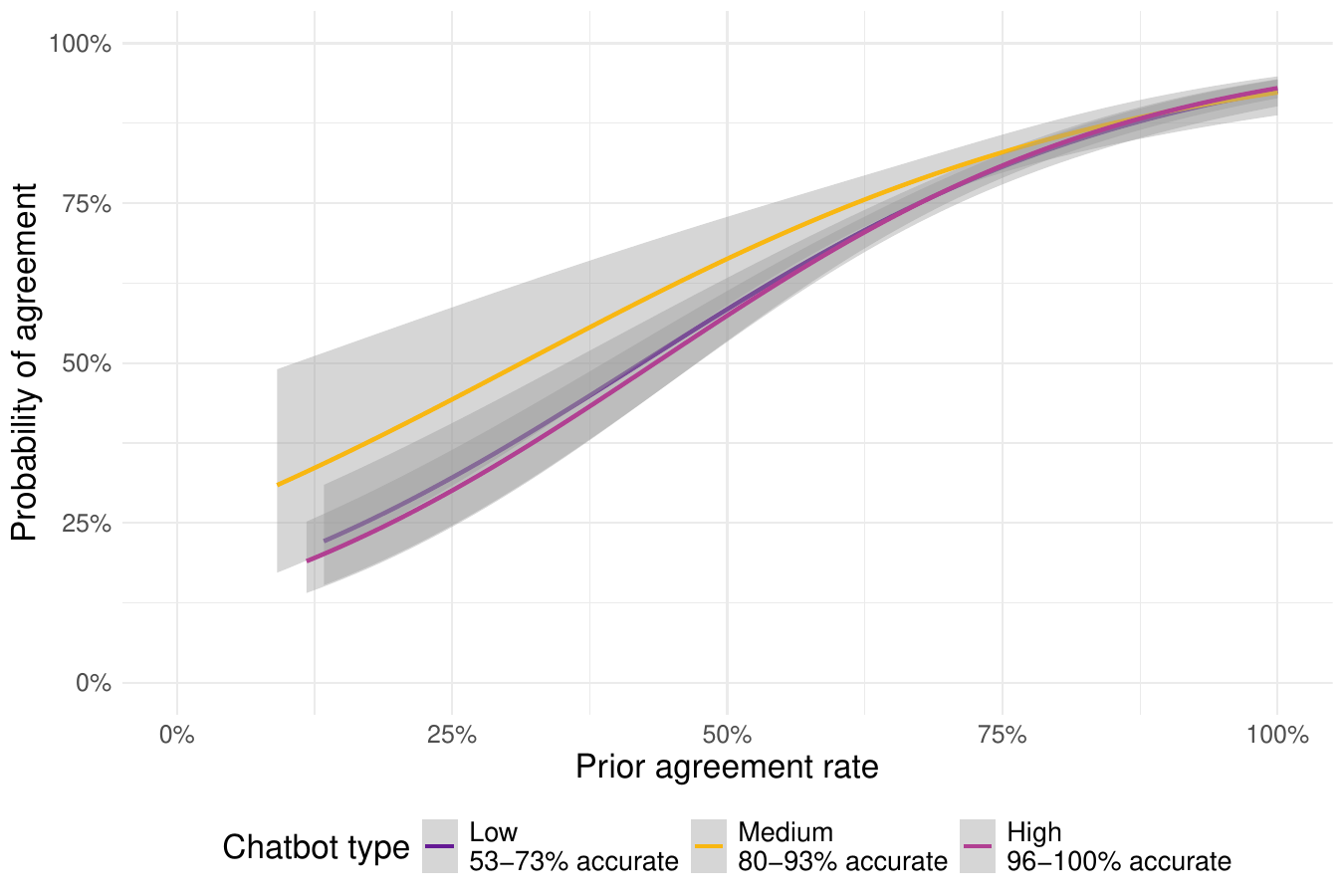}
\end{minipage}
\caption{\footnotesize{Prior agreement rate is highly predictive of probability of agreement. On the x-axis, we show the prior agreement rate, calculated for a given question and participant. The probability of agreement, estimated via a logistic regression model, is on the y-axis. We restrict to questions that occur after the fifth question position in the survey. We do not observe significant differences in agreement rates by chatbot quality. However, the probability of agreement for the lowest quality chatbot is consistently lower than the other chatbot types on average.}}
    \label{fig:agreements_prob}
\end{figure}

In Figures~\ref{fig:varying_attributes}-\ref{fig:preamble_changes}, we examine whether client attributes affect caseworker question-level accuracy. We first present Figure~\ref{fig:varying_attributes}, which illustrates the null effects on caseworker question-level accuracy of changing client attributes like age or gender. In Figure~\ref{fig:preamble_changes}, we highlight how questions where the correct answer depends on client attributes are more challenging for caseworkers, resulting in reduced question-level accuracy on average. Figure~\ref{fig:compliance_effort_all} compares self-reported mental effort between caseworkers who saw chatbot suggestions and those who did not.
Figures~\ref{fig:agreements}-\ref{fig:agreements_prob} present the results from our analysis of question-agreements where we find that caseworker agreement is largely driven by prior behavior, i.e. higher rates of agreement in the past are associated with increased probability of agreeing on future questions.

\newpage 

\newpage

\newpage

\begin{figure}[htbp!]
    \centering
    \includegraphics[width=0.6\linewidth]{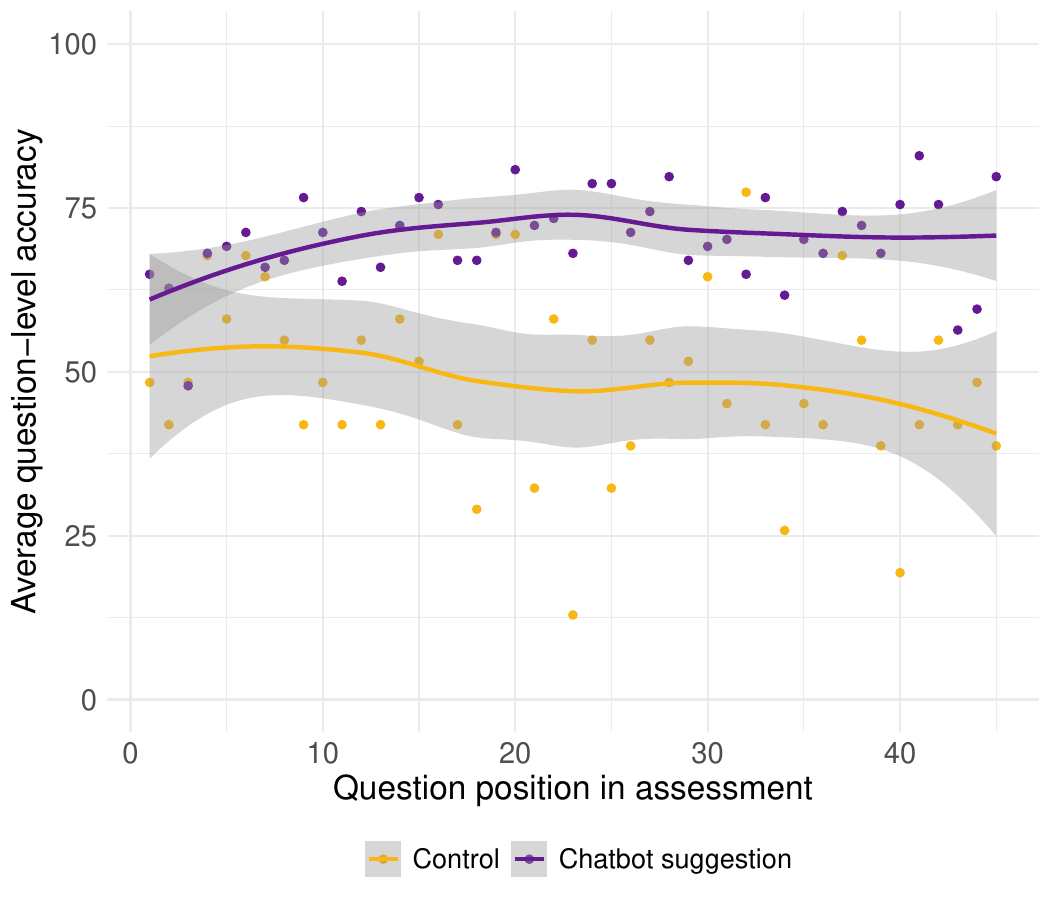}
    \caption{\footnotesize{Average question-level caseworker accuracy by question position in the assessment (from 1 to 45). Average question-level accuracy appears on the y-axis; the question position on the x-axis. Different line colors denote whether caseworkers saw chatbot suggestions or not. Caseworker accuracy is consistently higher with chatbot suggestions and increases slightly over the course of the assessment. As shown in Table~\ref{tab:regression_results_accuracy_X_position}, a question with chatbot suggestions that appears later in the assessment is associated with a 0.1 percentage point increase in accuracy ($p<0.05$) while a question without chatbot suggestions that appears later in the assessment is associated with a 0.3 percentage point decrease in caseworker accuracy ($p<0.05$).}}
    \label{fig:accuracyX_question_position}
\end{figure}

\newpage 
\FloatBarrier
\section{Results: Participants passing the attention check}
\label{sec:full_results}

\small{In this section, we replicate the main results from our analysis, restricting to participants who passed our attention check. As the attention check was challenging, only around 63\% of participants passed the attention check. The findings are broadly similar and directionally the same when compared to the results we report in the main text.}

\begin{figure*}[h!]
    \centering
    \begin{minipage}{\textwidth}
    \centering
    \includegraphics[width=0.95\linewidth]{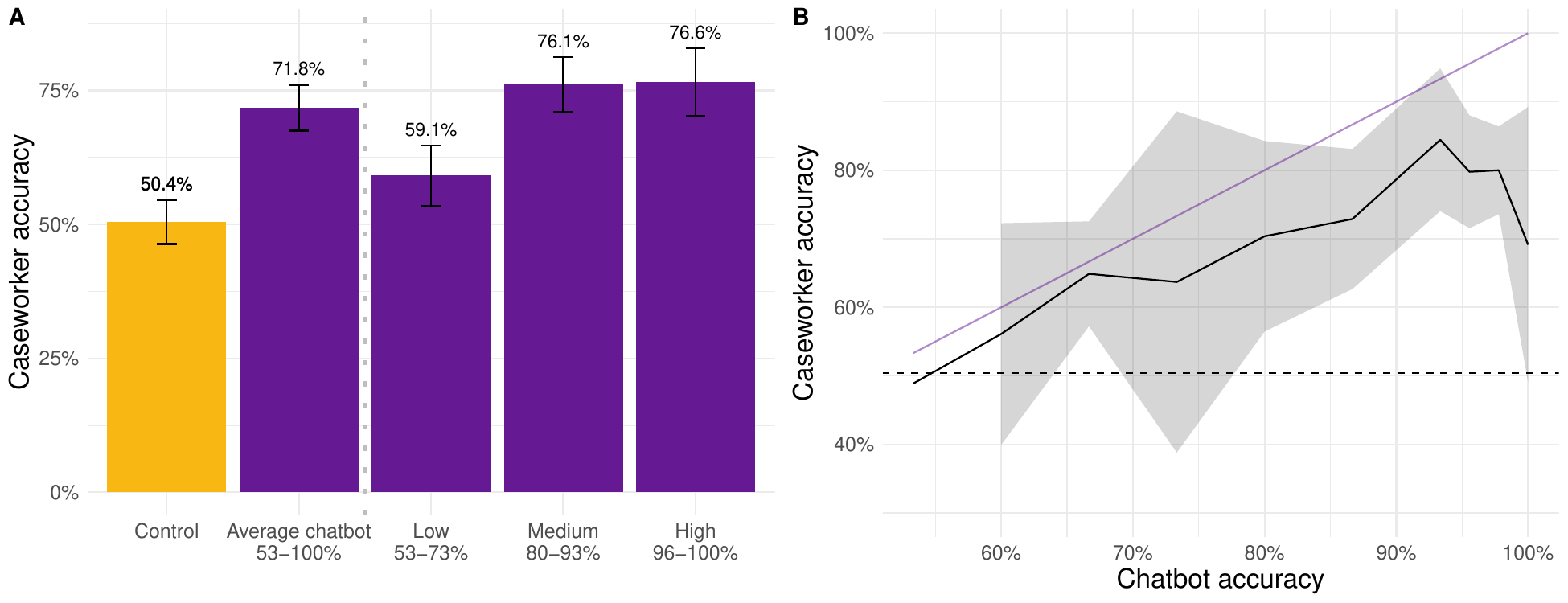}
    \caption{\footnotesize{(A) Caseworker participant-level accuracy for any chatbot on average (``Average chatbot'') and by chatbot quality: a low quality chatbot (53-73\% accurate), a medium quality chatbot (80-93\% accurate), and a high-quality chatbot (96-100\%). Overall, caseworker accuracy improves with chatbot suggestions by 21 percentage points, regardless of chatbot quality. High quality chatbots lead to the largest improvements in caseworker accuracy. (B) Human accuracy increases alongside chatbot accuracy up to a point. Improvements in accuracy plateau around 80\% accuracy. The dashed horizontal line shows accuracy from participants in the control group, which is roughly comparable to the worst-performing chatbot. The black line illustrates human accuracy if respondents followed 100\% of the chatbot suggestions. We include the 95\% confidence interval for each estimate in light gray. Due to small sample sizes for some chatbot accuracy conditions, we estimate 95\% confidence intervals using the student t-distribution.}}
    \label{fig:accuracy_panel_passed}
    \end{minipage}

    \begin{minipage}{\textwidth}
    \centering
    \includegraphics[width=0.95\linewidth]{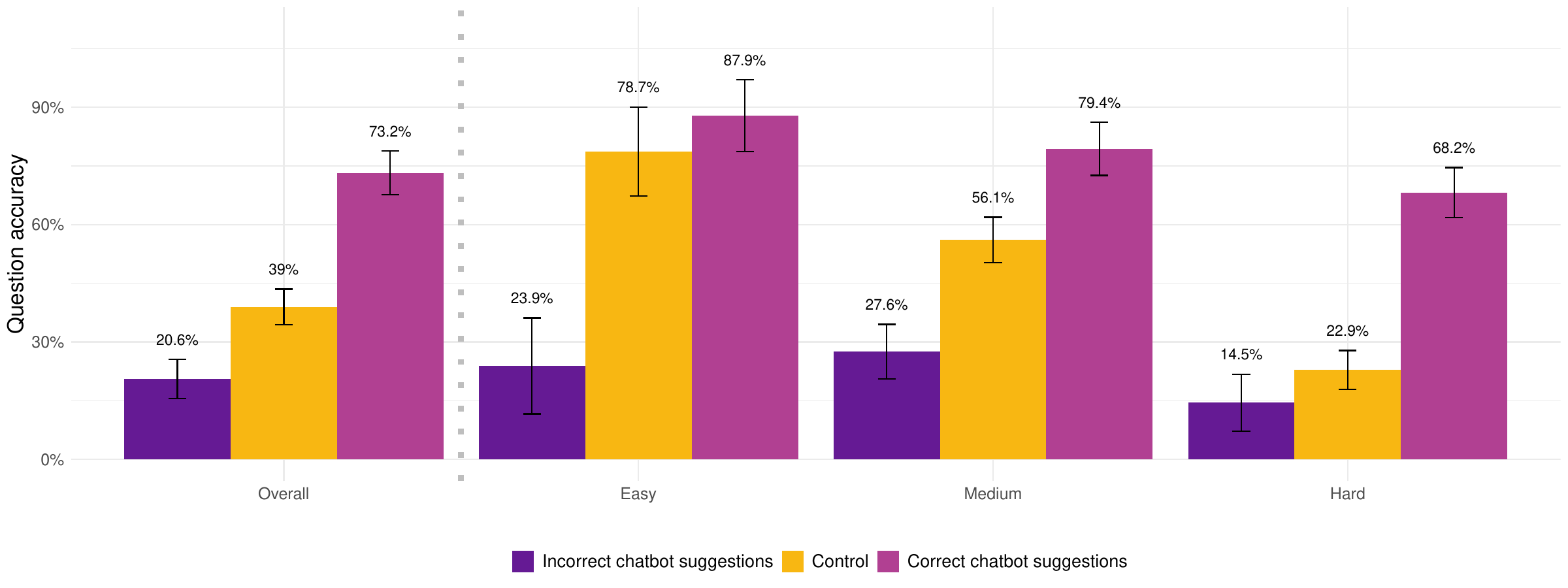}
    \caption{\footnotesize{Incorrect chatbot suggestions lead to the largest reductions in caseworker accuracy on easy and medium questions relative to accuracy in the control group, wherein no chatbot suggestions are seen. However, correct suggestions can be helpful, particularly on hard questions where there is a large increase in caseworker accuracy relative to the control group. Question difficulty appears on the x-axis. Different colors (purple, yellow, and light purple) indicate the type of chatbot suggestion. Caseworker question-level accuracy appears on the y-axis. We include error bars with the 95\% confidence intervals for all estimates.}}
\label{fig:question_level_effect_passed}
\end{minipage}
\end{figure*}

\begin{figure*}[t!]
    \begin{minipage}{\textwidth}
    \centering
    \includegraphics[width=\linewidth]{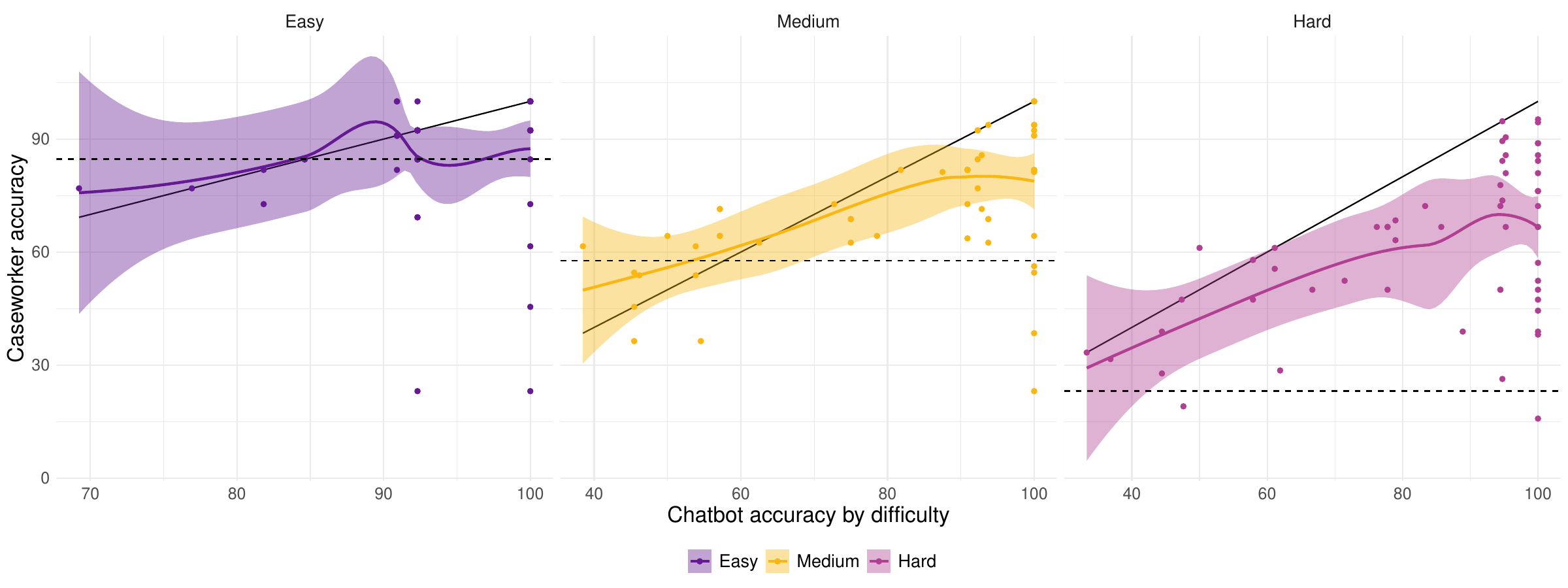}
    \caption{The gap between caseworker and chatbot accuracy is larger on hard and medium difficulty questions. The horizontal dashed lines show accuracy from participants in the control group for questions from each difficulty level. The black lines illustrate human accuracy if respondents followed $100\%$ of the chatbot suggestions. We present both respondent accuracy and linear trend lines with $95$\% confidence intervals.}
    \label{fig:lineplot_accuracy_x_difficulty_passed}
\end{minipage}

\begin{minipage}{\textwidth}
    \centering
    \includegraphics[width=0.6\linewidth]{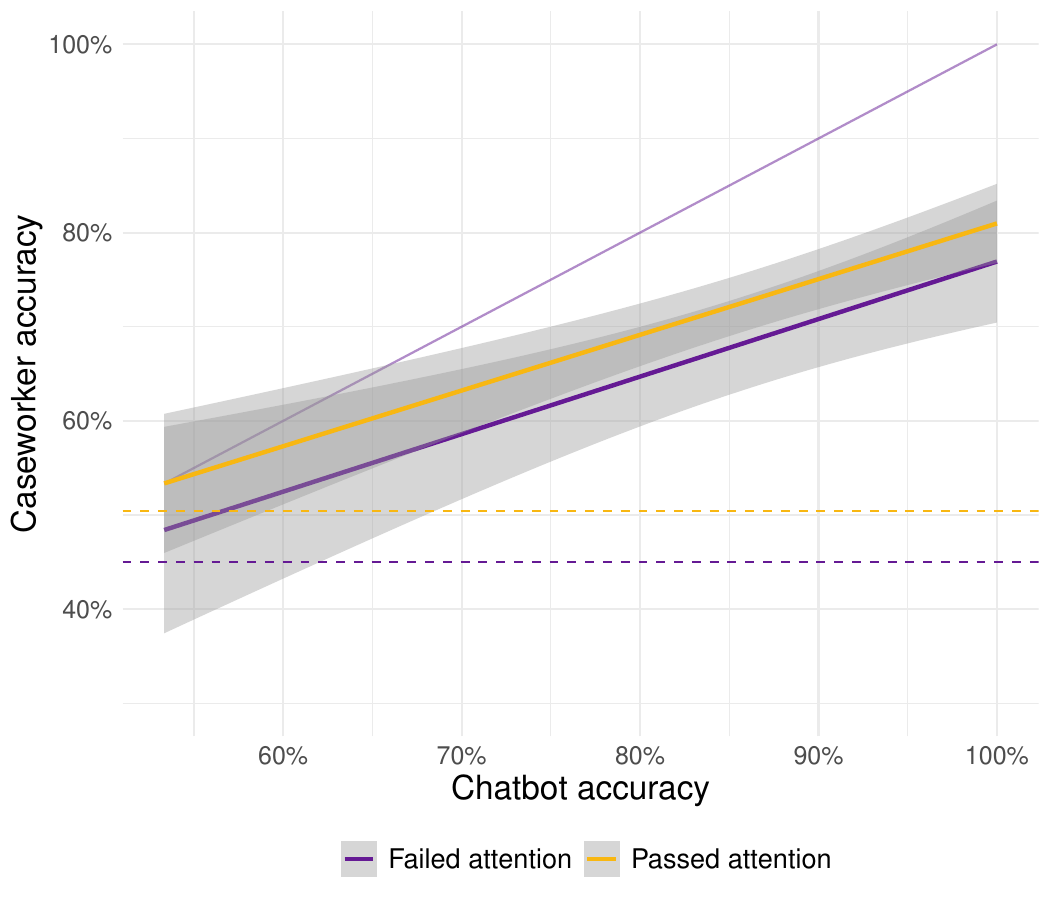}
    \caption{Comparison of caseworker accuracy by whether caseworkers pass or fail the attention check. Caseworker accuracy is on average slightly lower for participants who fail the attention check, though there is no statistically significant difference in trends (95\% confidence intervals, which are denoted in gray, fully overlap). The dashed lines indicate the average accuracy for caseworkers who pass or fail the attention check in the control group. }
    \label{fig:lineplot_X_attentioncheck}
    \end{minipage}
\end{figure*}

\newpage
\begin{figure*}[t!]
    \centering
    \begin{minipage}{0.6\textwidth}
        \includegraphics[width=\linewidth]{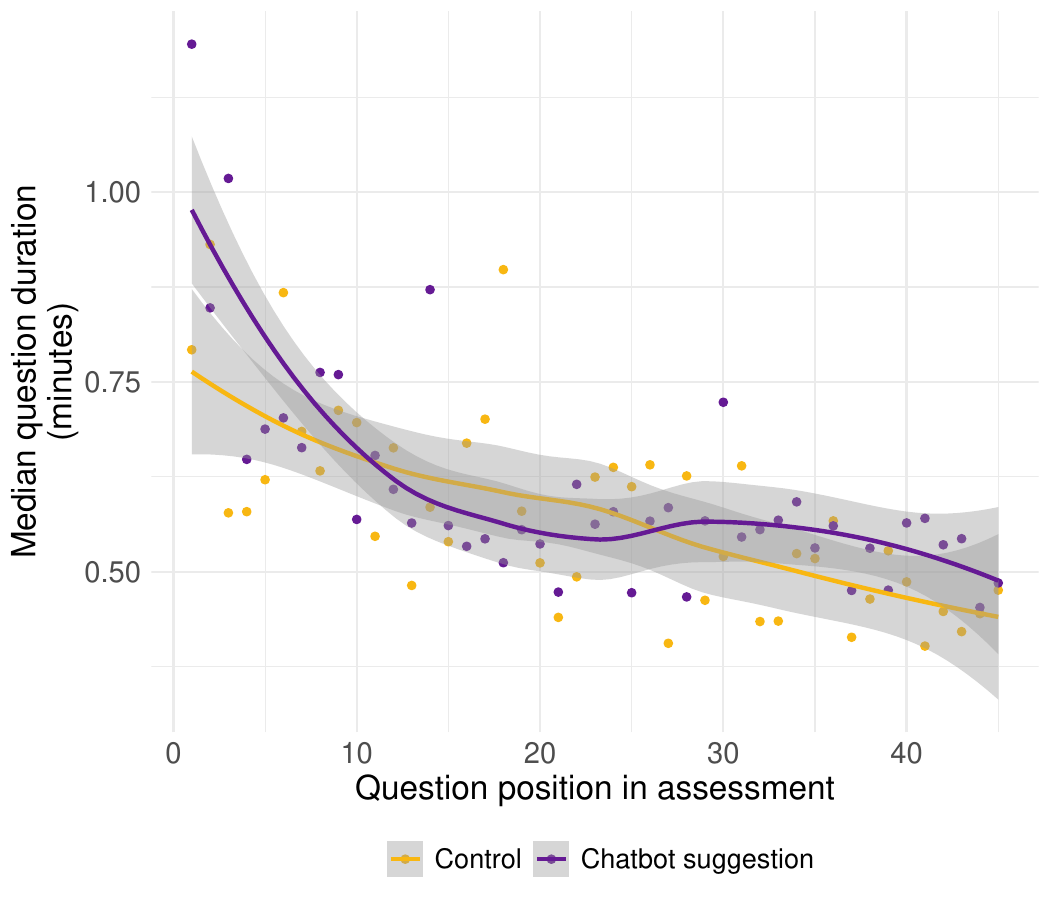}
    \end{minipage}
    \caption{\footnotesize{Median caseworker time to answer each question is similar with and without chatbot suggestions, and decreases over the course of the assessment. The position in the survey (from 1 to 45) is on the x-axis. The median question duration is on the y-axis. We use the median question duration, as large outliers may skew results. The line color indicates whether participants saw LLM suggestions (treatment group) or no LLM suggestions (control group). Additionally, we include a trend line with 95\% confidence intervals. If we were to observe significant differences between the treatment and control groups, we would expect that the lines would not overlap.}}
    \label{fig:lineplot_duration_passed}
    \begin{minipage}{0.75\textwidth}
    \centering 
    \includegraphics[width=\linewidth]{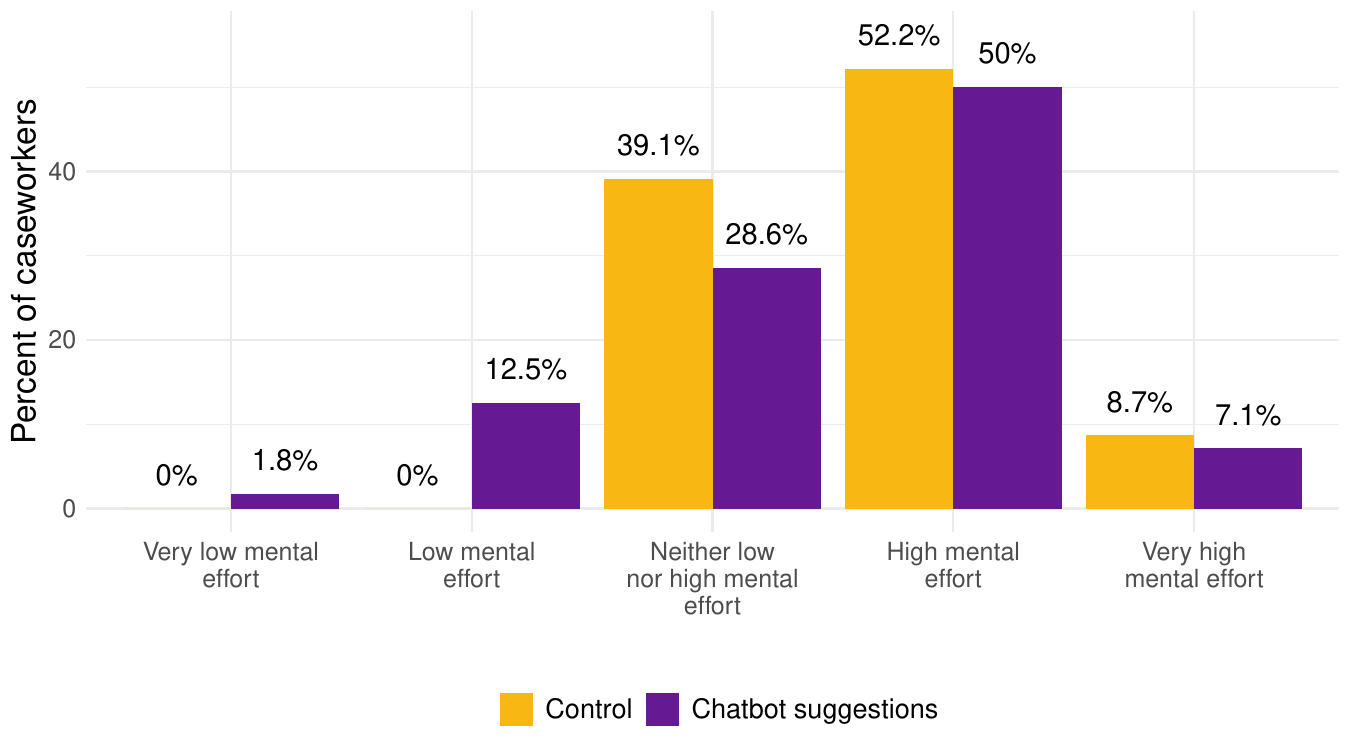}
\end{minipage}
\caption{\footnotesize{Differences in the self-reported mental effort by the percentage of caseworker responses to a post-assesment question. After finishing the assessment, caseworkers were asked to select the overall level of mental effort invested in the assessment. Caseworkers who saw chatbot suggestions reported investing high and very high mental effort at slightly higher rates. Some also reported lower levels of mental effort whereas no caseworkers in the control group reported very low and low mental effort. However,  using a chi-squared test for independence, we do not observe any statistically significant differences in the question responses between caseworkers who saw chatbot suggestions and those who did not.}}
    \label{fig:compliance_effort_passed}
\end{figure*}

\newpage 
\begin{figure}[t!]
\centering
\begin{minipage}{0.75\textwidth}
     \centering
        \includegraphics[width=\linewidth]{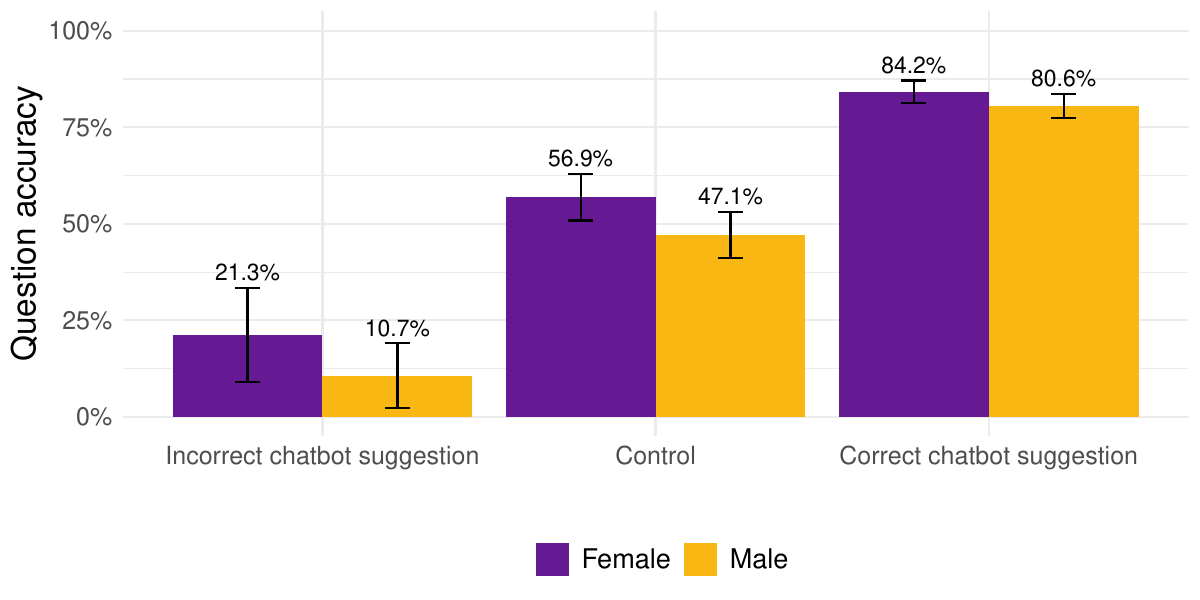}
    \end{minipage}
    \hfill
    \begin{minipage}{0.75\textwidth}
        \centering
        \includegraphics[width=\linewidth]{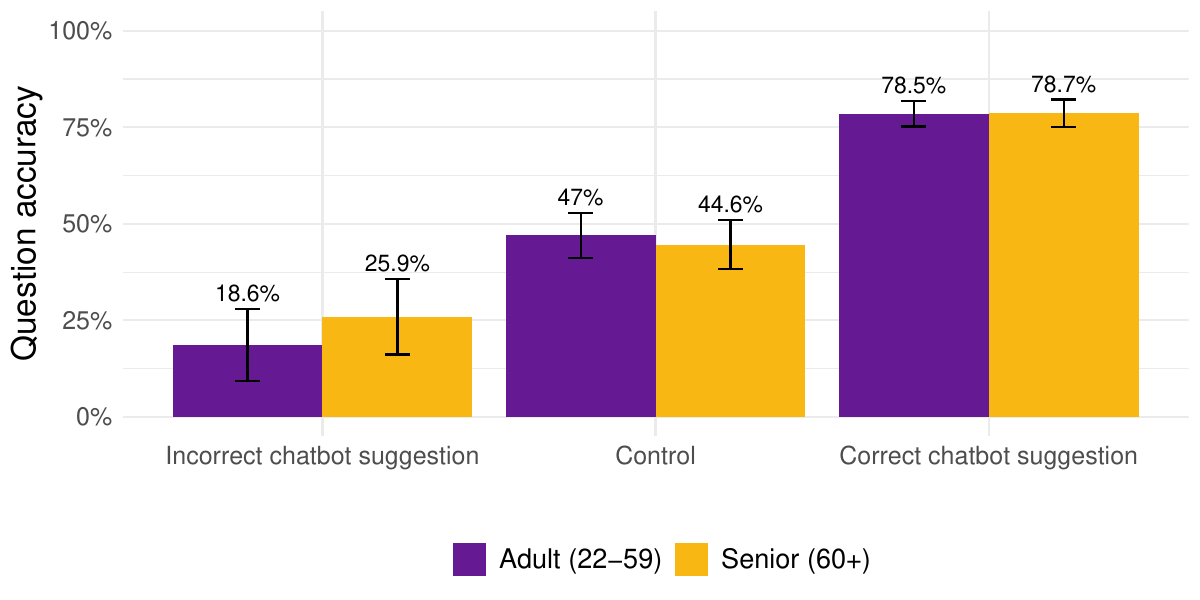}
\end{minipage}
\caption{Varying either age or gender largely has null effects for caseworker question-level accuracy regardless of the treatment condition or suggestion type. In this figure, the type of suggestion appears on the x-axis (correct, incorrect, or a question for the control group). The colors of the bars indicate the varied attribute value (Female versus Male, Adult versus Senior). Caseworker question-level accuracy appears on the y-axis. We provide error bars with 95\% confidence intervals.}
    \label{fig:varying_attributes_passed}
\end{figure}

\begin{figure}[h!]
\centering
\begin{minipage}{0.75\textwidth}
        \includegraphics[width=\linewidth]{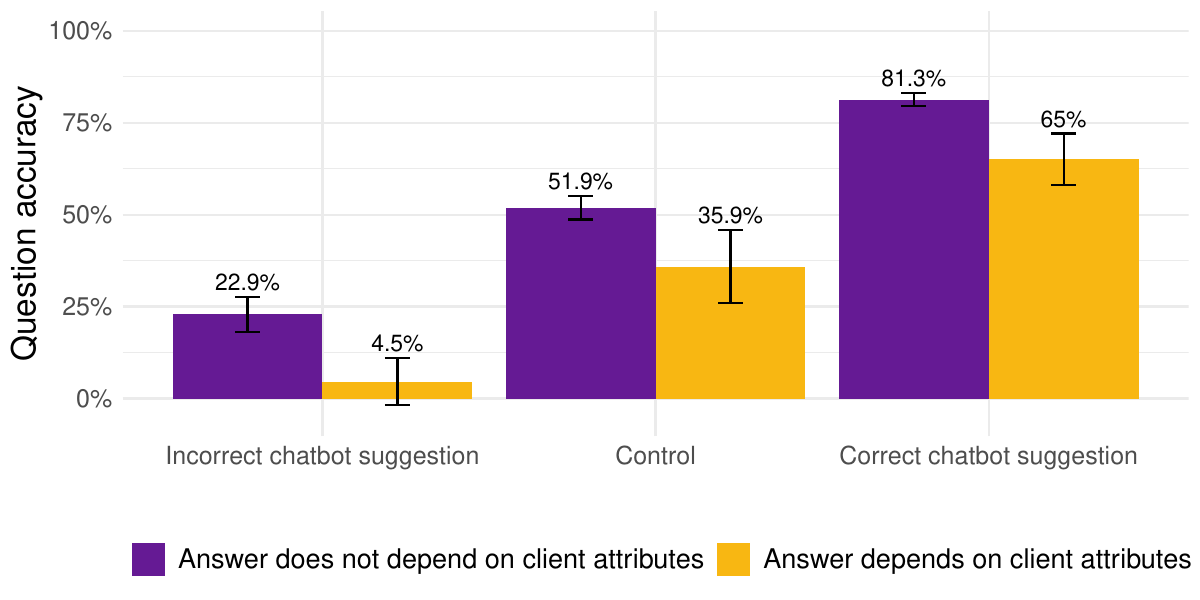}
\end{minipage}
\caption{Questions where the correct answer depends on client attributes were more difficult for caseworkers compared to questions whether the correct answer is independent of client attributes. These differences persist regardless of whether caseworkers saw chatbot suggestions or not, but the difference is largest for questions with an incorrect chatbot suggestion. Chatbot suggestion types appear on the x-axis (correct, incorrect, or a question for the control group). Caseworker question-level accuracy is on the y-axis. We provide error bars with 95\% confidence intervals.}
    \label{fig:preamble_changes_passed}
\end{figure}

\newpage 
\begin{figure*}[htbp!]
\centering
\begin{minipage}{\textwidth}
    \centering
    \includegraphics[width=\linewidth]{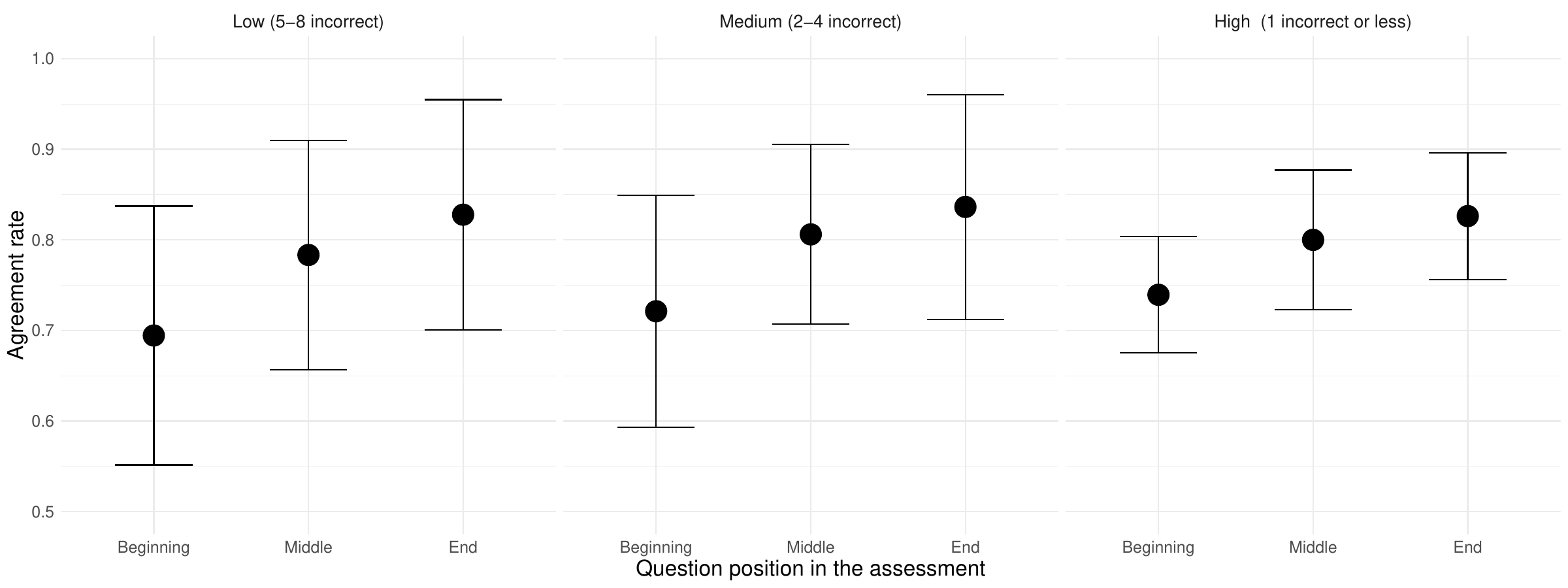}
    \caption{\small{Caseworker agreement based on the survey section and the number of incorrect suggestions within each section of the survey. The x-axis distinguishes between the beginning (questions 1-15), middle (questions 16-30), and end (questions 31-45) of the survey. On the y-axis, we show the percent of questions where caseworkers \emph{agreed} with the chatbot suggestions. We provide three distinct subplots based on whether a given section was low quality (5-8 incorrect chatbot suggestions), medium quality (2-4 incorrect suggestions), and high quality (less than 1 incorrect suggestion). Notably, we do not observe significant differences in agreement rates. }}
    \label{fig:agreements_passed}
    \end{minipage}
    \vfill
    
    \begin{minipage}{\textwidth}
        \includegraphics[width=\linewidth]{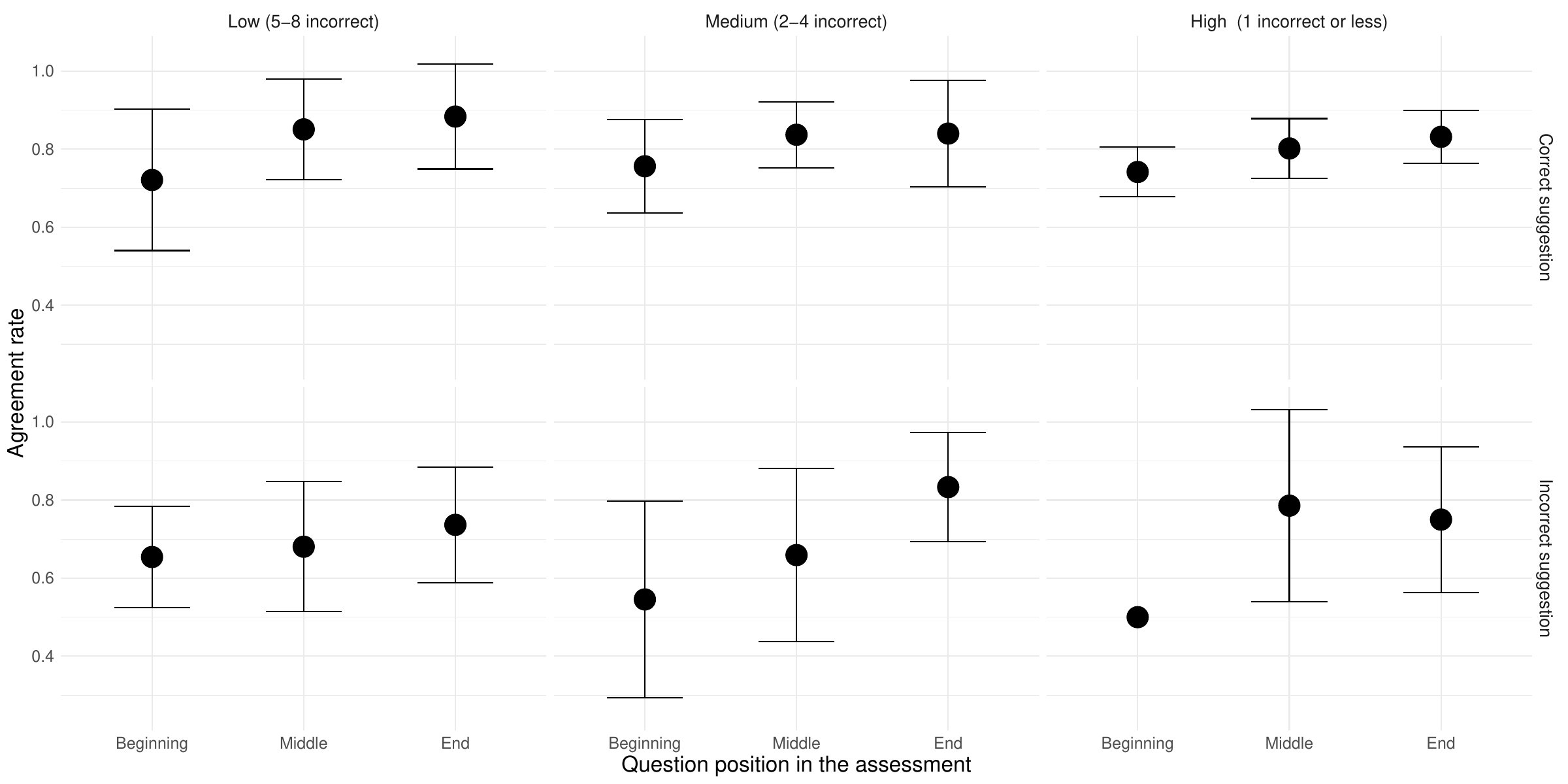}
    \caption{\small{Caseworker agreement based on survey section, suggestion type, and the number of incorrect suggestions shown. The x-axis distinguishes between the beginning (questions 1-15), middle (questions 16-30), and end (questions 31-45) of the survey. On the y-axis, we show the percent of questions where caseworkers \emph{agreed} with the chatbot suggestions. We provide three distinct subplots based on whether a given section was low quality (5-8 incorrect chatbot suggestions), medium quality (2-4 incorrect suggestions), and high quality (less than 1 incorrect suggestion). Notably, we do not observe significant differences in agreement rates. Error bars indicate the 95\% confidence intervals for our estimates; we exclude the error bars in one case where the full length would distort the y-axis.}}
    \end{minipage}
    
    \label{fig:agreementsXtype_passed_passed}
\end{figure*}

\newpage 
\begin{figure}[htbp!]
    \centering
    \begin{minipage}{\textwidth}
     \centering
        \centering 
        \includegraphics[width=0.6\linewidth]{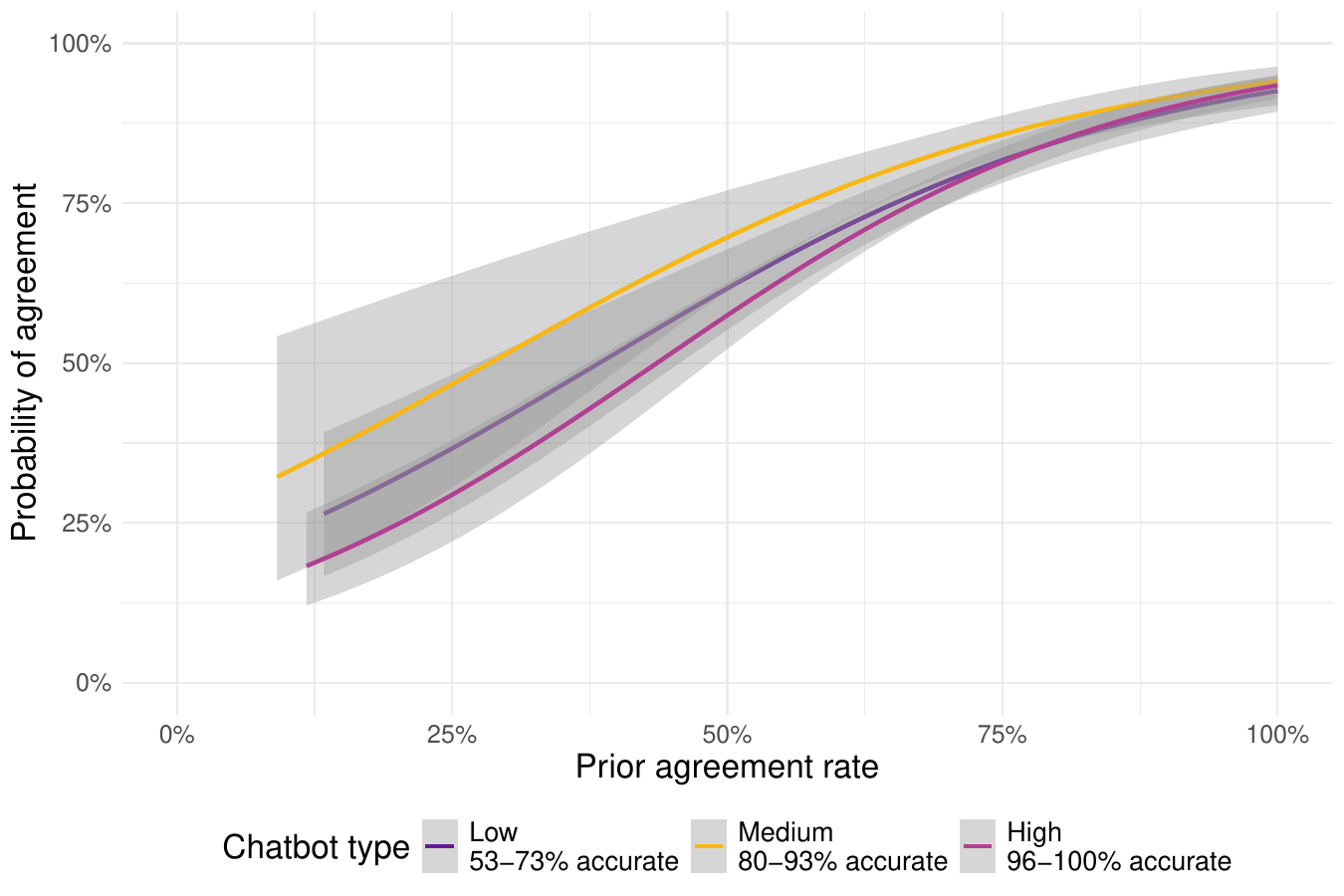}
    \caption{\small{Prior agreement rate is highly predictive of probability of agreement. On the x-axis, we show the prior agreement rate, calculated for a given question and participant. The probability of agreement, estimated via a logistic regression model, is on the y-axis. We restrict to questions that occur after the fifth question position in the survey. We do not observe significant differences in agreement rates by chatbot quality. However, the probability of agreement for the lowest quality chatbot is consistently lower than the other chatbot types on average.}}
    \label{fig:agreements_prob_passed}
\end{minipage}

\begin{minipage}{\textwidth}
    \centering
    \includegraphics[width=0.6\linewidth]{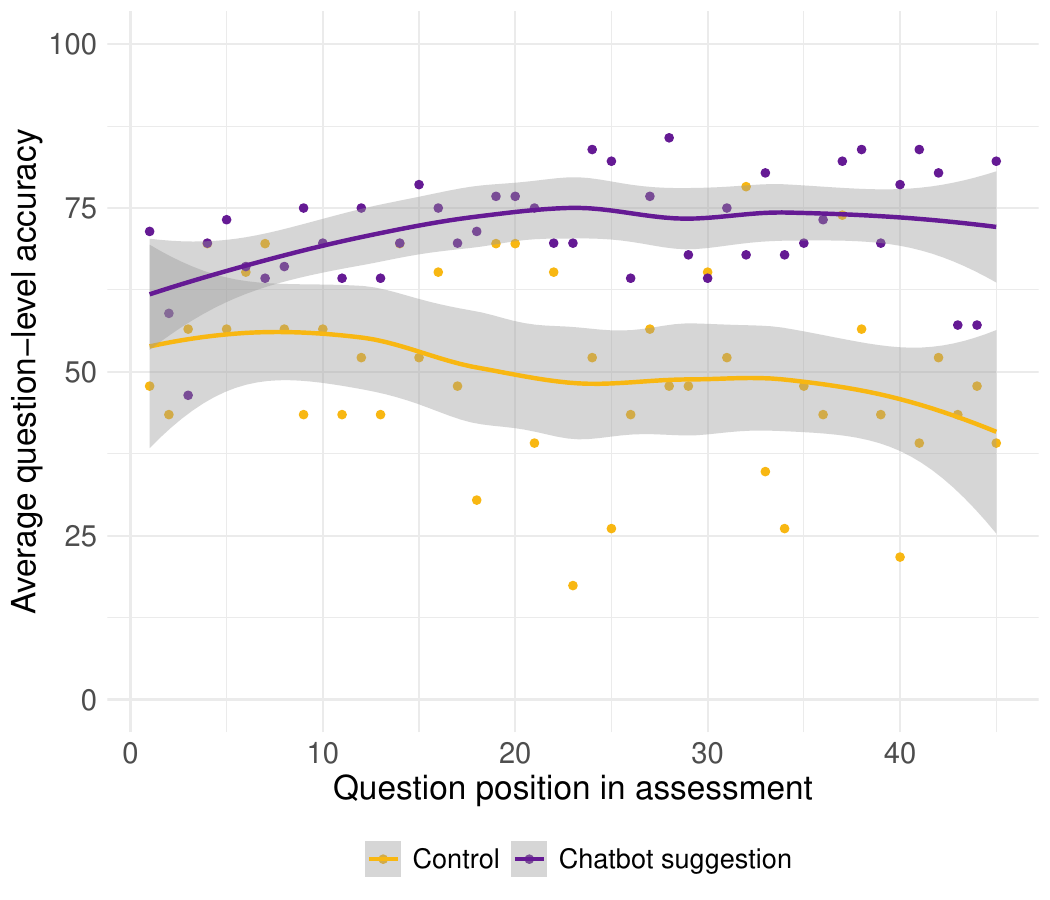}
    \caption{\small{Average question-level caseworker accuracy by question position in the assessment (from 1 to 45). Average question-level accuracy appears on the y-axis; the question position on the x-axis. Different line colors denote whether caseworkers saw chatbot suggestions or not. Caseworker accuracy is consistently higher with chatbot suggestions and increases slightly over the course of the assessment. As shown in Table~\ref{tab:regression_results_accuracy_X_position}, a question with chatbot suggestions that appears later in the assessment is associated with a 0.2 percentage point increase in accuracy ($p<0.01$) while a question without chatbot suggestions that appears later in the assessment is associated with a 0.3 percentage point decrease in caseworker accuracy ($p<0.05$).}}
    \label{fig:accuracyX_question_position_passed}
    \end{minipage}
\end{figure}

\begin{figure}[t!]
    \centering
    \begin{minipage}{\textwidth}
    \centering
    \includegraphics[width=0.5\linewidth]{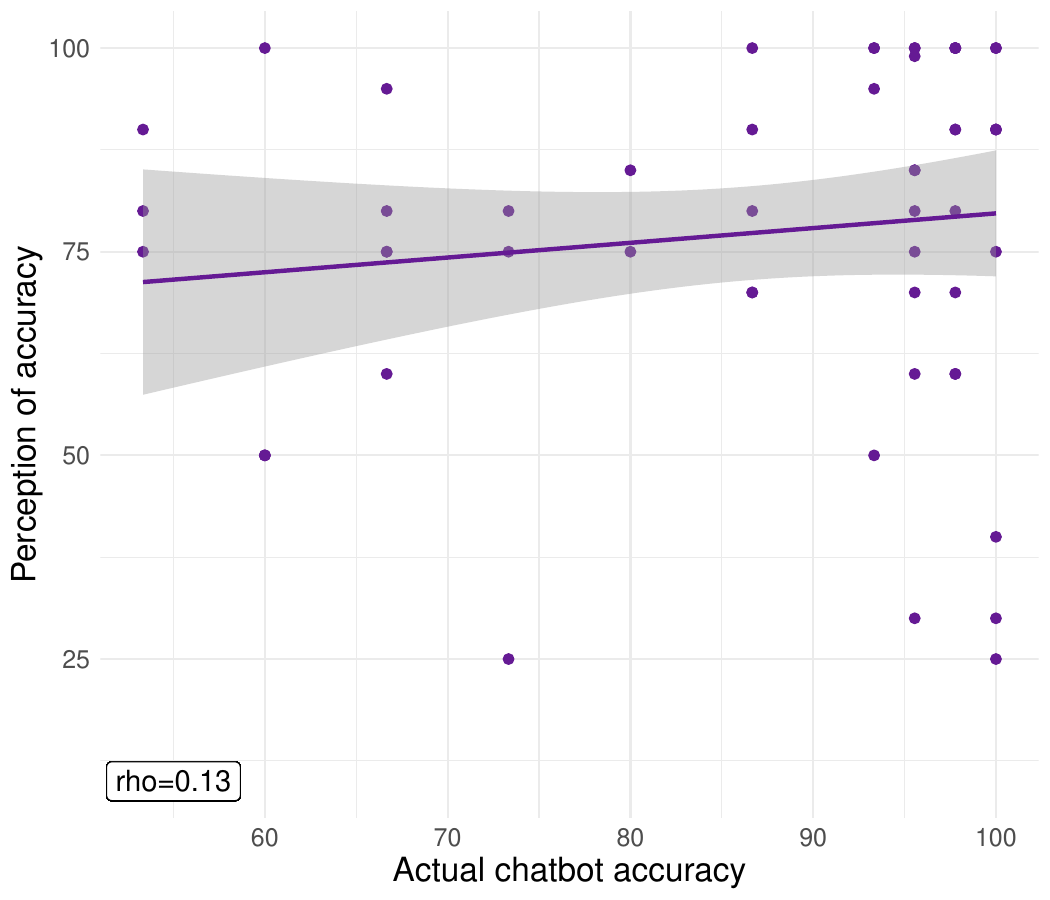}
    \caption{Relationship between caseworker perception of accuracy (y-axis) and actual chatbot accuracy (x-axis). While the two variables should be highly correlated, the Pearson correlation coefficient is 0.013 (annotated in the lower lefthand corner of the figure). Caseworkers appear to struggle in evaluating the accuracy of the chatbot.}
    \label{fig:accuracy_perception_passed}
    \end{minipage}

\begin{minipage}{\textwidth}
    \centering
    \includegraphics[width=0.65\linewidth]{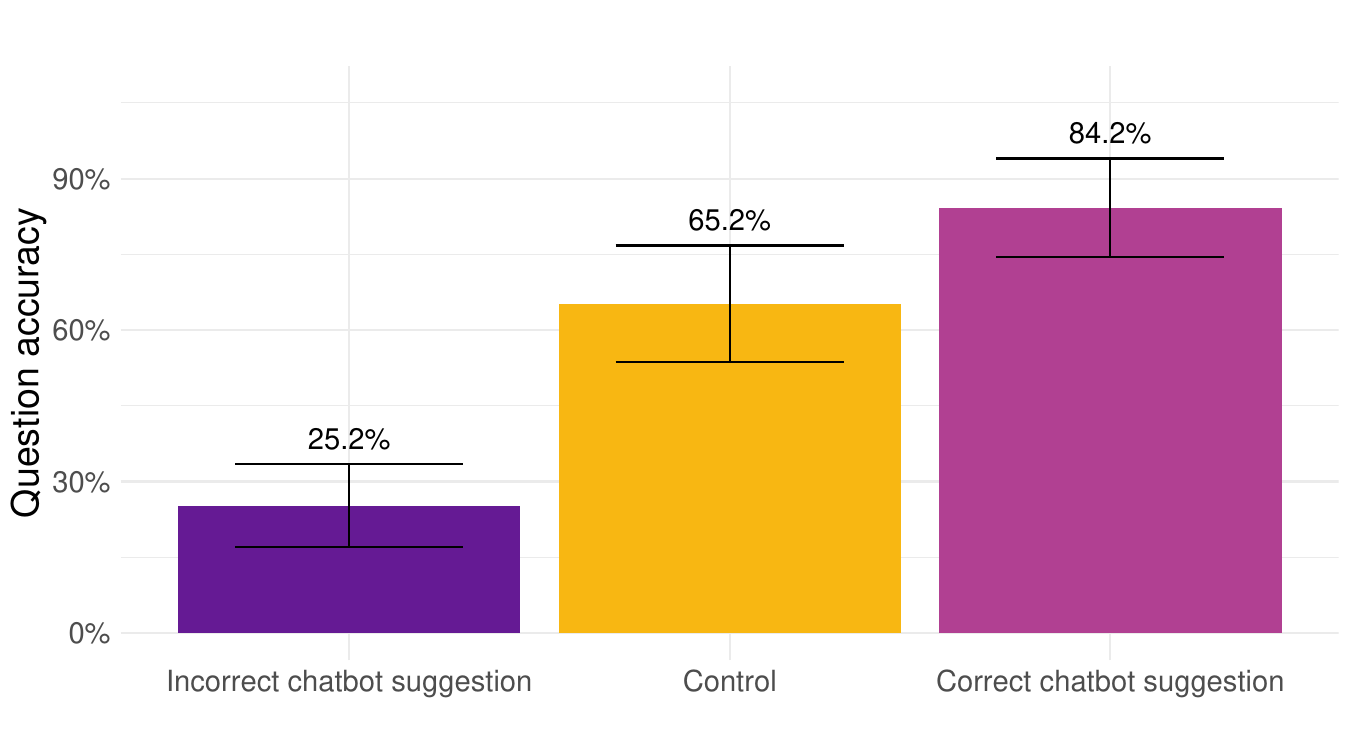}
    \caption{Question-level accuracy is shown for the three questions that appear in all assessments with the same set of client attributes. Similar to Figure~\ref{fig:question_level_effect_passed}, incorrect chatbot suggestions lead to large reductions in caseworker accuracy relative to accuracy in the control group (which is higher, on average, than the control-group accuracy shown for all questions in Figure~\ref{fig:question_level_effect_passed}). Correct suggestions also lead to improvements in caseworker accuracy relative to the control group. Different colors (purple, yellow, and light purple) indicate the type of chatbot suggestion. Caseworker question-level accuracy appears on the y-axis. We include error bars with the 95\% confidence intervals for all estimates.}
    \label{fig:special_qs_passed}
    \end{minipage}
\end{figure}

\begin{figure}[t!]
    \centering
    \includegraphics[width=0.8\linewidth]{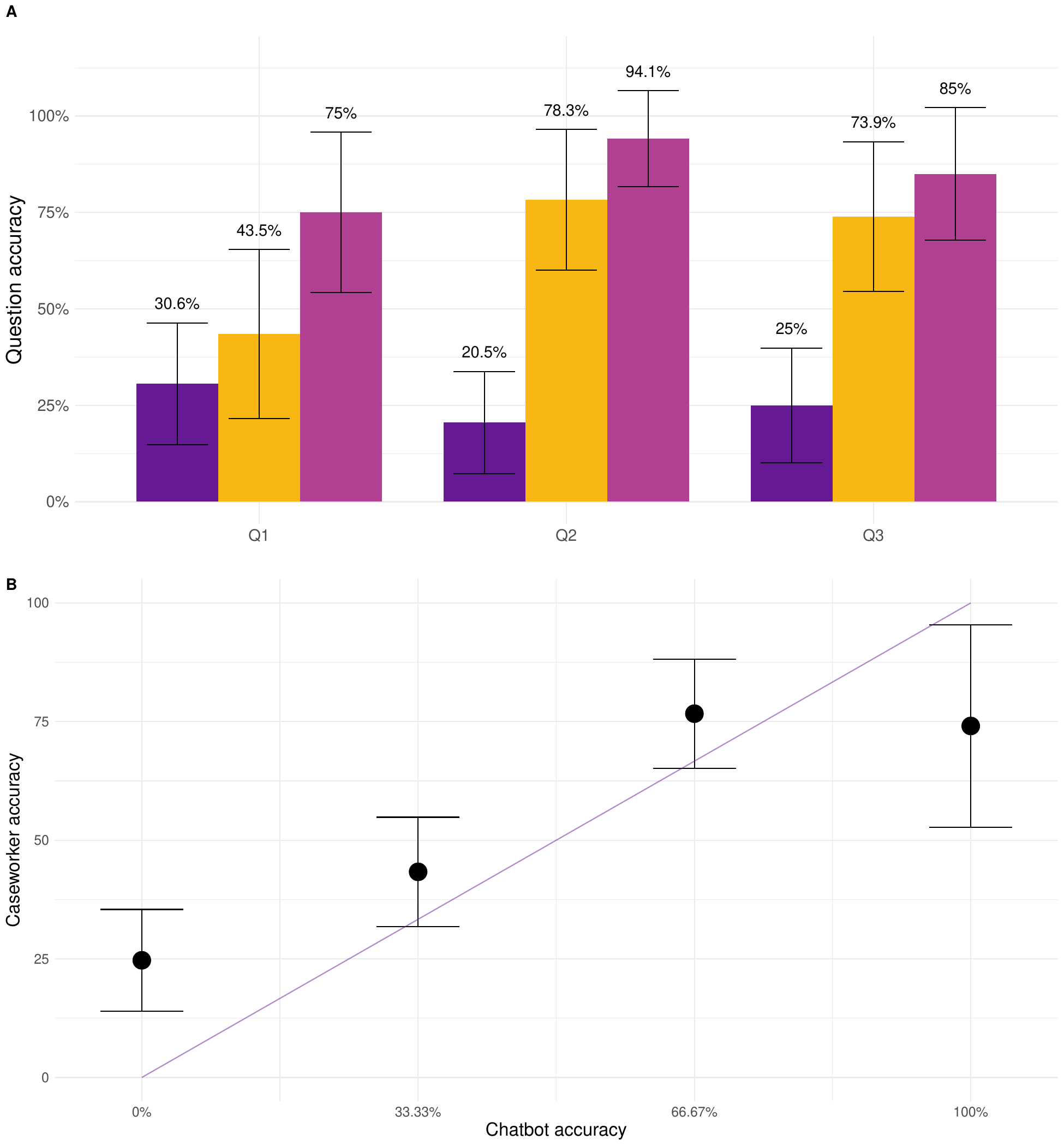}
    \caption{(A) Question-level accuracy is shown for the three questions that appear in all assessments with the same set of client attributes. In contrast to Figure~\ref{fig:special_qs_passed}, accuracy is shown for each specific question (denoted ``Q1,'' ``Q2,'' and ``Q3''). Incorrect chatbot suggestions lead to large reductions in caseworker accuracy relative to accuracy in the control group for Q2 and Q3, where accuracy in the control group is high. In Q1, correct suggestions lead to improvements in accuracy, but the confidence intervals overlap. Different colors (purple, yellow, and light purple) indicate the type of chatbot suggestion. Caseworker question-level accuracy appears on the y-axis. We include error bars with the 95\% confidence intervals for all estimates.  (B) Caseworker participant-level accuracy is shown only for the three questions that appear in all assessments with the same set of client attributes. The simulated chatbot accuracy for the three questions is shown on the x-axis. Simulated chatbot accuracy ranges from 0\% (all three questions have incorrect chatbot suggestions) to 100\% (all three questions have correct chatbot suggestions). Similar to Figure~\ref{fig:accuracy_panel_passed}, there is slight evidence for a plateau effect between the medium (66.67\%) and high (100\%) accuracy conditions. Error bars denote the 95\% confidence intervals for all estimates.}
    \label{fig:special_qs_X_question_passed}
\end{figure}

\FloatBarrier
\newpage 
\section{Regression Tables}

\begin{table}[!htbp] \centering 
\adjustbox{max width=\textwidth}{
\begin{tabular}{@{\extracolsep{5pt}}lcccccc} 
\\[-1.8ex]\hline 
\hline \\[-1.8ex] 
 & \multicolumn{6}{c}{\textit{Dependent variable:}} \\ 
\cline{2-7} 
\\[-1.8ex] & \multicolumn{6}{c}{Caseworker participant-level accuracy} \\ 
\\[-1.8ex] & \multicolumn{3}{c}{All participants (n=125)} & \multicolumn{3}{c}{Passed attention check (n=79)} \\ 
\\[-1.8ex] & (1) & (2) & (3) & (4) & (5) & (6)\\ 
\hline \\[-1.8ex] 
Treatment & 21.393$^{***}$ &  & $-$27.347$^{***}$ & 21.351$^{***}$ &  & $-$24.028$^{*}$ \\ 
  & (2.443) &  & (6.974) & (2.963) &  & (10.188) \\ 
  & & & & & & \\ 
 Low quality (53-73\% accurate) &  & 8.062$^{**}$ &  &  & 8.676$^{*}$ &  \\ 
  &  & (2.906) &  &  & (3.514) &  \\ 
  & & & & & & \\ 
 Medium quality (80-93\% accurate) &  & 24.170$^{***}$ &  &  & 25.676$^{***}$ &  \\ 
  &  & (3.249) &  &  & (3.290) &  \\ 
  & & & & & & \\ 
 High quality (96-100\% accurate) &  & 27.264$^{***}$ &  &  & 26.117$^{***}$ &  \\ 
  &  & (2.936) &  &  & (3.802) &  \\ 
  & & & & & & \\ 
 Treatment:Chatbot accuracy &  &  & 0.565$^{***}$ &  &  & 0.526$^{***}$ \\ 
  &  &  & (0.085) &  &  & (0.124) \\ 
  & & & & & & \\ 
 Constant & 49.032$^{***}$ & 49.032$^{***}$ & 49.032$^{***}$ & 50.435$^{***}$ & 50.435$^{***}$ & 50.435$^{***}$ \\ 
  & (1.713) & (1.727) & (1.720) & (2.050) & (2.077) & (2.064) \\ 
  & & & & & & \\ 
\hline \\[-1.8ex] 
Observations & 125 & 125 & 125 & 79 & 79 & 79 \\ 
R$^{2}$ & 0.268 & 0.431 & 0.454 & 0.315 & 0.455 & 0.463 \\ 
Adjusted R$^{2}$ & 0.262 & 0.417 & 0.445 & 0.306 & 0.433 & 0.449 \\ 
Residual Std. Error & 15.401 (df = 123) & 13.686 (df = 121) & 13.358 (df = 122) & 14.488 (df = 77) & 13.099 (df = 75) & 12.910 (df = 76) \\ 
F Statistic & 44.981$^{***}$ (df = 1; 123) & 30.571$^{***}$ (df = 3; 121) & 50.654$^{***}$ (df = 2; 122) & 35.410$^{***}$ (df = 1; 77) & 20.836$^{***}$ (df = 3; 75) & 32.777$^{***}$ (df = 2; 76) \\ 
\hline 
\hline \\[-1.8ex] 
\textit{Note:}  & \multicolumn{6}{r}{$^{*}$p$<$0.05; $^{**}$p$<$0.01; $^{***}$p$<$0.001} \\ 
\end{tabular}
}
  \caption{Results for effects of chatbot suggestions at the participant-level. We estimate all effects with robust standard errors. Column (1) is the overall treatment effect with a single covariate indicating whether caseworkers saw chatbot suggestions or not (``Treatment'').
  The coefficient on chatbot suggestions indicates that chatbot suggestions led to a 21.4 percentage point increase in accuracy on average.
  Column (2) is the effect of chatbot suggestions by  low, medium, and high quality chatbots. For example, a ``High quality chatbot'' is associated with a 27 percentage point increase in caseworker accuracy on average, relative to the accuracy of caseworkers in the control group. Column (3) interacts the treatment effect with a continuous variable for chatbot accuracy (out of 100\%). An increase in chatbot accuracy by one percentage point leads to a 0.6 percentage point increase in caseworker accuracy on average.
  Columns (4)-(6) present the same results only for caseworkers who pass the attention check (n=79). Results are directionally similar to columns (1)-(3).
  } 
  \label{tab:results} 
\end{table}

\newpage 
\begin{table}[!htbp]
\centering 
\begin{tabular}{@{\extracolsep{5pt}}lcc} 
\\[-1.8ex]\hline 
\hline \\[-1.8ex] 
 & \multicolumn{2}{c}{\textit{Dependent variable:}} \\ 
\cline{2-3} 
\\[-1.8ex] & \multicolumn{2}{c}{Caseworker participant-level accuracy} \\ 
\\[-1.8ex] & (1) & (2)\\ 
\hline \\[-1.8ex] 
Passed attention check & 1.222 & 5.435 \\ 
  & (3.410) & (3.348) \\ 
  & & \\ 
 Chatbot suggestions &  & 23.421$^{***}$ \\ 
  &  & (3.958) \\ 
  & & \\ 
 Chatbot suggestions:Passed attention check &  & $-$2.070 \\ 
  &  & (4.951) \\ 
  & & \\ 
 Constant & 64.348$^{***}$ & 45.000$^{***}$ \\ 
  & (2.790) & (2.641) \\ 
  & & \\ 
\hline \\[-1.8ex] 
Observations & 125 & 125 \\ 
R$^{2}$ & 0.001 & 0.279 \\ 
Adjusted R$^{2}$ & $-$0.007 & 0.261 \\ 
Residual Std. Error & 17.988 (df = 123) & 15.413 (df = 121) \\ 
F Statistic & 0.134 (df = 1; 123) & 15.577$^{***}$ (df = 3; 121) \\ 
\hline 
\hline \\[-1.8ex] 
\textit{Note:}  & \multicolumn{2}{r}{$^{*}$p$<$0.05; $^{**}$p$<$0.01; $^{***}$p$<$0.001} \\ 
\end{tabular} 
  \caption{Results for two linear regression models with robust standard errors that test whether there are statistically significant differences in average accuracy for participants who passed the attention check compared to those who did not (denoted by the binary covariate ``Passed attention check''). Column (1) tests overall whether there are  differences in caseworker accuracy. Column (2) tests whether there are any differences in the effect of chatbot suggestions. In both models, we do not observe any statistically significant differences in average caseworker accuracy or in the effect of chatbot suggestions on caseworker accuracy between participants who passed the attention check and those who did not. } 
  \label{tab:mod_attention} 
\end{table}

\begin{table}[!htbp] \centering 
  \begin{tabular}{@{\extracolsep{5pt}}lcc} 
\\[-1.8ex]\hline 
\hline \\[-1.8ex] 
 & \multicolumn{2}{c}{\textit{Dependent variable:}} \\ 
\cline{2-3} 
\\[-1.8ex] & \multicolumn{2}{c}{Caseworker median question-level duration} \\ 
\\[-1.8ex] & (1) & (2)\\ 
\hline \\[-1.8ex] 
 Chatbot suggestions & 2.147 & 2.776 \\ 
  & (3.923) & (4.045) \\ 
  & & \\ 
 Constant & 36.715$^{***}$ & 36.444$^{***}$ \\ 
  & (3.331) & (3.301) \\ 
  & & \\ 
\hline \\[-1.8ex] 
Observations & 125 & 79 \\ 
R$^{2}$ & 0.002 & 0.006 \\ 
Adjusted R$^{2}$ & $-$0.006 & $-$0.007 \\ 
Residual Std. Error & 19.726 (df = 123) & 17.021 (df = 77) \\ 
F Statistic & 0.276 (df = 1; 123) & 0.433 (df = 1; 77) \\ 
\hline 
\hline \\[-1.8ex] 
\textit{Note:}  & \multicolumn{2}{r}{$^{*}$p$<$0.05; $^{**}$p$<$0.01; $^{***}$p$<$0.001} \\ 
\end{tabular} 
\caption{Results for the effect of chatbot suggestions on caseworker median question-level duration. Column (1) tests for differences across all participants (n=125). Column (2) tests for differences only among participants who passed the attention check (n=79). Overall, we do not observe evidence that there are statistically significant differences in the median question-level duration based on whether caseworkers saw chatbot suggestions or not. } 
  \label{tab:duration_models} 
\end{table} 

\begin{table}[!htbp]
\centering 
\adjustbox{max width=\textwidth}{
\begin{tabular}{@{\extracolsep{5pt}}lcccc} 
\\[-1.8ex]\hline 
\hline \\[-1.8ex] 
 & \multicolumn{4}{c}{\textit{Dependent variable:}} \\ 
\cline{2-5} 
\\[-1.8ex] & \multicolumn{4}{c}{Caseworker participant-level accuracy} \\ 
\\[-1.8ex] & (1) & (2) & (3) & (4)\\ 
\hline \\[-1.8ex] 
Chatbot accuracy (\%) & 0.554$^{***}$ & 0.569$^{***}$ & 0.465$^{***}$ & 0.501$^{**}$ \\ 
  & (0.084) & (0.089) & (0.130) & (0.154) \\ 
  & & & & \\ 
 Perception of chatbot accuracy (\%) & 0.269$^{***}$ & 0.225$^{*}$ & 0.342$^{**}$ & 0.342$^{**}$ \\ 
  & (0.073) & (0.090) & (0.116) & (0.110) \\ 
  & & & & \\ 
 Percent used chatbot suggestions (\%)  &  & 0.070 &  & 0.050 \\ 
  &  & (0.067) &  & (0.076) \\ 
  & & & & \\ 
 Percent reflect situations with clients (\%) &  & $-$0.001 &  & 0.032 \\ 
  &  & (0.049) &  & (0.058) \\ 
  & & & & \\ 
 Slightly familiar &  & 10.704 &  & $-$2.137 \\ 
  &  & (8.699) &  & (5.441) \\ 
  & & & & \\ 
Somewhat familiar &  & 16.644$^{*}$ &  & 8.272 \\ 
  &  & (8.015) &  & (4.509) \\ 
  & & & & \\ 
 Familiar &  & 21.191$^{**}$ &  & 9.979 \\ 
  &  & (8.172) &  & (5.959) \\ 
  & & & & \\ 
Very &  & 17.320 &  & 5.957 \\ 
  &  & (10.303) &  & (6.594) \\ 
  & & & & \\ 
 Years of experience &  & $-$0.429 &  & $-$0.398 \\ 
  &  & (0.461) &  & (0.517) \\ 
  & & & & \\ 
 Constant & 2.234 & $-$14.571 & 5.183 & $-$7.740 \\ 
  & (7.509) & (11.128) & (10.860) & (10.994) \\ 
  & & & & \\ 
\hline \\[-1.8ex] 
Observations & 92 & 86 & 55 & 54 \\ 
R$^{2}$ & 0.416 & 0.517 & 0.450 & 0.545 \\ 
Adjusted R$^{2}$ & 0.403 & 0.459 & 0.429 & 0.452 \\ 
Residual Std. Error & 13.155 (df = 89) & 12.425 (df = 76) & 12.163 (df = 52) & 12.004 (df = 44) \\ 
F Statistic & 31.695$^{***}$ (df = 2; 89) & 9.024$^{***}$ (df = 9; 76) & 21.293$^{***}$ (df = 2; 52) & 5.858$^{***}$ (df = 9; 44) \\ 
\hline 
\hline \\[-1.8ex] 
\textit{Note:}  & \multicolumn{4}{r}{$^{*}$p$<$0.05; $^{**}$p$<$0.01; $^{***}$p$<$0.001} \\ 
\end{tabular}} 
  \caption{Regression results on the relationship between caseworker characteristics and accuracy only for participants who see chatbot suggestions. Columns (1) and (2) show results for all caseworkers; columns (3) and (4) show results only for caseworkers who pass the attention check. Caseworker perceptions of accuracy have the strongest positive relationship with actual caseworker accuracy on the assessment. A one percentage point increase in the perception of chatbot accuracy is associated with a 0.27 percentage point increase in accuracy on the assessment on average ($p < 0.001$). Other characteristics such as the percentage of chatbot suggestions used or years of experience are not meaningfully associated with average caseworker accuracy. Familiarity with the question content does lead to an increase in accuracy relative to caseworkers who are not at all familiar with the questions. However, this trend does not hold when restricting to caseworkers who pass the attention check in column (4). Missing responses to some questions reduce the sample sizes. We do not impute responses, though results are robust to using mean imputation for ``percent reflected situations with clients.''} 
  \label{tab:accuracy_perception} 
\end{table} 

\FloatBarrier
\begin{longtable}[c]{M{3in}cc}
\\[-1.8ex]\hline 
\hline \\[-1.8ex] 
 & \multicolumn{2}{c}{\textit{Dependent variable:}} \\ 
\cline{2-3} 
\\[-1.8ex] & \multicolumn{2}{c}{Caseworker participant-level accuracy} \\ 
\\[-1.8ex] & (1) & (2)\\ 
\hline \\[-1.8ex] 
Treatment & 25.988$^{***}$ & 20.055$^{*}$ \\ 
  & (3.976) & (9.011) \\ 
  & & \\ 
Chatbot accuracy (\%) & 0.596$^{***}$ & 0.501 \\ 
  & (0.115) & (0.276) \\ 
  & & \\ 
Know CalFresh : Moderately well & $-$6.683 & $-$1.451 \\ 
  & (4.238) & (7.102) \\ 
  & & \\ 
Know CalFresh : Very well & $-$12.124$^{**}$ & $-$10.269 \\ 
  & (4.371) & (11.849) \\ 
  & & \\ 
Know CalFresh : Extremely well & $-$11.868$^{*}$ & 7.486 \\ 
  & (5.320) & (7.999) \\ 
  & & \\ 
Existing chatbot use : Once a month or less & 0.987 & 5.974 \\ 
  & (5.365) & (9.308) \\ 
  & & \\ 
Existing chatbot use : A couple times a month & $-$8.445 & $-$6.823 \\ 
  & (6.149) & (13.023) \\ 
  & & \\ 
Existing chatbot use : At least once a week & $-$7.384 & 4.881 \\ 
  & (6.179) & (12.845) \\ 
  & & \\ 
Existing chatbot use : About once a day & $-$10.672 & 8.702 \\ 
  & (7.520) & (15.063) \\ 
  & & \\ 
Existing chatbot use : More than once a day & $-$31.886$^{***}$ & $-$19.832 \\ 
  & (8.656) & (16.170) \\ 
  & & \\ 
Baseline chatbot views : Somewhat negative & 8.754 & 5.214 \\ 
  & (13.594) & (19.840) \\ 
  & & \\ 
Baseline chatbot views : Neither positive nor negative & $-$0.794 & $-$8.648 \\ 
  & (8.959) & (16.481) \\ 
  & & \\ 
Baseline chatbot views : Somewhat positive & 5.051 & $-$1.241 \\ 
  & (10.497) & (18.222) \\ 
  & & \\ 
Baseline chatbot views : Very positive & 10.477 & $-$13.506 \\ 
  & (12.046) & (18.759) \\ 
  & & \\ 
Views on chatbot harm : About an equal mix of benefit and harm & 25.778 & 4.207 \\ 
  & (15.888) & (23.806) \\ 
  & & \\ 
Views on chatbot harm : More benefit than harm & 23.490 & $-$0.391 \\ 
  & (16.454) & (28.253) \\ 
  & & \\ 
Mental effort : Low mental effort & $-$18.272 & $-$47.828 \\ 
  & (14.647) & (30.088) \\ 
  & & \\ 
Mental effort : Neither low nor high mental effort & $-$14.675 & $-$38.654 \\ 
  & (9.677) & (22.863) \\ 
  & & \\ 
Mental effort : High mental effort & $-$15.751 & $-$33.126 \\ 
  & (10.263) & (24.527) \\ 
  & & \\ 
Mental effort : Very high mental effort & $-$16.077 & $-$51.508$^{*}$ \\ 
  & (10.730) & (25.864) \\ 
  & & \\ 
Difficulty : Slightly familiar & 7.755 & $-$41.685 \\ 
  & (10.969) & (38.696) \\ 
  & & \\ 
Difficulty : Somewhat familiar & 8.735 & $-$50.361 \\ 
  & (11.200) & (40.727) \\ 
  & & \\ 
Difficulty : Familiar & 13.931 & $-$33.298 \\ 
  & (11.477) & (38.118) \\ 
  & & \\ 
Difficulty : Very familiar & 26.025$^{*}$ & $-$38.087 \\ 
  & (12.688) & (37.266) \\ 
  & & \\ 
 Percent reflected situations with clients (\%) & $-$0.100 & 0.218 \\ 
  & (0.057) & (0.159) \\ 
  & & \\ 
Years of experience & $-$0.102 & $-$0.775 \\ 
  & (0.318) & (0.982) \\ 
  & & \\ 
Difficulty with questions : A couple times a month & 0.999 & $-$3.305 \\ 
  & (4.170) & (8.649) \\ 
  & & \\ 
Difficulty with questions : At least once a week & 1.891 & $-$2.712 \\ 
  & (4.360) & (8.306) \\ 
  & & \\ 
Difficulty with questions : About once a day & 1.743 & $-$3.557 \\ 
  & (5.714) & (9.217) \\ 
  & & \\ 
Difficulty with questions : More than once a day & $-$6.625 & $-$27.166 \\ 
  & (5.291) & (24.241) \\ 
  & & \\ 
Use Google : Once a month or less & $-$30.386 & 67.343 \\ 
  & (21.012) & (49.121) \\ 
  & & \\ 
Use Google : A couple times a month & $-$25.988 & 31.113 \\ 
  & (19.240) & (41.043) \\ 
  & & \\ 
Use Google : At least once a week & $-$31.543 & 28.047 \\ 
  & (19.174) & (40.634) \\ 
  & & \\ 
Use Google : About once a day & $-$16.311 & 44.884 \\ 
  & (18.848) & (40.487) \\ 
  & & \\ 
Use Google : More than once a day & $-$25.843 & 39.627 \\ 
  & (18.533) & (46.059) \\ 
  & & \\ 
Use Slack or email : Once a month or less & 5.970 & 20.316 \\ 
  & (9.660) & (22.481) \\ 
  & & \\ 
Use Slack or email : A couple times a month & $-$3.797 & 7.074 \\ 
  & (7.741) & (12.906) \\ 
  & & \\ 
Use Slack or email : At least once a week & $-$1.441 & $-$5.740 \\ 
  & (6.870) & (12.097) \\ 
  & & \\ 
Use Slack or email : About once a day & 2.293 & 2.237 \\ 
  & (9.516) & (23.021) \\ 
  & & \\ 
Use Slack or email : More than once a day & $-$6.654 & 0.657 \\ 
  & (8.625) & (20.207) \\ 
  & & \\ 
Use manual : Once a month or less & $-$2.464 & $-$11.210 \\ 
  & (6.231) & (11.169) \\ 
  & & \\ 
Use manual : A couple times a month & $-$1.855 & $-$5.399 \\ 
  & (7.044) & (12.077) \\ 
  & & \\ 
Use manual : At least once a week & $-$8.513 & $-$12.283 \\ 
  & (6.369) & (11.984) \\ 
  & & \\ 
Use manual : About once a day & $-$9.363 & $-$24.726 \\ 
  & (8.147) & (23.277) \\ 
  & & \\ 
Use manual : More than once a day & $-$5.297 & $-$22.468 \\ 
  & (9.341) & (17.374) \\ 
  & & \\ 
 Use phone : Once a month or less & 2.080 & 10.528 \\ 
  & (8.410) & (16.301) \\ 
  & & \\ 
Use phone : A couple times a month & 7.357 & 9.120 \\ 
  & (8.021) & (20.023) \\ 
  & & \\ 
Use phone : At least once a week & 6.616 & 20.533 \\ 
  & (6.038) & (20.988) \\ 
  & & \\ 
Use phone : About once a day & 5.834 & 27.759 \\ 
  & (8.215) & (27.360) \\ 
  & & \\ 
Use phone : More than once a day & 8.707 & 16.409 \\ 
  & (8.020) & (20.722) \\ 
  & & \\ 
Use chatbot : Once a month or less & $-$0.673 & $-$0.106 \\ 
  & (4.748) & (8.146) \\ 
  & & \\ 
Use chatbot : A couple times a month & 7.571 & 8.440 \\ 
  & (7.359) & (10.852) \\ 
  & & \\ 
Use chatbot : At least once a week & $-$4.586 & $-$5.238 \\ 
  & (7.433) & (16.655) \\ 
  & & \\ 
Use chatbot : About once a day & 7.414 & $-$1.432 \\ 
  & (9.554) & (22.366) \\ 
  & & \\ 
Use chatbot : More than once a day & 27.206$^{***}$ & 33.598$^{*}$ \\ 
  & (7.973) & (14.182) \\ 
  & & \\ 
In person : Once a month or less & 2.271 & $-$16.877 \\ 
  & (8.676) & (22.708) \\ 
  & & \\ 
In person : A couple times a month & $-$2.627 & $-$25.324 \\ 
  & (8.358) & (18.738) \\ 
  & & \\ 
In person : At least once a week & 3.559 & $-$23.919 \\ 
  & (7.857) & (18.133) \\ 
  & & \\ 
In person : About once a day & $-$5.472 & $-$39.458 \\ 
  & (8.823) & (24.394) \\ 
  & & \\ 
In person : More than once a day & 1.332 & $-$18.898 \\ 
  & (8.531) & (24.400) \\ 
  & & \\ 
 Constant & 9.485 & 61.558 \\ 
  & (20.961) & (46.640) \\ 
  & & \\ 
\hline \\[-1.8ex] 
Observations & 114 & 75 \\ 
R$^{2}$ & 0.811 & 0.900 \\ 
Adjusted R$^{2}$ & 0.596 & 0.469 \\ 
Residual Std. Error & 11.502 (df = 53) & 12.819 (df = 14) \\ 
F Statistic & 3.782$^{***}$ (df = 60; 53) & 2.090 (df = 60; 14) \\ 
\hline 
\hline \\[-1.8ex] 
\textit{Note:}  & \multicolumn{2}{r}{$^{*}$p$<$0.05; $^{**}$p$<$0.01; $^{***}$p$<$0.001} \\ 
\caption{\small{Overall regression results on the relationship between caseworker characteristics and accuracy. Column (1) shows results for all caseworkers; column (2) shows results only for caseworkers who pass the attention check. Higher levels of chatbot use (more than once a day) is associated with greater caseworker accuracy on average, even when controlling for chatbot accuracy. The positive effect holds even when restricting to caseworkers who pass the attention check, though the result is only significant at the 5\% level. Familiarity with CalFresh and with the question content is also associated with increased accuracy on average, but the trend is not statistically significant when restricting to caseworkers who pass the attention check.
Overall, chatbot suggestions do improve caseworker accuracy, though results are not as statistically significant when controlling for all participant characteristics and restricting to only caseworkers who pass the attention check, as shown in column (2).
Missing responses to some questions reduce the sample sizes. We do not impute responses, though results are robust to using mean imputation for perception of chatbot accuracy, percent used chatbot suggestions and percent reflected situations with clients.}}
  \label{tab:all_char}
  \end{longtable}

\FloatBarrier
\begin{table}[!htbp] \centering 
\adjustbox{max width=\textwidth}{
\begin{tabular}{@{\extracolsep{5pt}}lcc} 
\\[-1.8ex]\hline 
\hline \\[-1.8ex] 
 & \multicolumn{2}{c}{\textit{Dependent variable:}} \\ 
\cline{2-3} 
\\[-1.8ex] & \multicolumn{2}{c}{accept\_suggestion} \\ 
\\[-1.8ex] & (1) & (2)\\ 
\hline \\[-1.8ex] 
Prior agreement rate & 0.007$^{***}$ & 0.006$^{***}$ \\ 
  & (0.001) & (0.001) \\ 
  & & \\ 
Medium quality (80-93\% accurate) & 0.157$^{*}$ & 0.182 \\ 
  & (0.077) & (0.102) \\ 
  & & \\ 
High quality (96-100\% accurate) & $-$0.082 & $-$0.165 \\ 
  & (0.070) & (0.086) \\ 
  & & \\ 
Use chatbot : Once a month or less & 0.026 & 0.012 \\ 
  & (0.015) & (0.018) \\ 
  & & \\ 
Use chatbot : A couple times a month & $-$0.022 & $-$0.071$^{*}$ \\ 
  & (0.020) & (0.029) \\ 
  & & \\ 
Use chatbot : At least once a week & $-$0.082$^{**}$ & $-$0.123$^{**}$ \\ 
  & (0.025) & (0.040) \\ 
  & & \\ 
Use chatbot : About once a day & $-$0.004 & $-$0.024 \\ 
  & (0.018) & (0.050) \\ 
  & & \\ 
Use chatbot : More than once a day & 0.007 & $-$0.017 \\ 
  & (0.017) & (0.019) \\ 
  & & \\ 
 Prior agreement rate : Medium quality chatbot (80-93\% accurate) & $-$0.002 & $-$0.002 \\ 
  & (0.001) & (0.001) \\ 
  & & \\ 
 Prior agreement rate : High quality chatbot (96-100\% accurate) & 0.001 & 0.002$^{*}$ \\ 
  & (0.001) & (0.001) \\ 
  & & \\ 
\hline \\[-1.8ex] 
Observations & 3,680 & 2,200 \\ 
R$^{2}$ & 0.126 & 0.133 \\ 
Adjusted R$^{2}$ & 0.114 & 0.113 \\ 
F Statistic & 52.369$^{***}$ (df = 10; 3630) & 33.001$^{***}$ (df = 10; 2150) \\ 
\hline 
\hline \\[-1.8ex] 
\textit{Note:}  & \multicolumn{2}{r}{$^{*}$p$<$0.05; $^{**}$p$<$0.01; $^{***}$p$<$0.001} \\ 
\end{tabular}} 
  \caption{Regression results predicting whether caseworkers will agree with a chatbot suggestion. Each observation in the model reflects a caseworker's response to a question at different positions in the assessment.   We exclude the first 5 questions in the assessment. All models have random effects for different question positions in the assessment (1-45). Column (1) includes responses from all caseworkers. Column (2) excludes caseworkers who failed the attention check.
  Prior agreement rate is the percentage of questions in which a caseworker has agreed with the chatbot suggestions on the same assessment prior to this question. We also include covariates for chatbot quality and self-reported chatbot usage to answer client questions.
  A one percentage point increase in a caseworker's prior agreement rate is associated with a 0.7 percentage point increase in the likelihood that a caseworker will follow the chatbot suggestion on a new question for all caseworkers and a 0.6 percentage point increase for caseworkers who pass the attention check.
  } 
  \label{tab:regression-agreement} 
\end{table}

\newpage 

\begin{table}[!htbp] \centering 
\adjustbox{max width=\textwidth}{
\begin{tabular}{@{\extracolsep{5pt}}lccccc} 
\\[-1.8ex]\hline 
\hline \\[-1.8ex] 
 & \multicolumn{5}{c}{\textit{Dependent variable:}} \\ 
\cline{2-6} 
\\[-1.8ex] & \multicolumn{5}{c}{Caseworker answered correctly} \\ 
\\[-1.8ex] & \multicolumn{2}{c}{\textit{OLS}} & \multicolumn{3}{c}{\textit{Random effects}} \\ 
\\[-1.8ex] & (1) & (2) & (3) & (4) & (5)\\ 
\hline \\[-1.8ex] 
 Incorrect chatbot suggestion & $-$0.305$^{***}$ & $-$0.542$^{***}$ & $-$0.550$^{***}$ & $-$0.501$^{***}$ & $-$0.511$^{***}$ \\ 
  & (0.025) & (0.068) & (0.069) & (0.051) & (0.051) \\ 
  & & & & & \\ 
 Correct chatbot suggestion & 0.297$^{***}$ & 0.061$^{*}$ & 0.061$^{*}$ & 0.058$^{**}$ & 0.058$^{**}$ \\ 
  & (0.025) & (0.031) & (0.031) & (0.020) & (0.020) \\ 
  & & & & & \\ 
 Chatbot accuracy &  & $-$0.0001 & $-$0.0001 & $-$0.0001 & $-$0.0001 \\ 
  &  & (0.001) & (0.001) & (0.0005) & (0.0005) \\ 
  & & & & & \\ 
 Difficulty=medium &  & $-$0.289$^{***}$ & $-$0.288$^{***}$ & $-$0.292$^{***}$ & $-$0.290$^{***}$ \\ 
  &  & (0.025) & (0.025) & (0.026) & (0.026) \\ 
  & & & & & \\ 
 Difficulty=hard &  & $-$0.615$^{***}$ & $-$0.615$^{***}$ & $-$0.617$^{***}$ & $-$0.616$^{***}$ \\ 
  &  & (0.025) & (0.025) & (0.027) & (0.027) \\ 
  & & & & & \\ 
 Correct chatbot suggestion:Difficulty=medium &  & 0.207$^{***}$ & 0.209$^{***}$ & 0.209$^{***}$ & 0.212$^{***}$ \\ 
  &  & (0.028) & (0.028) & (0.025) & (0.025) \\ 
  & & & & & \\ 
 Correct chatbot suggestion:Difficulty=hard &  & 0.400$^{***}$ & 0.400$^{***}$ & 0.405$^{***}$ & 0.404$^{***}$ \\ 
  &  & (0.032) & (0.032) & (0.030) & (0.030) \\ 
  & & & & & \\ 
 Incorrect chatbot suggestion:Difficulty=medium &  & 0.214$^{**}$ & 0.217$^{**}$ & 0.174$^{*}$ & 0.178$^{*}$ \\ 
  &  & (0.068) & (0.070) & (0.083) & (0.083) \\ 
  & & & & & \\ 
 Incorrect chatbot suggestion:Difficulty=hard &  & 0.450$^{***}$ & 0.463$^{***}$ & 0.400$^{***}$ & 0.415$^{***}$ \\ 
  &  & (0.069) & (0.072) & (0.058) & (0.058) \\ 
  & & & & & \\ 
 Constant & 0.490$^{***}$ & 0.845$^{***}$ & 0.844$^{***}$ & 0.850$^{***}$ & 0.850$^{***}$ \\ 
  & (0.017) & (0.108) & (0.108) & (0.054) & (0.054) \\ 
  & & & & & \\ 
\hline \\[-1.8ex] 
Observations & 5,625 & 5,625 & 5,625 & 5,625 & 5,625 \\ 
R$^{2}$ & 0.180 & 0.276 & 0.254 & 0.272 &  \\ 
Adjusted R$^{2}$ & 0.180 & 0.275 & 0.252 & 0.271 &  \\ 
Log Likelihood &  &  &  &  & $-$2,768.213 \\ 
Akaike Inf. Crit. &  &  &  &  & 5,562.425 \\ 
Bayesian Inf. Crit. &  &  &  &  & 5,648.680 \\ 
Residual Std. Error & 0.432 (df = 5622) & 0.406 (df = 5615) &  &  &  \\ 
F Statistic & 617.636$^{***}$ (df = 2; 5622) & 238.233$^{***}$ (df = 9; 5615) & 1,908.589$^{***}$ & 1,696.412$^{***}$ &  \\ 
\hline 
\hline \\[-1.8ex] 
\textit{Note:}  & \multicolumn{5}{r}{$^{*}$p$<$0.05; $^{**}$p$<$0.01; $^{***}$p$<$0.001} \\ 
\end{tabular} }
  \caption{
  Results for question-level effects of chatbot suggestions by type (i.e., incorrect chatbot suggestion, control, and correct chatbot suggestion). We present results for all questions that appear in the assessment. Column (1) is a simple linear model with binary covariates, denoting whether the chatbot suggestion is incorrect or correct. Notably, an incorrect chatbot suggestion decreases caseworker accuracy by around 31 percentage points on average ($p < 0.001$) while a correct chatbot suggestion increases caseworker accuracy by around 30 percentage points on average ($p<0.001$).
  Column (2) presents results for an expanded number of features, including the chatbot accuracy that the caseworker saw, the question difficulty, and interactions with the chatbot suggestion type. The negative effects for incorrect chatbot suggestions and the positive effects for correct chatbot suggestions still hold; there are also statistically significant differences by question difficulty. Column (3) includes person-level random effects, column (4) includes question-level random effects, and column (5) includes both. Results are directionally similar to column (2).
  Except for column (5), all standard errors are robust and clustered by participants.} 
  \label{tab:question_all} 
\end{table}

\begin{table}[!htbp] \centering 
\adjustbox{max width=\textwidth}{
\begin{tabular}{@{\extracolsep{5pt}}lccccc} 
\\[-1.8ex]\hline 
\hline \\[-1.8ex] 
 & \multicolumn{5}{c}{\textit{Dependent variable:}} \\ 
\cline{2-6} 
\\[-1.8ex] & \multicolumn{5}{c}{Caseworker answered correctly} \\ 
\\[-1.8ex] & \multicolumn{2}{c}{\textit{OLS}} & \multicolumn{3}{c}{\textit{Random effects}} \\  
\\[-1.8ex] & (1) & (2) & (3) & (4) & (5)\\ 
\hline \\[-1.8ex] 
 Incorrect chatbot suggestion & $-$0.195$^{***}$ & $-$0.449$^{***}$ & $-$0.451$^{***}$ & $-$0.444$^{***}$ & $-$0.445$^{***}$ \\ 
  & (0.026) & (0.083) & (0.084) & (0.059) & (0.059) \\ 
  & & & & & \\ 
 Correct chatbot suggestion & 0.348$^{***}$ & 0.105 & 0.111 & 0.094$^{***}$ & 0.100$^{***}$ \\ 
  & (0.029) & (0.063) & (0.061) & (0.021) & (0.021) \\ 
  & & & & & \\ 
 Chatbot accuracy &  & 0.0001 & 0.0002 & $-$0.0002 & $-$0.0002 \\ 
  &  & (0.001) & (0.001) & (0.001) & (0.001) \\ 
  & & & & & \\ 
 Difficulty=medium &  & $-$0.211$^{***}$ & $-$0.208$^{***}$ & $-$0.215$^{***}$ & $-$0.213$^{***}$ \\ 
  &  & (0.047) & (0.045) & (0.049) & (0.049) \\ 
  & & & & & \\ 
 Difficulty=hard &  & $-$0.526$^{***}$ & $-$0.525$^{***}$ & $-$0.529$^{***}$ & $-$0.528$^{***}$ \\ 
  &  & (0.052) & (0.051) & (0.047) & (0.047) \\ 
  & & & & & \\ 
 Correct chatbot suggestion:Difficulty=medium &  & 0.148$^{**}$ & 0.142$^{**}$ & 0.160$^{***}$ & 0.154$^{***}$ \\ 
  &  & (0.057) & (0.054) & (0.029) & (0.029) \\ 
  & & & & & \\ 
 Correct chatbot suggestion:Difficulty=hard &  & 0.349$^{***}$ & 0.341$^{***}$ & 0.361$^{***}$ & 0.354$^{***}$ \\ 
  &  & (0.065) & (0.062) & (0.036) & (0.036) \\ 
  & & & & & \\ 
 Incorrect chatbot suggestion:Difficulty=medium &  & 0.137 & 0.135 & 0.114 & 0.112 \\ 
  &  & (0.078) & (0.079) & (0.090) & (0.090) \\ 
  & & & & & \\ 
 Incorrect chatbot suggestion:Difficulty=hard &  & 0.362$^{***}$ & 0.367$^{***}$ & 0.337$^{***}$ & 0.341$^{***}$ \\ 
  &  & (0.082) & (0.082) & (0.065) & (0.065) \\ 
  & & & & & \\ 
 Constant & 0.380$^{***}$ & 0.736$^{***}$ & 0.733$^{***}$ & 0.769$^{***}$ & 0.767$^{***}$ \\ 
  & (0.019) & (0.117) & (0.116) & (0.083) & (0.083) \\ 
  & & & & & \\ 
\hline \\[-1.8ex] 
Observations & 2,994 & 2,994 & 2,994 & 2,994 & 2,994 \\ 
R$^{2}$ & 0.204 & 0.251 & 0.199 & 0.244 &  \\ 
Adjusted R$^{2}$ & 0.203 & 0.249 & 0.197 & 0.241 &  \\ 
Log Likelihood &  &  &  &  & $-$1,693.786 \\ 
Akaike Inf. Crit. &  &  &  &  & 3,413.571 \\ 
Bayesian Inf. Crit. &  &  &  &  & 3,491.628 \\ 
Residual Std. Error & 0.445 (df = 2991) & 0.432 (df = 2984) &  &  &  \\ 
F Statistic & 382.820$^{***}$ (df = 2; 2991) & 110.972$^{***}$ (df = 9; 2984) & 741.499$^{***}$ & 925.596$^{***}$ &  \\ 
\hline 
\hline \\[-1.8ex] 
\textit{Note:}  & \multicolumn{5}{r}{$^{*}$p$<$0.05; $^{**}$p$<$0.01; $^{***}$p$<$0.001} \\ 
\end{tabular} }
  \caption{
  \small{Results for question-level effects of chatbot suggestions by type (i.e., incorrect chatbot suggestion, control, and correct chatbot suggestion). 
  Here, we present results only for questions that appear in at least one assessment with an incorrect chatbot suggestion. These questions are harder on average, and the average caseworker accuracy in the control group is lower on these questions compared to the full set of assessment questions.
  As in Table~\ref{tab:question_all}, column (1) is a simple linear model with binary covariates, denoting whether the chatbot suggestion is incorrect or correct.
  Notably, an incorrect chatbot suggestion decreases caseworker accuracy by around 20 percentage points on average ($p < 0.001$) while a correct chatbot suggestion increases caseworker accuracy by around 34 percentage points on average ($p<0.001$).
  Column (2) presents results for an expanded number of features, including the chatbot accuracy that the caseworker saw, the question difficulty, and interactions with the chatbot suggestion type. The negative effects for incorrect chatbot suggestions and the positive effects for correct chatbot suggestions still hold, though the positive effects for correct chatbot suggestions on easy questions and medium questions are not statistically significant at the 5\% level. The negative effect of incorrect chatbot suggestions is also reduced on medium and hard questions, compared to easy questions. Column (3) includes person-level random effects, column (4) includes question-level random effects, and column (5) includes both. Results are directionally similar to column (2).
  Except for column (5), all standard errors are robust and clustered by participants.}} 
  \label{tab:question_all_filteredset}
\end{table} 

\begin{table}[!htbp] \centering
\adjustbox{max width=\textwidth}{
\begin{tabular}{@{\extracolsep{5pt}}lccccc} 
\\[-1.8ex]\hline 
\hline \\[-1.8ex] 
 & \multicolumn{5}{c}{\textit{Dependent variable:}} \\ 
\cline{2-6} 
\\[-1.8ex] & \multicolumn{5}{c}{Caseworker answered correctly} \\ 
\\[-1.8ex] & \multicolumn{2}{c}{\textit{OLS}} & \multicolumn{3}{c}{\textit{Random effects}} \\ 
\\[-1.8ex] & (1) & (2) & (3) & (4) & (5)\\ 
\hline \\[-1.8ex] 
 Incorrect chatbot suggestion & $-$0.184$^{***}$ & $-$0.567$^{***}$ & $-$0.577$^{***}$ & $-$0.575$^{***}$ & $-$0.586$^{***}$ \\ 
  & (0.034) & (0.097) & (0.098) & (0.102) & (0.102) \\ 
  & & & & & \\ 
 Correct chatbot suggestion & 0.342$^{***}$ & 0.085 & 0.088 & 0.078 & 0.082 \\ 
  & (0.037) & (0.078) & (0.076) & (0.043) & (0.043) \\ 
  & & & & & \\ 
 Chatbot accuracy &  & $-$0.001 & $-$0.001 & $-$0.001 & $-$0.001 \\ 
  &  & (0.001) & (0.001) & (0.001) & (0.001) \\ 
  & & & & & \\ 
 Difficulty=medium &  & $-$0.226$^{***}$ & $-$0.226$^{***}$ & $-$0.230$^{***}$ & $-$0.228$^{***}$ \\ 
  &  & (0.058) & (0.058) & (0.068) & (0.068) \\ 
  & & & & & \\ 
 Difficulty=hard &  & $-$0.559$^{***}$ & $-$0.558$^{***}$ & $-$0.560$^{***}$ & $-$0.558$^{***}$ \\ 
  &  & (0.061) & (0.060) & (0.062) & (0.062) \\ 
  & & & & & \\ 
 Correct chatbot suggestion:Difficulty=medium &  & 0.142$^{*}$ & 0.140$^{*}$ & 0.150$^{**}$ & 0.148$^{**}$ \\ 
  &  & (0.071) & (0.068) & (0.057) & (0.057) \\ 
  & & & & & \\ 
 Correct chatbot suggestion:Difficulty=hard &  & 0.362$^{***}$ & 0.360$^{***}$ & 0.368$^{***}$ & 0.365$^{***}$ \\ 
  &  & (0.079) & (0.076) & (0.052) & (0.052) \\ 
  & & & & & \\ 
 Incorrect chatbot suggestion:Difficulty=medium &  & 0.258$^{**}$ & 0.270$^{**}$ & 0.238 & 0.250 \\ 
  &  & (0.091) & (0.091) & (0.140) & (0.140) \\ 
  & & & & & \\ 
 Incorrect chatbot suggestion:Difficulty=hard &  & 0.459$^{***}$ & 0.470$^{***}$ & 0.437$^{***}$ & 0.447$^{***}$ \\ 
  &  & (0.098) & (0.097) & (0.106) & (0.106) \\ 
  & & & & & \\ 
 Constant & 0.390$^{***}$ & 0.867$^{***}$ & 0.863$^{***}$ & 0.915$^{***}$ & 0.914$^{***}$ \\ 
  & (0.023) & (0.150) & (0.148) & (0.120) & (0.120) \\ 
  & & & & & \\ 
\hline \\[-1.8ex] 
Observations & 1,890 & 1,890 & 1,890 & 1,890 & 1,890 \\ 
R$^{2}$ & 0.187 & 0.243 & 0.204 & 0.243 &  \\ 
Adjusted R$^{2}$ & 0.186 & 0.240 & 0.200 & 0.239 &  \\ 
Log Likelihood &  &  &  &  & $-$1,092.739 \\ 
Akaike Inf. Crit. &  &  &  &  & 2,211.479 \\ 
Bayesian Inf. Crit. &  &  &  &  & 2,283.555 \\ 
Residual Std. Error & 0.450 (df = 1887) & 0.435 (df = 1880) &  &  &  \\ 
F Statistic & 216.355$^{***}$ (df = 2; 1887) & 67.195$^{***}$ (df = 9; 1880) & 481.264$^{***}$ & 578.627$^{***}$ &  \\ 
\hline 
\hline \\[-1.8ex] 
\textit{Note:}  & \multicolumn{5}{r}{$^{*}$p$<$0.05; $^{**}$p$<$0.01; $^{***}$p$<$0.001} \\ 
\end{tabular} }
  \caption{
  \small{Results for question-level effects of chatbot suggestions by type (i.e., incorrect chatbot suggestion, control, and correct chatbot suggestion). 
  Here, we present results only for questions that appear in at least one assessment with an incorrect chatbot suggestion and only for participants who passed the attention check.
  Column (1) is a simple linear model with binary covariates, denoting whether the chatbot suggestion is incorrect or correct.
  An incorrect chatbot suggestion decreases caseworker accuracy by around 18 percentage points on average ($p < 0.001$) while a correct chatbot suggestion increases caseworker accuracy by around 34 percentage points on average ($p<0.001$).
  Column (2) presents results for an expanded number of features, including the chatbot accuracy that the caseworker saw, the question difficulty, and interactions with the chatbot suggestion type. The negative effects for incorrect chatbot suggestions and the positive effects for correct chatbot suggestions still hold, though the positive effects for correct chatbot suggestions on easy questions and medium questions are not statistically significant at the 1\% level. The negative effect of incorrect chatbot suggestions is also reduced on medium and hard questions, compared to easy questions. Column (3) includes person-level random effects, column (4) includes question-level random effects, and column (5) includes both. Results are directionally similar to column (2).
  Except for column (5), all standard errors are robust and clustered by participants.}} 
  \label{tab:question_passed_filteredset}
\end{table} 

\FloatBarrier
\begin{table}[!htbp] \centering 
\adjustbox{max width=\textwidth}{
\begin{tabular}{@{\extracolsep{5pt}}lccccc} 
\\[-1.8ex]\hline 
\hline \\[-1.8ex] 
 & \multicolumn{5}{c}{\textit{Dependent variable:}} \\ 
\cline{2-6} 
\\[-1.8ex] & \multicolumn{5}{c}{Caseworker answered correctly} \\ 
\\[-1.8ex] & \multicolumn{2}{c}{\textit{Logistic regression}} & \multicolumn{3}{c}{\textit{Logistic regression with random effects}} \\ 
\\[-1.8ex] & (1) & (2) & (3) & (4) & (5)\\ 
\hline \\[-1.8ex] 
 Incorrect chatbot suggestion & $-$1.442$^{***}$ & $-$2.458$^{***}$ & $-$2.878$^{***}$ & $-$2.225$^{***}$ & $-$2.674$^{***}$ \\ 
  & (0.140) & (0.577) & (0.594) & (0.389) & (0.622) \\ 
  & & & & & \\ 
 Correct chatbot suggestion & 1.347$^{***}$ & 0.598 & 0.569 & 0.583 & 0.512 \\ 
  & (0.126) & (0.664) & (0.557) & (0.306) & (0.568) \\ 
  & & & & & \\ 
 Difficulty=medium &  & $-$1.446$^{***}$ & $-$1.514$^{***}$ & $-$1.489$^{***}$ & $-$1.568$^{***}$ \\ 
  &  & (0.147) & (0.174) & (0.189) & (0.199) \\ 
  & & & & & \\ 
 Difficulty=hard &  & $-$2.891$^{***}$ & $-$3.063$^{***}$ & $-$2.956$^{***}$ & $-$3.145$^{***}$ \\ 
  &  & (0.171) & (0.177) & (0.189) & (0.200) \\ 
  & & & & & \\ 
 experiment\_typecontrol:Chatbot accuracy &  &  &  &  &  \\ 
  &  &  &  &  &  \\ 
  & & & & & \\ 
 experiment\_typetreatment:Chatbot accuracy &  & $-$0.001 & 0.002 & $-$0.001 & 0.002 \\ 
  &  & (0.007) & (0.006) & (0.003) & (0.006) \\ 
  & & & & & \\ 
 Correct chatbot suggestion:Difficulty=medium &  & 0.751$^{***}$ & 0.776$^{***}$ & 0.792$^{***}$ & 0.820$^{***}$ \\ 
  &  & (0.190) & (0.220) & (0.214) & (0.222) \\ 
  & & & & & \\ 
 Correct chatbot suggestion:Difficulty=hard &  & 1.478$^{***}$ & 1.474$^{***}$ & 1.540$^{***}$ & 1.546$^{***}$ \\ 
  &  & (0.230) & (0.214) & (0.207) & (0.216) \\ 
  & & & & & \\ 
 Incorrect chatbot suggestion:Difficulty=medium &  & 1.050$^{**}$ & 1.150$^{**}$ & 0.790$^{*}$ & 0.871$^{*}$ \\ 
  &  & (0.354) & (0.364) & (0.388) & (0.406) \\ 
  & & & & & \\ 
 Incorrect chatbot suggestion:Difficulty=hard &  & 1.868$^{***}$ & 2.098$^{***}$ & 1.563$^{***}$ & 1.762$^{***}$ \\ 
  &  & (0.408) & (0.366) & (0.389) & (0.408) \\ 
  & & & & & \\ 
 Constant & $-$0.039 & 1.638$^{***}$ & 1.727$^{***}$ & 1.694$^{***}$ & 1.801$^{***}$ \\ 
  & (0.068) & (0.165) & (0.197) & (0.150) & (0.210) \\ 
  & & & & & \\ 
\hline \\[-1.8ex] 
Observations & 5,625 & 5,625 & 5,625 & 5,625 & 5,625 \\ 
Log Likelihood & $-$3,133.824 & $-$2,838.605 & $-$2,698.455 & $-$2,825.347 & $-$2,681.837 \\ 
Akaike Inf. Crit. & 6,273.649 & 5,697.211 & 5,418.910 & 5,672.694 & 5,387.673 \\ 
Bayesian Inf. Crit. &  &  & 5,491.894 & 5,745.679 & 5,467.293 \\ 
\hline 
\hline \\[-1.8ex] 
\textit{Note:}  & \multicolumn{5}{r}{$^{*}$p$<$0.05; $^{**}$p$<$0.01; $^{***}$p$<$0.001} \\ 
\end{tabular} }
  \caption{
  Results for question-level effects of chatbot suggestions by type (i.e., incorrect chatbot suggestion, control, and correct chatbot suggestion) with logistic regression models. We present results for all questions that appear in the assessment. Results are directionally similar to Table~\ref{tab:question_all}.
  Column (1) is a simple linear model with binary covariates, denoting whether the chatbot suggestion is incorrect or correct.
  Column (2) presents results for an expanded number of features, including the chatbot accuracy that the caseworker saw, the question difficulty, and interactions with the chatbot suggestion type. Column (3) includes person-level random effects, column (4) includes question-level random effects, and column (5) includes both.
  Columns (1) and (2) use cluster robust standard errors; (3) - (5) use robust standard errors.
  } 
  \label{tab:question_all_glm} 
\end{table} 

\begin{table}[!htbp] \centering  
\adjustbox{max width=\textwidth}{
\begin{tabular}{@{\extracolsep{5pt}}lccccc} 
\\[-1.8ex]\hline 
\hline \\[-1.8ex] 
 & \multicolumn{5}{c}{\textit{Dependent variable:}} \\ 
\cline{2-6} 
\\[-1.8ex] & \multicolumn{5}{c}{Caseworker answered correctly} \\ 
\\[-1.8ex] & \multicolumn{2}{c}{\textit{Logistic regression}} & \multicolumn{3}{c}{\textit{Logistic regression with random effects}} \\ 
\\[-1.8ex] & (1) & (2) & (3) & (4) & (5)\\ 
\hline \\[-1.8ex] 
 Incorrect chatbot suggestion & $-$0.993$^{***}$ & $-$2.018$^{***}$ & $-$2.288$^{***}$ & $-$1.863$^{***}$ & $-$2.120$^{**}$ \\ 
  & (0.146) & (0.531) & (0.648) & (0.504) & (0.662) \\ 
  & & & & & \\ 
 Correct chatbot suggestion & 1.477$^{***}$ & 0.593 & 0.587 & 0.681 & 0.686 \\ 
  & (0.139) & (0.621) & (0.693) & (0.563) & (0.712) \\ 
  & & & & & \\ 
 Difficulty=medium &  & $-$0.942$^{***}$ & $-$0.979$^{**}$ & $-$0.981$^{**}$ & $-$1.029$^{**}$ \\ 
  &  & (0.244) & (0.318) & (0.373) & (0.395) \\ 
  & & & & & \\ 
 Difficulty=hard &  & $-$2.342$^{***}$ & $-$2.467$^{***}$ & $-$2.412$^{***}$ & $-$2.554$^{***}$ \\ 
  &  & (0.282) & (0.316) & (0.364) & (0.386) \\ 
  & & & & & \\ 
 experiment\_typecontrol:Chatbot accuracy &  &  &  &  &  \\ 
  &  &  &  &  &  \\ 
  & & & & & \\ 
 experiment\_typetreatment:Chatbot accuracy &  & 0.001 & 0.002 & $-$0.001 & 0.001 \\ 
  &  & (0.006) & (0.006) & (0.004) & (0.006) \\ 
  & & & & & \\ 
 Correct chatbot suggestion:Difficulty=medium &  & 0.509 & 0.493 & 0.596 & 0.598 \\ 
  &  & (0.350) & (0.424) & (0.414) & (0.430) \\ 
  & & & & & \\ 
 Correct chatbot suggestion:Difficulty=hard &  & 1.316$^{***}$ & 1.308$^{**}$ & 1.426$^{***}$ & 1.437$^{***}$ \\ 
  &  & (0.398) & (0.414) & (0.404) & (0.420) \\ 
  & & & & & \\ 
 Incorrect chatbot suggestion:Difficulty=medium &  & 0.554 & 0.601 & 0.419 & 0.453 \\ 
  &  & (0.398) & (0.448) & (0.445) & (0.462) \\ 
  & & & & & \\ 
 Incorrect chatbot suggestion:Difficulty=hard &  & 1.329$^{**}$ & 1.462$^{**}$ & 1.190$^{**}$ & 1.306$^{**}$ \\ 
  &  & (0.455) & (0.448) & (0.443) & (0.460) \\ 
  & & & & & \\ 
 Constant & $-$0.488$^{***}$ & 1.099$^{***}$ & 1.154$^{***}$ & 1.127$^{***}$ & 1.190$^{**}$ \\ 
  & (0.081) & (0.262) & (0.312) & (0.335) & (0.374) \\ 
  & & & & & \\ 
\hline \\[-1.8ex] 
Observations & 2,994 & 2,994 & 2,994 & 2,994 & 2,994 \\ 
Log Likelihood & $-$1,747.796 & $-$1,667.018 & $-$1,620.551 & $-$1,653.953 & $-$1,605.092 \\ 
Akaike Inf. Crit. & 3,501.591 & 3,354.037 & 3,263.101 & 3,329.906 & 3,234.184 \\ 
Bayesian Inf. Crit. &  &  & 3,329.149 & 3,395.954 & 3,306.237 \\ 
\hline 
\hline \\[-1.8ex] 
\textit{Note:}  & \multicolumn{5}{r}{$^{*}$p$<$0.05; $^{**}$p$<$0.01; $^{***}$p$<$0.001} \\ 
\end{tabular} }
\caption{
  \small{Results for question-level effects of chatbot suggestions by type (i.e., incorrect chatbot suggestion, control, and correct chatbot suggestion) with logistic regression models. Results are directionally similar to Table~\ref{tab:question_all_filteredset}.
  We present results only for questions that appear in at least one assessment with an incorrect chatbot suggestion. These questions are harder on average, and the average caseworker accuracy in the control group is lower on these questions compared to the full set of assessment questions.
  Column (1) is a simple linear model with binary covariates, denoting whether the chatbot suggestion is incorrect or correct.
  Column (2) presents results for an expanded number of features, including the chatbot accuracy that the caseworker saw, the question difficulty, and interactions with the chatbot suggestion type.
  Column (3) includes person-level random effects, column (4) includes question-level random effects, and column (5) includes both.
  Columns (1) and (2) use cluster robust standard errors; (3) - (5) use robust standard errors.
  }}
  \label{tab:question_all_filteredset_glm}
\end{table} 

\begin{table}[!htbp] \centering 
\adjustbox{max width=\textwidth}{
\begin{tabular}{@{\extracolsep{5pt}}lccccc} 
\\[-1.8ex]\hline 
\hline \\[-1.8ex] 
 & \multicolumn{5}{c}{\textit{Dependent variable:}} \\ 
\cline{2-6} 
\\[-1.8ex] & \multicolumn{5}{c}{Caseworker answered correctly} \\ 
\\[-1.8ex] & \multicolumn{2}{c}{\textit{Logistic regression}} & \multicolumn{3}{c}{\textit{Logistic regression with random effects}} \\ 
\\[-1.8ex] & (1) & (2) & (3) & (4) & (5)\\ 
\hline \\[-1.8ex] 
 Incorrect chatbot suggestion & $-$0.904$^{***}$ & $-$2.127$^{**}$ & $-$2.422$^{**}$ & $-$1.972$^{**}$ & $-$2.263$^{**}$ \\ 
  & (0.184) & (0.695) & (0.828) & (0.666) & (0.847) \\ 
  & & & & & \\ 
 Correct chatbot suggestion & 1.453$^{***}$ & 1.084 & 1.039 & 1.279 & 1.258 \\ 
  & (0.175) & (0.784) & (0.895) & (0.753) & (0.918) \\ 
  & & & & & \\ 
 Difficulty=medium &  & $-$1.064$^{**}$ & $-$1.114$^{**}$ & $-$1.092$^{*}$ & $-$1.145$^{*}$ \\ 
  &  & (0.337) & (0.397) & (0.444) & (0.466) \\ 
  & & & & & \\ 
 Difficulty=hard &  & $-$2.525$^{***}$ & $-$2.657$^{***}$ & $-$2.585$^{***}$ & $-$2.729$^{***}$ \\ 
  &  & (0.364) & (0.394) & (0.435) & (0.458) \\ 
  & & & & & \\ 
 experiment\_typecontrol:Chatbot accuracy &  &  &  &  &  \\ 
  &  &  &  &  &  \\ 
  & & & & & \\ 
 experiment\_typetreatment:Chatbot accuracy &  & $-$0.004 & $-$0.003 & $-$0.007 & $-$0.005 \\ 
  &  & (0.008) & (0.008) & (0.006) & (0.008) \\ 
  & & & & & \\ 
 Correct chatbot suggestion:Difficulty=medium &  & 0.435 & 0.448 & 0.497 & 0.517 \\ 
  &  & (0.498) & (0.573) & (0.562) & (0.580) \\ 
  & & & & & \\ 
 Correct chatbot suggestion:Difficulty=hard &  & 1.309$^{*}$ & 1.341$^{*}$ & 1.386$^{*}$ & 1.430$^{*}$ \\ 
  &  & (0.562) & (0.561) & (0.549) & (0.567) \\ 
  & & & & & \\ 
 Incorrect chatbot suggestion:Difficulty=medium &  & 1.230$^{*}$ & 1.355$^{*}$ & 1.119$^{*}$ & 1.242$^{*}$ \\ 
  &  & (0.503) & (0.572) & (0.569) & (0.588) \\ 
  & & & & & \\ 
 Incorrect chatbot suggestion:Difficulty=hard &  & 1.876$^{**}$ & 2.050$^{***}$ & 1.772$^{**}$ & 1.935$^{**}$ \\ 
  &  & (0.604) & (0.577) & (0.572) & (0.591) \\ 
  & & & & & \\ 
 Constant & $-$0.446$^{***}$ & 1.308$^{***}$ & 1.376$^{***}$ & 1.331$^{**}$ & 1.403$^{**}$ \\ 
  & (0.097) & (0.346) & (0.388) & (0.405) & (0.444) \\ 
  & & & & & \\ 
\hline \\[-1.8ex] 
Observations & 1,890 & 1,890 & 1,890 & 1,890 & 1,890 \\ 
Log Likelihood & $-$1,121.282 & $-$1,060.400 & $-$1,036.278 & $-$1,052.887 & $-$1,027.480 \\ 
Akaike Inf. Crit. & 2,248.565 & 2,140.800 & 2,094.557 & 2,127.774 & 2,078.961 \\ 
Bayesian Inf. Crit. &  &  & 2,155.544 & 2,188.762 & 2,145.493 \\ 
\hline 
\hline \\[-1.8ex] 
\textit{Note:}  & \multicolumn{5}{r}{$^{*}$p$<$0.05; $^{**}$p$<$0.01; $^{***}$p$<$0.001} \\ 
\end{tabular} }
  \caption{
  Results for question-level effects of chatbot suggestions by type (i.e., incorrect chatbot suggestion, control, and correct chatbot suggestion) with logistic regression models.
  W present results only for questions that appear in at least one assessment with an incorrect chatbot suggestion and only for participants who passed the attention check.
  Results are directionally similar to Table~\ref{tab:question_passed_filteredset}.
  Column (1) is a simple linear model with binary covariates, denoting whether the chatbot suggestion is incorrect or correct.
  Column (2) presents results for an expanded number of features, including the chatbot accuracy that the caseworker saw, the question difficulty, and interactions with the chatbot suggestion type.
  Column (3) includes person-level random effects, column (4) includes question-level random effects, and column (5) includes both. Results are directionally similar to column (2).
  Columns (1) and (2) use cluster robust standard errors;  (3) - (5) use robust standard errors.
  } 
  \label{tab:question_passed_filteredset_glm}
\end{table}

\begin{table}[!htbp] \centering 
\adjustbox{max width=\textwidth}{
\begin{tabular}{@{\extracolsep{5pt}}lcc} 
\\[-1.8ex]\hline 
\hline \\[-1.8ex] 
 & \multicolumn{2}{c}{\textit{Dependent variable:}} \\ 
\cline{2-3} 
\\[-1.8ex] & \multicolumn{2}{c}{Caseworker answered correctly} \\ 
\\[-1.8ex] & (1) & (2)\\ 
\hline \\[-1.8ex] 
 Question position & $-$0.003$^{*}$ & $-$0.003$^{*}$ \\ 
  & (0.001) & (0.001) \\ 
  & & \\ 
 Chatbot suggestions & 0.129$^{***}$ & 0.097$^{*}$ \\ 
  & (0.038) & (0.045) \\ 
  & & \\ 
 Question position:Chatbot suggestions & 0.004$^{**}$ & 0.005$^{***}$ \\ 
  & (0.001) & (0.002) \\ 
  & & \\ 
 Constant & 0.548$^{***}$ & 0.572$^{***}$ \\ 
  & (0.032) & (0.036) \\ 
  & & \\ 
\hline \\[-1.8ex] 
Observations & 5,625 & 3,555 \\ 
R$^{2}$ & 0.040 & 0.046 \\ 
Adjusted R$^{2}$ & 0.039 & 0.045 \\ 
Residual Std. Error & 0.467 (df = 5621) & 0.464 (df = 3551) \\ 
F Statistic & 77.057$^{***}$ (df = 3; 5621) & 56.950$^{***}$ (df = 3; 3551) \\ 
\hline 
\hline \\[-1.8ex] 
\textit{Note:}  & \multicolumn{2}{r}{$^{*}$p$<$0.05; $^{**}$p$<$0.01; $^{***}$p$<$0.001} \\ 
\end{tabular} }
  \caption{Regression results for the effect of question position on caseworker accuracy by whether caseworkers saw chatbot suggestions or not. Column (1) presents results for all participants. Column (2) only includes participants who passed the attention check. Question position is positively associated with increases in accuracy for caseworkers in the treatment group who see chatbot suggestions ($p<0.01$) and associated with reductions in accuracy for those in the control group ($p<0.05$). Both models in columns (1) and (2) use cluster robust standard errors.} 
  \label{tab:regression_results_accuracy_X_position} 
\end{table} 

\end{singlespacing}

\end{document}